\PassOptionsToPackage{unicode}{hyperref}
\PassOptionsToPackage{hyphens}{url}
\PassOptionsToPackage{dvipsnames,svgnames,x11names}{xcolor}
\documentclass[
  american,
  11pt,
  a4paper,
]{article}
\usepackage{xcolor}
\usepackage[margin=25mm]{geometry}
\usepackage{amsmath,amssymb}
\usepackage{iftex}
\ifPDFTeX
  \usepackage[T1]{fontenc}
  \usepackage[utf8]{inputenc}
  \usepackage{textcomp} 
\else 
  \usepackage{unicode-math} 
  \defaultfontfeatures{Scale=MatchLowercase}
  \defaultfontfeatures[\rmfamily]{Ligatures=TeX,Scale=1}
\fi
\usepackage{lmodern}
\ifPDFTeX\else
\fi
\IfFileExists{upquote.sty}{\usepackage{upquote}}{}
\IfFileExists{microtype.sty}{
  \usepackage[]{microtype}
  \UseMicrotypeSet[protrusion]{basicmath} 
}{}
\makeatletter
\@ifundefined{KOMAClassName}{
  \IfFileExists{parskip.sty}{%
    \usepackage{parskip}
  }{
    \setlength{\parindent}{0pt}
    \setlength{\parskip}{6pt plus 2pt minus 1pt}}
}{
  \KOMAoptions{parskip=half}}
\makeatother
\usepackage{longtable,booktabs,array}
\usepackage{calc} 
\usepackage{etoolbox}
\makeatletter
\patchcmd\longtable{\par}{\if@noskipsec\mbox{}\fi\par}{}{}
\makeatother
\IfFileExists{footnotehyper.sty}{\usepackage{footnotehyper}}{\usepackage{footnote}}
\makesavenoteenv{longtable}
\ifLuaTeX
\usepackage[bidi=basic,shorthands=off]{babel}
\else
\usepackage[bidi=default,shorthands=off]{babel}
\fi
\ifLuaTeX
  \usepackage{selnolig} 
\fi
\providecommand{\tightlist}{%
  \setlength{\itemsep}{0pt}\setlength{\parskip}{0pt}}
\usepackage{graphicx}
\usepackage{needspace}
\usepackage{newunicodechar}
\newunicodechar{→}{\ensuremath{\rightarrow}}
\newunicodechar{−}{\ensuremath{-}}
\PassOptionsToPackage{mathlines}{lineno}
\usepackage{xurl}
\usepackage{float}
\usepackage{fvextra}
\DefineVerbatimEnvironment{Highlighting}{Verbatim}{commandchars=\\\{\},breaklines,breakanywhere,fontsize=\footnotesize}
\usepackage{setspace}
\usepackage{lineno}
\usepackage{pdflscape}
\AtBeginEnvironment{longtable}{\small\singlespacing}
\usepackage{fancyhdr}
\AtBeginDocument{\hypersetup{pdfauthor={Yanzhe Zhao, Yongzhang Ji}}}

\usepackage{ragged2e}
\usepackage{bookmark}
\hypersetup{
  pdftitle={History-bearing thalamocortical signals reconstruct and redirect later decisions in a circuit model},
  pdflang={en-US},
  colorlinks=true,
  linkcolor={blue},
  filecolor={Maroon},
  citecolor={Blue},
  urlcolor={blue},
  pdfcreator={LaTeX}}

\makeatletter
\renewcommand{\paragraph}{\@startsection{paragraph}{4}{\z@}%
  {1.8ex \@plus 0.5ex \@minus 0.2ex}%
  {-0.85em}%
  {\normalfont\normalsize\bfseries}}
\let\orig@paragraph\paragraph
\renewcommand{\paragraph}[1]{\orig@paragraph{#1.}}
\makeatother

\title{History-bearing thalamocortical signals reconstruct and redirect
later decisions in a circuit model}
\author{Yanzhe Zhao\textsuperscript{1} \and Yongzhang Ji\textsuperscript{2}}
\date{}

\newcommand{\mainfigurepage}[3]{%
  \clearpage
  \pdfbookmark[2]{Figure #1. #2}{main-figure-#1}%
  \hypertarget{main-figure-link-#1}{}%
  \begin{minipage}[t]{\linewidth}
    \noindent\includegraphics[width=\linewidth]{figures/Figure_#1.pdf}\par
    \vspace{0.6em}
    \begin{singlespace}
    \small\textbf{Figure #1. #2.}\enspace #3\par
    \end{singlespace}
  \end{minipage}\par
}

\begin{document}
\maketitle
\vspace{-8mm}
\thispagestyle{fancy}
{\centering\small
\begin{singlespace}
\textsuperscript{1}School of Future Technology, Tianjin University, Tianjin, China\par
\textsuperscript{2}Faculty of Psychology, Southwest University, Chongqing, China\par
\vspace{0.4em}
Preprint contact: Yanzhe Zhao, \href{mailto:justinzhaoyz@outlook.com}{justinzhaoyz@outlook.com}, \href{https://orcid.org/0009-0003-5342-6611}{ORCID 0009-0003-5342-6611}\par
Yongzhang Ji: \href{mailto:royceji0605@email.swu.edu.cn}{royceji0605@email.swu.edu.cn}, \href{https://orcid.org/0009-0006-4111-2000}{ORCID 0009-0006-4111-2000}\par
\end{singlespace}
}
\vspace{0.35em}

\begingroup\small
\subsection{Abstract}\label{abstract}

Recent sensory experience can help resolve ambiguous evidence, but it can also bias decisions when current evidence points in a different direction. We asked how thalamocortical circuits retain information about previous stimuli while regulating its influence across successive decisions. Reanalysis of published data from adult male mice showed that suppression of the posterior parietal cortex (PPC) preferentially improved sensory judgments when past and present stimuli favored opposing choices, consistent with reduced history weighting. Suppression of the pulvinar (PUL) showed a similar, more variable effect. To determine whether thalamic feedback carries information about specific past stimuli, we replayed recorded PUL-to-PPC inputs from opposing sensory histories into trained circuits initialized in identical neural states. The recorded inputs reconstructed distinct history-dependent responses. In sequential decisions, exchanging thalamic inputs redirected later choices even after the circuits correctly judged an intervening strong stimulus. Much of the difference between opposing histories was captured by a single direction in PUL population activity, which could generate effective cortical inputs under the model's firing-rate constraints. Redistributing thalamic input among individual cortical neurons selectively weakened the influence of history on later choices in several circuits, even when the total input to each cortical cell class and the magnitude of the history-dependent current difference were preserved. Strong sensory evidence greatly reduced the behavioral influence of history while leaving substantial history-dependent state differences intact. Under ambiguity, these differences were more strongly expressed through neuronal firing-rate thresholds and became more closely aligned with the decision readout. Small shifts in cortical states relative to neuronal firing thresholds produced modest changes in historical influence. Briefly weakening the history-dependent component of recurrent input consistently reduced the influence of history on later choices while preserving current sensory judgments. Recurrent propagation also improved predictions of post-stimulus output trajectories relative to passive decay. Together, our findings identify three complementary operations supporting history-guided choice: thalamic feedback determines which past is reconstructed, recurrent dynamics sustain its influence across decisions, and current sensory evidence regulates when it affects behavior.

\subsection{Significance Statement}\label{significance-statement}

Past experience provides a useful basis for decisions when sensory evidence is uncertain, yet it can interfere with accurate judgments when current evidence is clear. How neural circuits preserve this information while regulating its influence remains a central question. In published mouse behavioral data, suppressing the posterior parietal cortex reduced the accuracy cost associated with conflicting sensory history, with a similar pattern following pulvinar suppression. In trained thalamocortical circuits, we found that input from the pulvinar to posterior parietal cortex could reconstruct information about specific past stimuli and redirect subsequent choices. The reconstructed history remained available even after a correct judgment about a new stimulus. Strong sensory evidence temporarily reduced its influence on behavior, whereas ambiguity allowed the retained neural activity to guide choice again. Recurrent interactions helped sustain this influence across decisions. These findings provide a circuit-level account of how past experience can remain available for future behavior while its influence is regulated by current sensory demands.

\endgroup

\subsection{Introduction}\label{introduction}

Recent sensory experience provides a context for interpreting new information and making decisions. When current sensory evidence is ambiguous, information about previously encountered stimuli can help resolve uncertainty. However, when the current stimulus clearly favors a different interpretation, relying on the same experience can introduce a systematic bias. Recent stimuli can even bias perception and choice in opposite directions (Fritsche et al., 2017). In rodents, perturbations of the posterior parietal cortex (PPC) alter the influence of recent sensory inputs and actions on subsequent decisions (Akrami et al., 2018; Hwang et al., 2017). These observations raise a fundamental question about the organization of neural circuits: how can information about the past remain available across successive decisions while its influence on behavior is adjusted to the demands of the present?

One possible mechanism for preserving information across successive decisions is the recurrent interaction between the cortex and higher-order thalamus. Thalamocortical feedback can sustain cortical preparatory activity (Guo et al., 2017), regulate functional connectivity between cortical regions, and support transitions between task representations (Schmitt et al., 2017; Rikhye et al., 2018). The pulvinar (PUL), in particular, coordinates communication between cortical areas as attentional demands change (Saalmann et al., 2012). Studies in mice have also demonstrated contributions of the thalamus to sensory selection and to different phases of working memory (Wimmer et al., 2015; Bolkan et al., 2017).

Hosford et al. (2026) linked these thalamocortical interactions to the use of sensory history. Transient suppression of either PPC or PUL impaired decisions that depended on previous stimuli while improving judgments of the current stimulus, suggesting that the circuit contributes to the influence of past experience on choice. There are at least two ways in which thalamic feedback could support this function. It could provide a general drive that helps cortical neurons maintain an existing representation of sensory history. Alternatively, the feedback signal itself could carry information about which stimuli were previously encountered, thereby helping to reconstruct a particular historical representation.

Distinguishing these possibilities requires controlling the content of the returning thalamic signal. If inputs recorded from different sensory histories can generate different history-dependent responses in otherwise identical circuits, the inputs themselves must carry information capable of specifying those responses. A further test is whether the reconstructed history can persist through an accurate judgment about a new sensory stimulus and influence a subsequent choice.

Even if thalamic feedback carries information about previous stimuli, the circuit faces a further computational challenge. It must preserve this information while processing new sensory evidence, allowing history to influence a later decision without overwhelming a current judgment. Work on latent working memory has shown that information can remain available within a neural system even when it has little immediate effect on behavior, and can subsequently be reactivated to guide action (Rose et al., 2016; Wolff et al., 2017). Persistent neural activity and activity-silent processes can interact to produce serial biases in recurrent circuits (Masse et al., 2019; Barbosa et al., 2020).

The geometry of neural population activity provides a possible mechanism for regulating the influence of retained information. For example, two past experiences may produce substantially different patterns of population activity while generating nearly identical behavioral outputs. This can occur when the difference between the two activity patterns lies in directions that contribute little to the neural readout governing the current behavior (Kaufman et al., 2014; Muhle-Karbe et al., 2021). Changes in the organization of population activity can subsequently make the same information more relevant to behavior. Indeed, stable population codes can coexist with heterogeneous dynamics in individual neurons, and rotations of population activity can reduce interference between memories and new sensory inputs (Murray et al., 2017; Libby and Buschman, 2021).

These observations suggest that the retention and behavioral expression of sensory history may depend on different aspects of circuit dynamics. In a thalamocortical loop, the influence of history may depend both on the information carried by returning thalamic input and on how cortical and thalamic neurons transform that input as current sensory evidence changes. This raises the question of how current sensory input changes the sensitivity of neuronal firing to retained history, and how recurrent interactions preserve and transform that history across successive decisions.

Here, we combine a reanalysis of published mouse behavioral data with targeted interventions in the trained thalamocortical circuit models of Hosford et al. (2026). We first examine when the influence of sensory history helps or impairs a decision, establishing the behavioral conditions under which regulating that influence is particularly important. We then ask whether returning PUL-to-PPC input carries information about specific past stimuli. By replacing this input while controlling the initial state of the receiving circuit, we test whether different thalamic signals can reconstruct different historical representations and redirect later choices.

Having established that a reconstructed history can survive an intervening sensory judgment, we investigate which features of the returning input support this function. We examine how historical information is distributed across neurons, whether it can be represented by a low-dimensional pattern of activity, and which alternative input patterns can be generated through the existing thalamocortical projection.

Finally, we investigate how the receiving circuit regulates the influence of the reconstructed history. By varying sensory evidence independently of trial position, examining neuronal activation thresholds, and perturbing recurrent interactions, we identify how retained history-dependent neural activity contributes to behavior under different sensory demands. A complementary conductance-based analysis examines how history-dependent synaptic input can reconstruct an intrinsic neuronal state (Extended Theory ET7). Together, these experiments distinguish the processes that determine which history is reconstructed, how its influence persists across decisions, and when it is expressed in behavior.

\subsection{Materials and Methods}\label{materials-and-methods}

\subsubsection{Published behavioral and model data}\label{published-data-and-source-provenance}

We used published behavioral data and trained thalamocortical circuit models from Hosford et al. (2026) to investigate complementary aspects of history-guided decision-making. The behavioral data allowed us to determine when information about previous sensory stimuli helps or impairs an animal's current choice. The trained circuit models allowed us to manipulate specific neural states and pathways, thereby investigating the mechanisms through which sensory history can influence subsequent decisions. The original study used adult male mice; animal care, experimental procedures, and ethical approvals are described in the original article.

In the behavioral task, mice judged auditory stimuli containing mixtures of high-pass and low-pass tone pips. On regular trials, the rewarded choice depended on the relative abundance of the two types of pips in the current stimulus. On history probes, the two types were equally represented, and the rewarded choice instead depended on the recent sensory history. This distinction allowed us to examine judgments based on current sensory evidence separately from judgments that required information about preceding stimuli.

The PPC cohort contains 14,611 observations from nine mice across 83 sessions. The PUL cohort contains 14,189 observations from ten mice across 80 sessions. The TRN-to-PUL cohort contains 14,824 observations from six mice across 50 sessions. We analyzed the cohorts separately.

The source study describes a 500-ms intertrial intervention delivered with a probability of 33\%. Because intervention sequences in the released tables vary across animals, we used within-session row order to index sequential trials for carryover analyses. Figure 1-1 shows the intervention sequences and the resulting overlap between exposure conditions.

Reconstructing trial order was necessary because an intervention could influence both the current decision and the neural or behavioral conditions under which subsequent decisions were made. We therefore examined the temporal organization of intervention exposure before interpreting differences between intervention and control trials.

\subsubsection{Historical weighting in mouse choices}\label{behavioral-estimands-and-sequential-sensitivity-analyses}

To quantify how current and previous stimuli jointly influence behavior, we represented both sources of sensory evidence on a common signed scale. Positive values favor the high-pass category, negative values favor the low-pass category, and values near zero indicate weak or balanced evidence.

Stimulus evidence is \(s_t=2p_t-1\), where \(p_t\) is the high-pass pip fraction. Within each session, \(h_t=\frac15\sum_{k=1}^{5}s_{t-k}\) is the preceding five-trial mean. The product \(s_th_t\) identifies whether current and historical evidence favor the same choice. Trials with \(s_th_t>0\) were classified as agreement trials, whereas trials with \(s_th_t<0\) were classified as conflict trials. History probes contained equal numbers of high-pass and low-pass pips and were designated as probe trials in the published dataset. Accuracy was determined from the recorded trial feedback. The conflict analysis included regular trials for which current and historical evidence each favored a particular choice, the animal made a binary response, and the recorded target agreed with the direction of current sensory evidence. The choice models included regular trials with recorded binary choices and targets, together with the preceding trial's intervention condition and choice. Both correct and incorrect trials enter these analyses. The evidence-magnitude comparison also includes eligible balanced trials (Figure 1-3).

An intervention could improve sensory judgments by increasing sensitivity to the current stimulus, reducing the influence of past stimuli, or changing the animal's general preference for one response. To distinguish these possibilities, we fitted a logistic choice model for each mouse, with session-specific intercepts and separate terms for current evidence, historical evidence, and response bias:

\[
\operatorname{logit}P(Y_t=1)=a_{\mathrm{session}}+\beta_s s_t+\beta_hh_t
+L_t(\beta_L+\beta_{sL}s_t+\beta_{hL}h_t). \tag{1}
\]

In this model, \(\beta_s\) and \(\beta_h\) quantify the dependence of choice on current and historical evidence, respectively. The interaction coefficients \(\beta_{sL}\) and \(\beta_{hL}\) quantify how the intervention changes these dependencies. In particular, a negative \(\beta_{hL}\) indicates that the intervention reduces the influence of previous stimuli on choice.

The indicator \(L_t\) marks current intervention. Intervention contrasts are standardized over the same observed covariates.

Because changes in response bias, sensory sensitivity, and history weighting can jointly influence choice probabilities, we used a Shapley decomposition to estimate how much each component contributed to the additional accuracy benefit observed on conflict trials. The contribution of each component was averaged over all six possible orders in which the three components could be introduced into the fitted model.

To test whether the inferred contribution of history depended on assuming a linear relationship between current sensory evidence and the log odds of choice, we also fitted a more flexible sensory response using a cubic B-spline with knots at -0.5, 0, and 0.5, together with its interactions with intervention (Figure 1-2).

Recent intervention exposure may influence the state of the circuit at the beginning of a new trial. Similarly, recent choices and outcomes may affect subsequent behavior. We therefore extended the base model to account for these potential carryover effects.

The carryover model included two mean-centered covariates: intervention on the preceding trial and average intervention exposure over the preceding second to fifth trials. Each covariate was allowed to modify response bias, sensory sensitivity, history weighting, and the effects of current intervention on these three components. We also included mean-centered previous choice and correctness, together with their interactions with current intervention. The relationship between sensory evidence and behavior may also change gradually during a session, for example as the animal's engagement or response tendencies change. To account for such changes, we introduced an additional model containing within-session drift terms.

The drift model adds linear and quadratic session-position terms for bias, sensory slope, and history slope. Position is scaled to \([-1,1]\), with polynomial terms centered within each session. All three models use the same 13,539 PPC, 13,281 PUL, and 13,520 TRN observations. The history-by-intervention coefficient, \(\beta_{hL}\), estimates the change in history weighting associated with current intervention at the mean levels of the preceding exposure and behavioral covariates. When an unpenalized fit was unstable, all mice in that cohort and specification were fitted with the same \(10^{-4}\) ridge penalty on non-intercept coefficients. A separate Firth-penalized analysis examines the comparison after five control trials, where four mice have only treated observations (Figure 1-1; Firth, 1993).

We treated individual mice as the independent sampling units for population-level inference, accounting for the shared biological and experimental characteristics of trials from the same animal.

Intervals for mouse behavior use 100,000 resamples of mice within each cohort. Exact signed-rank tests compare effects across mice, with Holm adjustment across intervention cohorts. Figure 1-1 provides the session and trial counts and the definitions of the carryover models.

\subsubsection{Circuit architecture, dynamics and decision readout}\label{inherited-circuit-model-and-native-choices}

The trained models represent recurrent interactions among three neural populations: 400 units in PPC, 70 units in PUL, and 30 units in the thalamic reticular nucleus (TRN). We analyzed all nine published trained circuit models: Gtune20, Gtune22, Gtune23, Gtune24, Gtune25 (hereafter G20, G22, G23, G24 and G25), and Sussano0--3 (hereafter S0--S3). These circuits were trained to make auditory decisions that depend on both current sensory evidence and recent stimulus history. We analyzed them as a fixed collection of nine published circuit models with partially documented training histories, preserving their trained connections, integration factors, and decision readouts while examining their responses to newly generated histories.

Each unit has an internal state that evolves over time in response to its previous state, recurrent input from other units, and external sensory input. The firing rate of each unit is obtained by applying its activation function to this internal state. The decision output is then calculated as a weighted combination of the firing rates of all 500 units.

State and output obey

\[
x_{t+1}=(I-A)x_t+A\{Wr(x_t)+W_{\mathrm{in}}u_t\},
\qquad y_t=w^\top r(x_t)+b_0. \tag{2}
\]

The diagonal matrix \(A\) contains the learned integration factors. The PPC and PUL rate function is \(\operatorname{clip}(x,0,20)\), and the TRN rate function is \(\operatorname{clip}(10(x-9),0,20)\).

Thus, a change in internal state produces a change in firing rate only to the extent permitted by the unit's activation function. The lower rectification threshold and upper saturation bound are particularly relevant to our later analyses of how retained state differences influence the decision output.

The network was integrated in 5-ms time steps. Each trial lasted 1 s and was followed by a 1-s intertrial interval, corresponding to 200 time steps each. Twenty 15-ms tone pips occupied time steps 50--109, and the response window extended from time step 120 to 199.

During training, the circuit made five successive nonambiguous sensory judgments followed by a sixth trial with balanced sensory input. The sixth decision therefore required information about the preceding stimuli. To examine whether a reconstructed history could survive an accurate intervening sensory judgment, we selected fifth-trial mixtures containing 0, 2, 18, or 20 positive pips. These four strong mixtures were selected before testing new histories because some circuits exhibited weak or nonmonotonic responses at intermediate stimulus strengths (Figure 3-1). Each circuit was subsequently evaluated on newly generated histories.

\subsubsection{Separating the information in the initial neural state from the information carried by thalamic input}\label{causal-contrast-between-a-stored-state-and-an-arriving-message}

Different sensory histories can influence a later choice through differences already present in the neural state, through differences in subsequent recurrent input, or through both mechanisms. To separate these contributions, we generated pairs of trajectories that experienced opposing sensory histories but received identical subsequent external stimuli. We then quantified how strongly the difference between their histories was expressed in the decision output.

For a matched history pair \(+\) and \(-\), all subsequent stimuli are
identical. In response window \(R\), let

\[
m_R^\pm=|R|^{-1}\sum_{t\in R}y_t^\pm,
\qquad H_R=\frac{m_R^+-m_R^-}{2}. \tag{3}
\]

We refer to \(H_R\) as the historical contrast. It measures half the difference between the mean decision outputs produced by the two histories during the response window. A positive value indicates that the original positive history produces a higher output than the original negative history. A reduction toward zero indicates weaker differentiation between the histories, whereas a negative value indicates a reversal of their relative influence on the decision.

Our main question was whether the input returning from PUL to PPC itself carries information that can determine the identity of a subsequent historical response. We therefore manipulated the current delivered through this specific projection while leaving the remainder of the circuit intact.

The manipulated pathway supplies

\[
b_t=W_{C,T}r_T(x_t), \tag{4}
\]

the PUL current arriving at PPC.

We first recorded the current delivered through the PUL-to-PPC pathway during the natural evolution of each history. During a subsequent simulation, we temporarily replaced the endogenous current along this pathway with one of the recorded input patterns, while allowing all other connections to operate normally.

We considered three replay conditions. In own-history replay, each recipient circuit received the input recorded from its original history. With intact recipient states, this reproduces the natural trajectory. In opposite-history replay, each circuit received the input recorded from the opposing history. In pair-mean replay, both circuits received the average of the two recorded inputs.

To distinguish the contribution of this input from the contribution of information already present in the recipient, we also manipulated the initial neural states. We separately averaged the PPC state, the non-PPC state, or the complete 500-dimensional network state across the two histories. Complete-state averaging gives both recipient circuits exactly the same initial state. Because they also receive identical external stimuli, any subsequent difference between their trajectories must arise from the difference in the replayed input.

Measurements of neural decay reveal the timescales over which activity persists within a circuit. Reconstructing a representation after a localized perturbation also depends on how information is distributed and transferred between interacting populations. Extended Theory ET6 provides a mathematical illustration of this distinction: reciprocal circuits with identical eigenvalue spectra and identical decay curves when either population is held fixed can exhibit different recovery trajectories when their population states are erased in sequence. This construction extends earlier work on the relationship between isospectrality and responses to perturbation (van den Berg et al., 2014). Our input-replay experiments examine the corresponding reconstruction process directly by controlling the information delivered through a particular neural pathway.

\subsubsection{History replay across successive choices}\label{frozen-history-and-sequential-experiments}

A base history is one underlying stimulus sequence from which the mirrored history pair and the corresponding stimulus conditions are constructed.

\textbf{History reconstruction assay.} To determine whether returning thalamic input can reconstruct sensory history, we generated 96 new base histories, each consisting of a five-trial stimulus sequence paired with its mirror image. The two histories were followed by an identical balanced sixth-trial probe. Input replay began after the fifth trial, at global time step 1800, and lasted 100, 250, 500, or 1,000 ms (20, 50, 100, or 200 time steps). Combining replay durations, input substitutions, and recipient-state averaging yielded 67 experimental conditions.

\textbf{Sequential decision assay.} To examine whether a reconstructed history remains available through a new sensory judgment, we generated 48 new four-trial base histories. Each sequence was drawn subject to the condition \(|\sum(q-10)|>10\), where \(q\) is the number of high-pass pips in a stimulus, and was then paired with its mirror image. This ensured that the paired histories retained opposite cumulative evidence signs after the same strong fifth stimulus. Each base history was combined with fifth-trial stimuli containing 0, 2, 18, or 20 high-pass pips and both history signs, yielding 384 sequences per condition in each circuit. A balanced sixth stimulus then probed the influence of the preceding history.

Replay began at global time step 1400 and lasted 500 or 1,000 ms, ending before the fifth stimulus. The fifth- and sixth-trial response windows comprised time steps 1720--1799 and 2120--2199, respectively. Combining replay durations, input substitutions, and initial recipient-state manipulations yielded 25 experimental conditions.

We evaluated the resulting decisions using both the mean response and the original model's decision thresholds. The sign of the response-window mean indicates the direction favored by the output, whereas decision commitment requires the output to reach the threshold corresponding to that choice (\(\pm0.75\)) without crossing the opposite threshold. The two measurements distinguish the direction of a decision from the strength and consistency required to meet the model's commitment criterion.

Base-history intervals use 10,000 resamples that keep the four stimuli and both signs together.

\subsubsection{Separating the strength, spatial pattern and dimensionality of thalamic input}\label{source-amplitude-spatial-pattern-and-matched-energy-geometry}

Information about previous stimuli could be carried by changes in the overall amount of input delivered to PPC, by changes in how this input is distributed across PPC neurons, or by both. To examine these possibilities separately, we decomposed the nonnegative PUL-to-PPC current into an overall amplitude and a normalized spatial pattern.

For nonnegative arrivals with positive mean, write

\[
b_t=g_tq_t,\qquad g_t=400^{-1}\mathbf1^\top b_t,
\qquad 400^{-1}\mathbf1^\top q_t=1. \tag{5}
\]

The scalar \(g_t\) represents the mean current delivered to the 400 PPC neurons, while \(\mathbf q_t\) describes its distribution across those neurons. In amplitude exchange, the recipient receives the amplitude recorded from the opposing history together with its original spatial pattern. In pattern exchange, it receives the spatial pattern recorded from the opposing history while retaining its original total current. These manipulations allow us to test whether either feature of the input can independently influence the reconstructed history. The Euclidean-norm alternative and the zero-current convention are described in Extended Theory.

Although the arriving current has 400 coordinates, differences between opposing histories may be organized along a much smaller number of directions. We therefore asked whether a single pattern of differential input could capture most of the variation associated with sensory history and whether this reduced signal would remain functionally effective.

We fitted a unit axis \(v\) to the leading eigenvector of the uncentered second moment of the arriving current differences in 12 histories. For pair mean \(\mu\) and difference \(d\), the projection is \(d_1=(v^\top d)v\).

The projection preserves the component of the input difference along the estimated history axis. Its scalar coefficient describes how strongly that spatial pattern is expressed at a particular time.

We retained the higher-dimensional pair-mean current as a common background for the projected historical difference. Because the projection can produce a negative current at a target receiving little background input, we examined three ways of preserving nonnegative arrival currents. Increasing the common background current preserves an exactly rank-one difference. At each receiving neuron \(i\), both histories receive the same minimal increment \(\max(|d_{1,i}|/2-\mu_i,0)\). Coordinatewise clipping preserves the original background while adjusting the projected difference at individual targets. Global attenuation also preserves the rank-one direction, with its magnitude determined by the most restrictive target bound.

We evaluated the reduced input on eight independent base histories combined with the four strong fifth-trial stimuli.

Projecting the history-dependent input onto a single direction can change both its spatial organization and its magnitude. Consequently, any reduction in sensory performance could arise either because information in the discarded dimensions is functionally important or because the overall differential current has become weaker. To distinguish these possibilities, we used 24 new base histories to compare graded restoration of the discarded components with rescaling of the reduced signal to recover the original differential-current magnitude. We performed this comparison in G23 and S2, which showed impaired sensory performance after dimensional reduction, and in G24, which preserved its performance. At each replay time, bisection determined the scalar projection before targetwise clipping so that the differential-current norm matched that of the full input at the original pair mean (Figure 4-1).

The leading direction characterizes how historical differences are organized in the naturally occurring input. We next examined its functional significance by comparing its effects on the receiving circuit with those of alternative directions of the same differential-current magnitude.

The matched-energy comparison uses 24 new base histories per circuit. We sampled 16 Gaussian directions within the subspace of PPC neurons receiving input through the trained PUL projection and orthogonalized them against the estimated history direction \(v\) before clipping. The alternatives keep the temporal sign of the learned scalar coefficient, with their magnitudes matched to the natural difference under the same clipping and norm-matching rule. Complete-state averaging gives every replay the same recipient start. Clipping can turn a direction away from its initial axis. Figure 4-2 relates the resulting geometry to its effect on history. We compare the learned mode with the mean absolute historical effect of the alternatives,

\[
G=H_{\mathrm{learned}}-\frac1{16}\sum_{j=1}^{16}|H_{\mathrm{random},j}|. \tag{6}
\]

Throughout these analyses, differential-current energy denotes the squared Euclidean norm of the difference between the two arriving current vectors. Matching this quantity controls the magnitude of the history-dependent input while allowing its spatial allocation to vary. A positive \(G\) indicates that the estimated history direction produces a stronger consistent historical contrast than the average magnitude produced by the sampled alternatives.

The absolute value makes Equation 6 independent of random-axis orientation. We first average the signed contrast over contexts within each condition. Bootstrap intervals resample base histories with the sampled directions fixed.

\subsubsection{Redistributing thalamic input across cortical neurons while preserving aggregate excitatory and inhibitory target drive}\label{fine-target-redistribution-at-fixed-excitatory-and-inhibitory-aggregate-drive}

Changing the spatial pattern of thalamic input can also change the total current delivered to excitatory and inhibitory cortical neurons. Such changes could influence the circuit through differences in the overall recruitment of these two populations. To isolate the contribution of finer spatial organization, we redistributed the incoming current among neurons of the same target class while preserving the aggregate input to each class and the magnitude of the history-dependent current difference.

PUL sends nonnegative excitatory currents to both PPC recipient classes. Excitatory and inhibitory here identify the receiving neurons.

For each circuit, we combined 24 base histories with four strong fifth-trial stimuli and both history signs. Within the populations of 280 excitatory and 120 inhibitory PPC neurons, we formed disjoint triplets separately from the targets in each class that received PUL input. On each sweep, we randomly permuted the eligible targets in each class and grouped consecutive targets into triplets. Any eligible targets left over after forming complete triplets kept their currents for that sweep. The component of a triplet's current-difference vector along the all-ones axis determines the sum of its three entries. Rotating the vector around this axis preserves both that sum and its squared Euclidean norm while redistributing the current difference among individual neurons. Each rotation was applied when the resulting difference satisfied the nonnegativity bounds \([-2\mu,2\mu]\). Each of the four redistribution maps used 64 successive sweeps. A map-specific seed fixed the sequence of triplets and angles across all times and contexts.

For each class \(G\), \(\sum_{i\in G}d'_i=\sum_{i\in G}d_i\) and \(\|d'_G\|^2=\|d_G\|^2\). Because the pair-mean input is fixed, these constraints preserve the total current delivered to each target class in each history and the combined squared current norm across the two histories. The remaining change concerns the allocation of current among individual neurons within a class. The attained redistribution depends on the feasible rotations for each time and context (Figure 5-1).

The paired recipients start in an identical state and receive the replay at time steps 1400--1599. We measured both the subsequent historical contrast and the intervening sensory decision to determine whether fine-scale redistribution selectively affects the use of sensory history or more generally disrupts the circuit's decision-making ability.

The historical loss is the own-replay contrast minus the mean contrast across the four maps. Decision commitment and absolute contrasts describe current preservation and the magnitude of the later effect. Intervals use 10,000 paired resamples of the 24 base histories, keeping their stimuli, signs, and maps together.

\subsubsection{Alternative input patterns that can be generated by bounded pulvinar activity}\label{source-reachable-pulvinar-messages-and-exactly-matched-geometry}

Direct manipulation of the currents arriving at PPC allows us to characterize the receiving circuit's sensitivity to different input patterns. In the intact model, however, these currents are generated by a population of 70 PUL neurons and transmitted through a fixed projection to PPC. The limited firing-rate range of the PUL neurons therefore restricts the input patterns that the projection can generate. To investigate this additional constraint, we manipulated the firing rates of the 70 PUL neurons and delivered the resulting currents to PPC through the original trained projection \(B=W_{C,T}\in\mathbb R^{400\times70}\). At each time, the two natural rates \(r_T^\pm\in[0,20]^{70}\) define the pair mean \(\mu_T=(r_T^++r_T^-)/2\) and difference \(d_T=r_T^+-r_T^-\). At this mean, the available differences satisfy \(|d'_{T,i}|\leq c_i\), where \(c_i=2\min(\mu_{T,i},20-\mu_{T,i})\). This box maps to the set of receiving differences \(Bd'_T\).

At a fixed pair mean, each PUL neuron has a limited range over which the two historical firing rates can differ. The collection of these ranges defines the feasible source differences, and multiplication by \(B\) determines which corresponding current differences can reach PPC.

Twelve fitting histories supply the leading eigenvector of the uncentered PUL rate-difference second moment. We evaluated the resulting direction on 12 independent test base histories, combined with fifth-trial stimuli containing 2 or 18 high-pass pips and both history signs. From the complete pair-averaged state, the recipients receive own, opposite, pair-mean, learned-axis, or random-axis messages through \(B\) at time steps \([1400,1600)\). For each circuit, three random axes are sampled orthogonal to the learned axis. The pair mean is held at \(\mu_T\). To compare the attainable strengths of different source directions, we increased the magnitude of the firing-rate difference along each direction while enforcing the rate bounds of every PUL neuron. This construction allowed us to determine whether the resulting cortical current could reach the magnitude of the natural historical input.

Along each axis \(q\), we follow \(d'_T(k)=\operatorname{clip}(kq,-c,c)\) with the temporal sign of the learned coefficient. The squared receiving norm is quadratic between clipping breakpoints \(c_i/|q_i|\). The first crossing of the natural norm defines the match. If the natural norm lies above the complete path, we use the largest attainable norm (Figure 4-6).

This comparison separates the strength that a source pattern could achieve through the trained projection from the behavioral effect of the pattern once delivered to PPC.

We next asked whether different patterns of PUL activity could generate cortical input differences of the same magnitude while preserving the aggregate drive to excitatory and inhibitory target neurons. We constructed these alternatives directly in the PUL firing-rate space. Geometrically, the procedure searches along a direction that preserves the two aggregate inputs and identifies a second point at which the cortical current-difference energy matches its original value.

To vary allocation at matched receiving energy and aggregate drive, let \(G=B^\top B\) and let the two rows of \(\mathcal C\) sum input to the excitatory and inhibitory PPC target classes. We selected directions \(z\) satisfying \(\mathcal Cz=0\), restricting their components to source coordinates with available room for changing the paired firing-rate difference within its admissible bounds. When \(z^\top Gz>0\), the energy-preserving displacement along the line \(d_T+\alpha z\) is given by

\[
\alpha=-\frac{2d_T^\top Gz}{z^\top Gz},\qquad
 d'_T=d_T+\alpha z. \tag{7}
\]

The reflection preserves \(\|Bd'_T\|_2^2=\|Bd_T\|_2^2\) and \(\mathcal C d'_T=\mathcal C d_T\). The transformed source difference must also remain within the firing-rate bounds. The procedure therefore explores the intersection of the source-rate constraints, the aggregate-input constraints, and the fixed receiving-energy surface.

For each time and context, each of two maps per circuit takes the largest relative receiving displacement \(\|B(d'_T-d_T)\|/\|Bd_T\|\) among 32 sampled directions satisfying the rate bounds. A map uses the natural difference when the candidates yield no feasible nontrivial reflection.

During replay, the transformed PUL firing rates determined the current delivered through the PUL-to-PPC projection. The internal PUL state and all other recurrent interactions continued to evolve according to the recipient circuit's endogenous dynamics.

For each circuit, we relate the attained source displacement to later historical contrast and current decision commitment. Intervals resample the twelve test base histories with their paired histories and stimuli together. Figure 4-3 compares the two maps, including contexts that retain the natural message.

\subsubsection{Tracking how history-dependent neural differences propagate and influence choice}\label{exact-trajectory-explanation-and-evidence-dependence}

The effect of a difference in neural state depends on the regions of the activation function occupied by the two trajectories. For example, a substantial internal state difference may produce no firing-rate difference if both states lie below the rectification threshold. The same state difference can generate different firing rates when the states lie within the active range or on opposite sides of a threshold. To describe this relationship exactly for finite differences between two trajectories, we defined a diagonal matrix of activation secants, \(D_{ii,t}=\Delta r_{i,t}/\Delta x_{i,t}\), with the local derivative convention specified in ET2 when the two states coincide.

Each diagonal entry measures the effective conversion of an internal state difference into a firing-rate difference for a particular neuron and trajectory pair. This construction accommodates finite differences that cross activation thresholds.

Let \(\widetilde W_t\) omit the endogenous
PUL-to-PPC block during replay; let \(E_C\) inject a 400-dimensional
cortical current. Then

\[
\Delta x_{t+1}=M_t\Delta x_t+A E_C\chi_t\Delta b_t,
\qquad M_t=I-A+A\widetilde W_tD_t, \tag{8}
\]

where \(\chi_t\) marks replay. The first term describes how the existing state difference is transformed by the circuit's dynamics. The second describes the additional difference introduced by the replayed input. Because the external sensory stimulus is identical in the two histories, its direct contribution cancels when the two state equations are subtracted.

The delivered current difference separates into amplitude and pattern contributions through the symmetric product identity

\[
\Delta b_t=\bar q_t\Delta g_t+\bar g_t\Delta q_t. \tag{9}
\]

The two terms respectively describe differences in overall input amplitude at the average spatial pattern and differences in spatial pattern at the average amplitude. Their sum equals the total difference between the delivered currents.

Applying Equation 8 to each input term and initial state component separates the contributions of initial PPC, initial PUL/TRN, incoming amplitude, and incoming pattern. All four components are propagated using the same activation secants calculated from the realized trajectory pair. Their signed contributions therefore sum exactly to the observed response difference for that pair. A different intervention can generate a different trajectory pair and hence a different set of activation secants. This pathwise construction is related to finite activation-difference rules used in neural attribution methods (Shrikumar et al., 2017).

A difference between the neural activities associated with two histories can remain large even when their decision outputs are nearly identical. This is possible because the output depends on the projection of the activity difference onto the trained readout. Moreover, projections with different signs at different times can partially cancel when the response is averaged. We therefore quantified three aspects of the historical difference: its population magnitude, its alignment with the readout, and the temporal consistency of its projection.

For response window \(R\), define

\[
R_x=\sqrt{\langle\|\Delta r_t\|^2\rangle_R},\qquad
a_R=\frac{\sqrt{\langle(w^\top\Delta r_t)^2\rangle_R}}{\|w\|R_x},\qquad
c_R=\frac{|\langle\Delta y_t\rangle_R|}{\sqrt{\langle\Delta y_t^2\rangle_R}}. \tag{10}
\]

These quantities satisfy

\[
|H_R|=\tfrac12\|w\|R_xa_Rc_R. \tag{11}
\]

Equation 11 provides an exact factorization of the magnitude of the historical contrast into these three components. It allows us to distinguish a weak behavioral influence caused by reduced history-dependent neural activity from a weak influence caused by poor readout alignment or temporal cancellation. The zero-norm conventions are in Extended Theory. We evaluate the relation on 16 base histories from the sequential experiment crossed with four fifth-trial strengths.

In the standard sequential task, the strong sensory judgment occurs on trial 5 and the ambiguous history-guided judgment occurs on trial 6. To separate the contributions of stimulus strength and trial position, we independently varied the fifth- and sixth-trial stimuli across all nine combinations of positive-pip counts in \(\{2,10,18\}^2\). We evaluated natural evolution and 500-ms own-history and opposite-history replay in 12 new base histories. At each trial position, the preceding neural states were held fixed across the stimulus conditions being compared. Ambiguous fifth trials and strong sixth trials extend the original training schedule and were evaluated using continuous historical contrasts.

\subsubsection{Testing alternative source patterns across histories and time}\label{fresh-history-source-confirmation-and-temporally-coherent-alternatives}

The initial source-space analysis established that the trained projection constrains the range of alternative cortical input patterns. We next examined these constraints across a larger set of histories and explored whether broader searches could identify alternative patterns that selectively alter the later use of sensory history.

To examine source flexibility on new histories, we used 48 base histories crossed with four strong fifth stimuli and mirrored histories. After complete-state averaging, the recipients receive replay at time steps 1400--1599. Current and later responses are measured at time steps 1720--1799 and 2120--2199. Three-coordinate directions in the null space of the E/I aggregate rows define reflections, including feasible inward movement at a rate bound. The searches using 64 and 256 candidate transformations followed the same ordered sequence of directions, applying reflections that satisfied the firing-rate bounds and increased receiving-current displacement. The larger search thus explored additional feasible transformations beyond those identified by the smaller one.

Searching for a separate feasible transformation at each time point may introduce rapid fluctuations that are absent from the natural source signal. We therefore included an additional transformation that reduces abrupt temporal changes while preserving the instantaneous receiving-current energy.

A third map smooths the transformed-minus-natural difference over five time steps and restores the natural energy along that direction.

To determine whether the source transformations selectively influence history, we assessed their effects on both the subsequent historical contrast and the intervening sensory response. We measured continuous signed sensory output in addition to threshold-defined commitment because a perturbation may alter the magnitude of the response even when the choice remains correct. An interval entirely within \(\pm0.02\) model output units defines equivalence in the mean sensory response. The paired 95\% intervals use base-history resampling with strengths, signs, and maps together. We assess the pooled comparison and each stimulus strength separately.

To examine whether an alternative source pattern can retain a consistent spatial organization throughout replay, we used 48 further base histories combined with the same four strong stimuli. Each of the 256 candidate reflections acted along a fixed three-coordinate direction across the entire replay interval. A reflection was accepted when the transformed source difference satisfied the firing-rate bounds at every time step, increased the integrated displacement of the receiving current, and preserved temporal roughness within 1.25 times its natural value.

Temporal roughness for a source difference \(d\) over the replay was measured as

\[
\mathcal R(d)=\sum_{t=1400}^{1598}\|d(t+1)-d(t)\|^2
\]

Temporal roughness quantifies the accumulated squared changes in the source firing-rate difference between successive time steps. The accepted reflections compose a single spatial transformation for the entire replay interval in each context. Contexts without an accepted transformation retain their original historical input. Figure 4-5 relates the achieved receiving-current displacement and temporal roughness to the subsequent historical contrast.

\subsubsection{Testing how neuronal operating states and recurrent interactions regulate historical influence}\label{calibrated-receiver-interventions-and-activation-regimes}

After the historical signal has been delivered to the circuit, its subsequent influence depends on how the receiving neurons transform and propagate the resulting state differences. We investigated two potentially complementary mechanisms. First, changes in the operating states of neurons can alter how internal state differences are converted into firing-rate differences. Second, recurrent interactions can sustain or transform those differences over time. We designed separate perturbations to examine these two mechanisms.

The initial receiver experiment used 12 reference and 24 test histories, combined with stimuli containing 2, 10, or 18 high-pass pips. All conditions received the same own-source reconstruction. Immediately after the sensory pip train, at time step 1710, we shifted the mean PPC state halfway toward the interval \([2,18]\), which lies within the linear activation range of PPC neurons, \([0,20]\). Both historical states received the same displacement, preserving their initial difference. An equal-norm Gaussian displacement tests direction dependence. A 100-ms pause in the 500-unit recurrent-drive difference tests propagation (Figures 6-3 and 6-4).

The smaller interventions use 12 independent calibration histories and 48 test base histories, combined with stimuli containing 2, 10, or 18 high-pass pips. Each history in a pair received the same shift, preserving their initial state difference.

At time step 1710, write \(s=\tfrac12[\operatorname{clip}(\bar x_C,2,18)-\bar x_C]\). We compare the inward displacement \(\epsilon s\) with eight Gaussian directions of norm \(\epsilon\|s\|\) held fixed across histories and strengths within each circuit. The random-direction controls distinguish effects associated with the magnitude of the displacement from effects associated with moving the states in a particular direction relative to the neuronal activation functions.

To examine the contribution of recurrent propagation, we manipulated the difference between the recurrent inputs received by the two history-dependent trajectories. At each time step, we preserved their mean recurrent drive and attenuated their recurrent-drive difference by a fraction \(\lambda\) under identical external input. The attenuation lasted for 20 time steps (100 ms), after which endogenous recurrent dynamics resumed.

Our aim was to identify circuit mechanisms that selectively regulate historical influence while preserving the processing of current sensory evidence. We therefore selected perturbation strengths using an independent set of 12 calibration histories, evaluating the unperturbed condition and the values \(\epsilon,\lambda\in\{0.001,0.003,0.01,0.03,0.1,0.3,1\}\).

For the common-state displacement, we selected the largest \(\epsilon\) for which the inward shift and all eight equal-norm random shifts preserved the sensory response under both strong stimulus conditions. Specifically, the paired 95\% confidence intervals for changes in sensory output had to lie within \(\pm0.02\) output units, with a reduction in decision commitment of at most 0.01 relative to own-history replay. We applied the same criteria to select the attenuation strength \(\lambda\).

The selected perturbation strengths were then held fixed across the independent test histories. Strong stimuli assessed the preservation of sensory judgments, while ambiguous stimuli revealed how the perturbations changed the influence of history.

We measured state and rate differences immediately after the shift, and measured response-window separation, alignment, and temporal consistency as in Equations 10--11. To examine how neuronal nonlinearities shape the contribution of retained states to the decision output, we classified each paired neuronal state according to its position on the activation function. The two states could occupy the same flat region, share the linear segment, or lie on opposite sides of the lower threshold, upper threshold, or both. The activation boundaries are 0 and 20 for PPC and PUL, and 9 and 11 for TRN, whose active segment has a slope of 10.

For each regime we compare \(|w_i|g_i|\Delta x_i|\), where \(g_i\) is the slope on the active segment, with the realized contribution \(|w_i\Delta r_i|\). The former quantity describes the readout contribution that the state difference would produce at the slope of the active segment. The latter is the contribution actually expressed through the neuron's activation function. Their comparison reveals how threshold crossings and saturation alter the conversion of retained state differences into the decision output.

Intervals preserve the strengths and fixed directions within each resampled base history.

\subsubsection{Predicting the evolution of history-dependent output across new sensory histories}\label{conditional-receiver-propagation-and-prediction-benchmarks}

The intervention experiments establish a causal contribution of recurrent interactions to the subsequent influence of history. We next asked how these interactions shape the evolution of history-dependent output following sensory stimulation. To address this question, we estimated effective neuronal activation gains by averaging finite-difference activation secants across 12 reference base histories at each sensory evidence strength. Individual secants were calculated as \(D_{ii,t}=\Delta r_{i,t}/\Delta x_{i,t}\), using the equal-state convention defined in ET2. Each forecast began from the measured state difference of the corresponding test pair at time step 1710 and predicted its evolution over the following 90 time steps using the trained integration factors and decision readout.

We constructed four propagation models by combining evidence-specific or pre-sensory activation gains with recurrent propagation or passive leak alone. Comparing the two gain estimates reveals how sensory-evoked changes in neuronal activation contribute to prediction. Comparing recurrent and leak-only propagation at matched gains identifies the contribution of recurrent interactions to the evolution of the retained state difference.

Current sensory evidence may generate a characteristic output trajectory shared across histories. To quantify this shared component, we constructed a reference template using the mean output-difference trajectory of the 12 reference base histories at each evidence strength. The template assigns the same prediction to every test history at that strength, while the propagation models use each test pair's measured initial state difference.

All five predictors were evaluated on the same 24 independent test base histories. We quantified predictive improvement relative to the reference template using

\[ 1- \frac{ \sum(\widehat{\Delta y}-\Delta y)^2 }{ \sum(\Delta y-\Delta y_{\mathrm{template}})^2 }. \]

Positive template-relative skill indicates a reduction in squared prediction error relative to the independent reference trajectory. We evaluated prediction errors both across evidence strengths and within each evidence condition, and estimated uncertainty in mean squared error differences by resampling complete base histories. Within each evidence condition, \(R^2\) uses the single mean observed output difference across all test histories and time steps as its baseline. This score reflects variation over time and across histories.

\subsubsection{Sampling units}\label{numerical-methods-and-reproducibility}

Statistical uncertainty was estimated at the level of the independent sampling unit appropriate to each analysis. For the animal behavioral data, we resampled mice within their respective intervention cohorts. For circuit simulations, we resampled base histories within each trained circuit, retaining the paired history signs, stimulus conditions, and transformation variants belonging to each base history. The resulting intervals quantify variability across animals or across histories, respectively. Table 1 summarizes the sampling units and comparisons underlying the principal findings.

\subsubsection{Data and code availability}\label{code-accessibility}

The published mouse data and original model checkpoints are available at
\url{https://doi.org/10.5281/zenodo.19465617}. The original model code is at
\url{https://github.com/DistribCogLab-1/history_based_rnn}. The manuscript and Extended Data describe the simulations used to obtain the reported outcomes. The analysis and simulation code for the mouse and trained-circuit studies, with instructions for obtaining the inputs and reproducing the reported comparisons, is available at \url{https://github.com/JustinZHAO-05/History-bearing-thalamocortical-code}.

\subsection{Results}\label{results}

\subsubsection{How does thalamocortical input determine which sensory history guides later choices?}\label{which-history-is-carried-forward}

\paragraph{Sensory history helps or impairs choices depending on current evidence}\label{historical-evidence-contributes-benefit-and-bias-according-to-present-evidence}

To understand why neural circuits must regulate the influence of sensory history, we first examined how information about previous stimuli affected decisions in the published mouse behavioral data. In the task, mice listened to auditory stimuli containing mixtures of high-pass and low-pass tone pips. On regular trials, the correct choice depended on the relative abundance of the two types of pips in the current stimulus. On history probes, the stimulus contained equal numbers of the two types, and the correct choice instead depended on the preceding sensory history (\hyperlink{main-figure-link-1}{Figure~1a},b). This task allowed us to examine the contributions of current and historical sensory evidence to behavior.

Suppressing either PPC or PUL improved judgments of the current stimulus while impairing judgments that required sensory history (\hyperlink{main-figure-link-1}{Figure~1c}). PPC suppression increased current accuracy by 4.9 percentage points (95\% mouse-bootstrap interval [1.5, 8.0]) and reduced history-probe accuracy by 14.0 percentage points [-21.9, -6.2]. PUL suppression produced a similar tradeoff, increasing current accuracy by 4.4 percentage points [1.9, 6.3] while reducing history-probe accuracy by 20.9 percentage points [-35.0, -7.6]. These findings are consistent with the reduced influence of sensory history reported by Hosford et al. (2026).

If the improvement in current judgments arises from weakening the influence of past stimuli, it should be particularly pronounced when the past and present favor opposing choices. We therefore divided regular trials according to whether the mean sensory evidence from the preceding five trials agreed or conflicted with the current stimulus. Under PPC suppression, accuracy improved by 10.3 percentage points when history conflicted with the current stimulus, compared with only 1.0 point when they agreed. The additional benefit under conflict was 9.3 percentage points [5.5, 13.0] and was observed in all nine mice (Holm-adjusted signed-rank \(p=0.012\); \hyperlink{main-figure-link-1}{Figure~1d}). PUL suppression showed a similar, more variable effect, with a conflict advantage of 6.1 percentage points [1.8, 11.2] across seven of ten mice (Holm-adjusted \(p=0.055\)). Activating the TRN-to-PUL pathway produced little consistent difference between conflict and agreement conditions (-0.8 percentage points [-5.9, 4.0]).

An intervention could also change how strongly animals respond to current sensory evidence or alter their general preference for one choice. To distinguish these possibilities, we fitted choice models that separately estimated the influence of current stimuli, preceding stimuli, and response bias. In the PPC cohort, changes in the fitted history terms accounted for 8.4 [5.1, 11.5] percentage points of the additional benefit under conflict, whereas changes in the sensory terms accounted for -0.04 points. This pattern was maintained when using a more flexible nonlinear model of sensory evidence and when including all eligible binary choices (Figure 1-2).

We further examined whether the reduction in history weighting persisted after accounting for the temporal organization of the experiment, including prior intervention exposure, recent choices and outcomes, and gradual within-session changes (Figure 1-1). After these adjustments, PPC suppression reduced the fitted influence of sensory history (\(\beta_{hL}=-0.70\) [-1.11, -0.27], negative in eight of nine mice). PUL suppression showed a weaker negative estimate (-0.47 [-1.07, 0.22], negative in seven of ten mice), while the effect of TRN-to-PUL activation remained close to zero (-0.08 [-0.50, 0.32]; \hyperlink{main-figure-link-1}{Figure~1e}). Additional analyses across stimulus strengths are shown in Figure 1-3.

Together, these behavioral results identify the conditions under which relying on recent sensory experience becomes costly. When past stimuli favor an incorrect interpretation of the present stimulus, reducing the influence of history improves sensory judgments. This effect was most consistently associated with PPC suppression, motivating our investigation of how the thalamocortical circuit carries information about past stimuli into subsequent decisions.

\paragraph{Returning pulvinar input determines which sensory history is reconstructed}\label{returning-feedback-selects-which-past-is-reconstructed}

The behavioral findings suggest that thalamocortical interactions contribute to the use of information about previous stimuli. We next asked whether the input returning from PUL to PPC itself carries this information. One possibility is that the returning input provides a general drive that helps sustain a history-dependent state already present in the recipient circuit. Alternatively, different patterns of returning input may specify different sensory histories and thereby reconstruct different neural representations.

To distinguish these possibilities, we exposed the same trained circuit to pairs of opposing sensory histories and recorded the current delivered from PUL to PPC. We then initialized both recipient circuits at the average of their complete 500-dimensional neural states, placing them in an identical starting state. From this common state, each circuit received its own recorded input, the input recorded from the opposite history, or the average of the two recorded inputs. The original PUL-to-PPC connection resumed after replay, and both circuits continued to receive identical external stimuli.

We applied this reconstruction procedure in two temporal settings: after five historical trials immediately preceding the balanced probe, and after four historical trials preceding a new sensory judgment. \hyperlink{main-figure-link-2}{Figure~2a} illustrates the latter arrangement.

In the sequential arrangement illustrated in \hyperlink{main-figure-link-2}{Figure~2}, the population firing rates began to diverge during own-history and opposite-history replay. This separation persisted after endogenous feedback resumed (\hyperlink{main-figure-link-2}{Figure~2b,c}). Supplying the same pair-mean input to both recipients maintained identical neural trajectories. The identity of the replayed signal therefore determined the history-dependent difference reconstructed from the common initial state.

To quantify how the reconstructed history influenced later decisions, we measured the historical contrast, H, defined as half the difference between the mean decision outputs associated with the original positive and negative histories. Positive values indicate that the original positive history produces a higher decision output, whereas negative values indicate a reversal of this relationship.

We first quantified reconstruction in the five-trial history assay using 96 new base histories in each of the nine trained circuits. One second of opposite-history replay reversed the historical contrast in every circuit, from natural values of +0.51 to +0.65 to values of -0.64 to -0.38 after replay (all nine 95\% confidence intervals below zero). Replaying the original input restored positive historical contrasts after either PPC-only or complete-state averaging.

The same reconstruction principle also operated when replay preceded an intervening sensory judgment, as illustrated in \hyperlink{main-figure-link-2}{Figure~2}. These results show that returning PUL-to-PPC input carries information capable of determining which historical distinction is reconstructed from an identical initial neural state.

Which features of the arriving input carry this historical information? The input recorded from one history may differ from that recorded from another in its overall amplitude, in the distribution of current across cortical neurons, or in both. We therefore separated the arriving current into its mean amplitude and a normalized spatial pattern and exchanged these components independently (Equation 5).

Both input amplitude and spatial pattern contributed to the reconstruction of sensory history, with their relative influence varying across circuits (Figure 2-2). During 500-ms replay, amplitude exchange produced a larger historical effect in G22, G23, G25, and S1, while spatial-pattern exchange produced a larger effect in G20, G24, S2, and S3. All eight 95\% confidence intervals excluded zero. S0 showed more comparable effects of the two exchanges (difference +0.036 [-0.025, +0.095]). Thus, different trained circuits convey historical information through different combinations of input amplitude and spatial organization.

Having established that the returning signal can reconstruct a particular sensory history, we next examined whether this signal can also redirect later behavior in circuits that retain their original history-dependent states.

\paragraph{Reconstructed sensory history persists through an accurate intervening decision}\label{a-reconstructed-past-survives-an-intervening-successful-judgment}

For information about previous stimuli to remain useful across successive decisions, it must survive the processing of new sensory evidence while allowing the circuit to make an accurate judgment about that evidence. We therefore asked whether a history reconstructed by returning thalamic input could remain available after the circuit completed an intervening sensory decision.

We generated pairs of opposing four-trial histories and exchanged their PUL-to-PPC input patterns before delivering the same strong fifth stimulus to both recipient circuits. The recipients retained their original internal states in this main exchange experiment. Following the fifth sensory judgment, both circuits received a balanced sixth stimulus, allowing us to determine which history influenced their later choice (\hyperlink{main-figure-link-3}{Figure~3a}; 48 new base histories per circuit).

Following 1 s of opposite-history replay, all nine circuits correctly judged the intervening strong stimulus in every sequence, as measured by the sign of the mean response (\hyperlink{main-figure-link-3}{Figure~3b}). Decision commitment remained at 99.5--100\%, with the outputs reaching the correct threshold and respecting the opposite decision bound.

When the same circuits subsequently encountered the balanced sixth stimulus, however, their choices were strongly redirected by the substituted history. The historical contrast changed from +0.52 to +0.67 under natural evolution to -0.66 to -0.35 after opposite-history replay, reversing its sign in all nine circuits. Relative to the original historical target, median mean-sign accuracy decreased from 93.5\% to 16.2\%. The median fraction of sequences reaching the decision threshold corresponding to the substituted history was 64.6\% (\hyperlink{main-figure-link-3}{Figure~3e}).

The different effects of replay on the two successive decisions were also evident in the continuous output. Exchanging the thalamic input changed the mean response by only 0.03--0.13 output units during the strong fifth-trial stimulus, but by 0.91--1.33 units during the subsequent history-guided decision (\hyperlink{main-figure-link-3}{Figure~3c}). The strength of this redirection depended in part on the duration of the input exchange. Shortening replay to 500 ms still reversed the later historical contrast in eight of the nine circuits; the remaining circuit, G20, showed a contrast close to zero (+0.030).

The main exchange experiment tested whether returning thalamic input could redirect choices in recipients retaining their original history-dependent states. We also examined whether this input could restore history following the selective removal of the initial cortical state difference. After averaging the PPC states alone, 500-ms own-history replay produced higher later mean-sign accuracy than pair-mean replay (\hyperlink{main-figure-link-3}{Figure~3d}). This comparison connects the complete-state reconstruction illustrated in \hyperlink{main-figure-link-2}{Figure~2} with the more localized manipulation of cortical history-dependent states. The substituted history thus remained available to influence a later choice even after the circuit had accurately processed a new stimulus. This separation between current sensory performance and the subsequent use of history raises two further questions: which features of the returning input allow a particular history to be reconstructed, and how does the circuit regulate the influence of that history as current sensory demands change?

\paragraph{A low-dimensional pattern of thalamic input can reconstruct sensory history}\label{a-low-dimensional-historical-message-is-shaped-by-source-and-target-geometry}

Although thalamic feedback is delivered to hundreds of cortical neurons, the difference between inputs associated with opposing histories may be organized along a much smaller number of spatial directions. We therefore asked whether this history-dependent component of the input could be described by a common pattern across different sensory histories, and whether such a reduced signal would remain sufficient to reconstruct the historical response.

We estimated the leading direction of the differences between arriving PUL-to-PPC currents using an independent set of histories. This single direction captured 77.7--98.8\% of the difference energy in held-out histories across the nine circuits. At each time point, we projected the history-dependent current difference onto this single direction while retaining the pair-mean current as a shared background. The projection can produce negative arrival currents at targets with little background input. To maintain the nonnegativity of the excitatory PUL-to-PPC projection while preserving an exactly rank-one historical difference, we increased the shared background current at the targets that required it, by an amount corresponding to 0.2--15.9\% of the natural mean-current norm. This single spatial direction was sufficient to reconstruct positive historical contrasts during own-history replay and to reverse their signs during opposite-history replay in all nine circuits. Thus, the information required to distinguish opposing histories can be carried by a compact, low-dimensional signal embedded within a higher-dimensional background input.

In two circuits, reducing the dimensionality of the historical input also impaired the intervening sensory decision: mean-sign accuracy decreased to 84.4\% in G23 and 97.9\% in S2. Projecting onto the leading direction removes both the components orthogonal to that direction and part of the original differential-current magnitude. We therefore considered two possible explanations for the impaired sensory performance: the discarded components could contribute independently to sensory processing, or the reduced magnitude of the remaining signal could be insufficient to maintain the circuit's usual dynamics.

To distinguish these possibilities, we compared restoring the discarded components with rescaling the reduced signal to match the magnitude of the original current difference. Restoring the magnitude alone recovered 100\% current mean-sign accuracy and nearly complete decision commitment (192/192 sequences in G23 and 190/192 in S2), while maintaining strong reversals of the later historical contrast (-0.49 and -0.57, respectively; Figure 4-1). Restoring the magnitude of the low-dimensional signal was therefore sufficient to recover the intervening sensory performance while preserving the substituted history in both circuits.

The estimated history direction captured a large fraction of the naturally occurring input differences. We next asked whether this direction was also particularly effective at influencing later decisions. If the distribution of current across neurons is important, then different input patterns of the same magnitude should produce different historical responses. We compared the learned direction with 16 randomly sampled directions that were initially orthogonal to it, while matching the pair-mean current and the differential-current norm (24 base histories per circuit).

The learned direction generated substantially stronger historical responses than the sampled alternative directions. Across the nine circuits, it produced positive historical contrasts of +0.50 to +0.69, while the mean magnitudes of the contrasts produced by random directions ranged from 0.02 to 0.35. The advantage of the learned direction ranged from 0.32 to 0.65 output units, with positive 95\% confidence intervals in all nine circuits (Figure 4-4). Only three of the 144 sampled alternatives produced historical contrasts at least as large in magnitude as those generated by the learned direction, comprising two directions in G23 and one in G24. Several random directions also disrupted the intervening sensory judgment, with commitment rates as low as 19.3\%. The learned direction preserved commitment at 99.5--100\% (Figure 4-4B).

These comparisons show that the historical effect of thalamic input depends on how its current is distributed across cortical neurons, in addition to the magnitude of the difference between opposing histories. The direction naturally expressed by the trained circuits was generally more effective at reconstructing history while preserving the intervening sensory judgment than the sampled alternatives.

\paragraph{The trained thalamocortical projection constrains which historical input patterns can reach PPC}

The preceding experiments directly manipulated the current arriving at PPC, allowing us to examine the receiving circuit's sensitivity to different spatial patterns. In the intact network, however, this current must be generated by the activity of 70 PUL neurons and transmitted through the existing connections to 400 PPC neurons. Each PUL neuron also has a limited firing-rate range. Together, these constraints determine which of the input patterns examined at the receiving end can actually be generated by the source population (\hyperlink{main-figure-link-4}{Figure~4a}).

We therefore examined the organization of sensory history directly in the PUL firing-rate space. A single direction captured 59.2--97.6\% of the rate-difference energy in the fitting histories and 59.4--97.5\% in independent test histories (\hyperlink{main-figure-link-4}{Figure~4b}). To test whether this reduced source signal remained functionally effective, we delivered the corresponding current through the original PUL-to-PPC projection, while enforcing the firing-rate bounds of the source neurons. Across all nine circuits, replaying own-history, opposite-history, or learned-axis messages preserved 100\% current decision commitment. The reduced own-history signal reconstructed historical contrasts of +0.52 to +0.71, and opposite-history replay reversed the subsequent historical contrast (Figure 4-6).

The firing-rate constraints also allowed us to ask whether alternative patterns of PUL activity could generate cortical currents as strong as those produced by the naturally occurring historical signal. For each candidate source direction, we increased the magnitude of the firing-rate difference while keeping the mean source activity fixed and respecting the bounds of every PUL neuron. We then determined whether the resulting difference in cortical input could match the magnitude of the natural current difference.

Along the learned source direction, PUL activity reproduced the magnitude of the natural cortical current difference to within 1\% in 78.8--100\% of the evaluated time--context combinations, depending on the circuit. The corresponding median fraction across the 27 random source directions was only 2.1\% (\hyperlink{main-figure-link-4}{Figure~4c}). S3 permitted all three sampled random directions to match the natural current magnitude throughout replay.

Thus, the learned historical direction is both functionally effective when transmitted to PPC and broadly attainable within the firing-rate constraints of the trained source population. The corresponding constraints sharply limit the attainable strength of many alternative source directions, particularly in circuits whose natural source activity leaves little flexibility for generating alternative patterns.

\paragraph{The distribution of thalamic input among individual cortical neurons shapes later history-guided choices}\label{fine-target-allocation-modifies-history-use-beyond-aggregate-input}

Changing the direction of an input vector can alter both the distribution of input within cortical populations and the total current delivered to excitatory and inhibitory neurons. Consequently, the weaker historical effects of alternative input directions could arise partly from changes in how strongly these two populations are recruited. We next asked whether the circuit is sensitive to the finer distribution of thalamic input among neurons of the same target class.

To address this question, we redistributed the history-dependent current difference separately within the 280 excitatory and 120 inhibitory PPC neurons. These transformations preserved the total current delivered to each target class in each history, the pair-mean input, and the differential-current energy. Thus, the same aggregate input could be delivered through different allocations of current among individual target neurons (\hyperlink{main-figure-link-5}{Figure~5a}; four redistribution maps and 24 base histories per circuit).

In three circuits, this finer redistribution selectively weakened the subsequent influence of history while leaving the intervening sensory decision virtually unchanged. Historical contrast decreased by 0.51 [0.44, 0.57] in G20, 0.29 [0.22, 0.36] in G24, and 0.24 [0.16, 0.33] in S0. All three circuits retained 100\% current mean-sign accuracy and 99.5--100\% decision commitment (\hyperlink{main-figure-link-5}{Figure~5c}).

The remaining circuits exhibited different relationships between input redistribution and behavior. In S2, redistribution reduced both the historical contrast (loss 0.56 [0.47, 0.64]) and current decision commitment (91.2--93.8\%). Five other circuits were less sensitive to the redistribution and retained 100\% current mean-sign accuracy. G23 exhibited a small historical loss of 0.012 [0.006, 0.018], while the mean effects in G22, G25, and S1 were close to zero, with confidence intervals spanning zero. S3 was particularly tolerant of changes in input organization: a relative displacement of 0.90 in the arriving current difference produced a historical loss of only 0.073 [0.049, 0.098] (Figures 5-1 and 5-2).

These results reveal that the function of thalamic input can depend on its allocation among individual cortical neurons, even when the total input delivered to excitatory and inhibitory target populations is preserved. They also reveal substantial diversity among trained circuits. Some solutions rely strongly on fine-scale input organization to guide later choices, whereas others can preserve similar historical behavior across substantially different spatial patterns of input.

\paragraph{The ability to generate alternative thalamic signals and their influence on behavior vary across circuits}\label{fresh-histories-reveal-source-specific-functional-sensitivity}

The preceding experiments demonstrate that some receiving circuits are sensitive to changes in the fine spatial distribution of arriving current. However, directly redistributing the current across PPC neurons allows a broader range of input patterns than can necessarily be produced by the trained PUL-to-PPC projection. We therefore asked whether similarly controlled reallocations could be generated by the source population itself, and whether those attainable alternatives would alter subsequent history-guided behavior.

We constructed alternative patterns of PUL activity while preserving the pair-mean source activity, individual firing-rate bounds, the magnitude of the resulting cortical current difference, and the aggregate input differences delivered to excitatory and inhibitory PPC targets (\hyperlink{main-figure-link-5}{Figure~5b}). Both the range of attainable input patterns and their behavioral effects varied markedly across circuits (12 base histories per circuit; Figure 4-3).

In a further comparison across 48 independent base histories per circuit, increasing the number of candidate transformations from 64 to 256 revealed additional feasible reallocations of cortical input (\hyperlink{main-figure-link-4}{Figure~4d}). In three circuits, these reallocations selectively weakened subsequent historical influence: historical contrast decreased by 0.044 [0.034, 0.054] in G20, 0.016 [0.009, 0.025] in S0, and 0.012 [0.008, 0.016] in S3.

The intervening sensory judgments remained stable, with decision commitment of 99.5--100\%. Changes in mean sensory output were small, with paired 95\% confidence intervals lying within \(\pm0.02\) output units in both the pooled and stimulus-specific comparisons (\hyperlink{main-figure-link-5}{Figure~5c} and Figure 5-3). These results reveal how alternative patterns of PUL activity can selectively regulate the later influence of sensory history while preserving the processing of strong current evidence.

The source transformations considered so far could vary from one time point to the next, potentially introducing rapid fluctuations into the historical signal. We therefore examined alternative patterns generated by a single spatial transformation maintained throughout replay. We also constrained temporal roughness, defined as the sum of squared changes in the source firing-rate difference between consecutive time steps, to at most 1.25 times its natural value. These comparisons used 48 further independent base histories per circuit (Figure 4-5).

Under these joint spatial and temporal constraints, the sampled transformations yielded substantially less freedom to redistribute thalamic input. Seven circuits retained the original input throughout the comparison, and G20 achieved a relative displacement of only 0.23\%. S3 permitted a considerably larger displacement of 14.2\%, with virtually unchanged historical contrast (loss -0.0001 [-0.0025, 0.0024]) and 100\% decision commitment.

The computational function of a thalamocortical pathway therefore depends on two complementary properties: the range of input patterns that can be generated by the thalamic source population and the sensitivity of the receiving circuit to those patterns. These properties vary across trained circuits. In particular, S3 demonstrates that substantially different patterns of thalamic input can remain feasible under spatial and temporal constraints while supporting nearly identical history-guided behavior.

\subsubsection{When does retained history influence
choice?}\label{when-does-retained-history-influence-choice}

\paragraph{History-dependent neural activity persists even when its influence on choice is weak}\label{persistent-neural-differences-acquire-different-behavioral-projection-across-judgments}

Our replay experiments showed that a reconstructed sensory history could guide a later ambiguous choice after the circuit had accurately judged a strong intervening stimulus. How does the circuit retain information about the past while limiting its influence during the intervening sensory decision?

One possibility is that strong sensory input substantially reduces the neural differences produced by opposing histories. Another is that these differences remain large but contribute little to the activity that determines the current decision. To distinguish these possibilities, we measured both the magnitude of the population firing-rate difference between the two histories and the projection of this difference onto the trained decision readout.

The decision output is a weighted combination of neuronal firing rates. Consequently, a large difference between two population activity patterns can produce a small output difference if that activity difference is poorly aligned with the readout direction. Furthermore, output differences with opposing signs at different times can partially cancel when averaged across the response window. We therefore decomposed the historical contrast into the magnitude of the population activity difference, its alignment with the fixed readout, and the temporal consistency of its projection (Equation 11).

History-dependent differences in population activity remained substantial during both the current sensory judgment and the later history-guided judgment. In G22, for example, the root-mean-square (RMS) magnitude of the population firing-rate difference was 27.1 during the current judgment and 24.7 during the later judgment. Over the same two response windows, its RMS alignment with the decision readout (Equation 10) increased more than sixfold, from 0.0017 to 0.011. Across all nine circuits, readout alignment increased by a factor of 6.2--17.9 (\hyperlink{main-figure-link-6}{Figure~6a,b}). The retained historical activity therefore exerted a much stronger influence on the later choice as its projection onto the decision readout increased.

We also asked how a historical signal entering PPC through the PUL-to-PPC pathway subsequently influences activity in other parts of the circuit. Because the trained decision readout receives activity from all three neural populations, information introduced through a single projection can ultimately affect the output through multiple populations. We therefore decomposed the difference in decision output between natural evolution and opposite-history replay into the contributions of PPC, PUL, and TRN. The PPC contribution ranged from 16.5\% to 98.9\% across circuits. In S3, for example, PUL accounted for 83.6\% of the final output change, PPC for 16.5\%, and TRN for -0.07\% (Figure 6-2).

The pathway through which historical information enters the circuit can therefore differ from the populations through which its behavioral effect is ultimately expressed. Recurrent interactions distribute the reconstructed historical distinction across the circuit, while the organization of the resulting activity relative to the trained readout determines its influence on choice.

\paragraph{The strength of current sensory evidence regulates the influence of retained history}\label{current-evidence-modulates-historical-expression-within-the-same-sequence-position}

In the sequential experiments, the circuit encountered a strong stimulus on trial 5 and a balanced stimulus on trial 6. The change in historical influence between these judgments could therefore depend on the strength of the sensory evidence, on the position of the trial within the sequence, or on both. To distinguish these possibilities, we varied the strength of the fifth and sixth stimuli independently while preserving the corresponding preceding network states across stimulus conditions (\hyperlink{main-figure-link-7}{Figure~7a},b; 12 new base histories per circuit).

Even when the trial position was held fixed, the influence of history changed markedly with current sensory evidence. On trial 5, the mean historical contrasts across circuits were only +0.01 and +0.03 for strong negative and positive stimuli containing 2 and 18 positive pips, respectively. When the same preceding states were followed by a balanced stimulus containing 10 positive pips, the mean historical contrast increased to +0.70. All nine circuits exhibited substantially stronger historical influence under ambiguity at both trial positions.

The change in output raised a question about what the circuit retained under strong sensory evidence. Strong fifth-trial stimuli reduced the historical output contrast to 2.0--7.7\% of its magnitude under ambiguity. However, the norm of the internal state difference at the end of the response window remained at 49.2--111.7\% of its ambiguous-condition value (Figures 7-1 and 7-2). Thus, strong sensory input greatly reduced the expression of history in the decision output while leaving substantial history-dependent state differences within the circuit.

When the fifth stimulus itself was ambiguous, 500-ms opposite-history replay immediately reversed the mean historical contrast to -0.43, with eight of nine circuits showing a reversal. The substituted history was already available to guide behavior at the fifth-trial position. Its influence on choice increased when the sensory evidence became ambiguous, demonstrating that current evidence can regulate historical expression within a fixed stage of the decision sequence.

How can the influence of sensory history change so substantially while its underlying state difference remains present? A minimal nonlinear model illustrates how this separation can arise even when the decision readout remains fixed. Consider a circuit that retains a scalar historical variable \(m=\pm a\) and combines it with current sensory evidence \(z\) through the saturating readout \(y=\tanh(gz+km)\). For the two opposing histories, the difference in their decision outputs can be expressed as the historical contrast

\[
H(z)=\frac{\sinh(2ka)}{\cosh(2gz)+\cosh(2ka)}, \tag{12}
\]

When current sensory evidence is weak, differences in the retained history can produce substantial differences in the decision output. As the magnitude of sensory evidence increases, both historical conditions approach the same saturated region of the readout, reducing their output difference even though the historical state difference remains fixed at \(2a\). Consequently, \(H(z)\) decreases monotonically with \(|z|\) for fixed positive \(ka\) and nonzero sensory gain. A complementary discrete model uses a single memory variable and two firing-rate units with piecewise-linear activation functions, reflecting the activation structure of the trained thalamocortical circuits. It reproduces the central sequential phenomenon: both histories yield the correct saturated response under strong sensory evidence, and the retained or substituted memory subsequently guides the choice under ambiguity (Figure 6-1). These minimal models provide an intuitive illustration of how a fixed nonlinear transformation can regulate the expression of retained information according to current sensory demands.

\paragraph{Neuronal activation thresholds help regulate the expression of retained history}\label{receiver-susceptibility-links-evidence-to-continued-historical-use}

The minimal models illustrate how nonlinear activation can separate the retention of sensory history from its influence on a decision. We next examined how the activation functions of individual neurons contribute to this separation in the trained thalamocortical circuits.

For a given neuron, the relationship between an internal state difference and a firing-rate difference depends on where the two states lie on its activation function. For example, two history-dependent states may differ substantially while both remain below the threshold for firing, producing no difference in firing rate. If the two states lie on opposite sides of the threshold, their difference becomes expressed in neuronal activity. Changes in current sensory input can therefore alter how history-dependent internal states are converted into firing rates and subsequently influence the decision readout.

To examine this mechanism, we classified paired neuronal states according to whether they occupied the same region of the activation function or crossed its lower or upper threshold. Under ambiguous evidence, neurons whose paired states lay on opposite sides of the lower rectification threshold accounted for an average of 69.9\% of the summed absolute PPC readout contributions. The corresponding fractions were 25.1\% under strong negative evidence and 33.8\% under strong positive evidence. Upper-threshold crossings contributed less than 1.0\% across all three conditions (\hyperlink{main-figure-link-6}{Figure~6c}).

Lower-threshold crossings thus constitute a major component of the history-dependent activity contributing to the PPC readout under ambiguity. The position of neuronal states relative to the firing threshold provides a mechanism through which current sensory input can regulate the expression of retained history.

To examine how neuronal operating states and recurrent interactions regulate the use of retained history, we introduced two complementary perturbations immediately after the fifth-trial sensory input. We selected their strengths using 12 independent calibration histories, with the aim of preserving the circuit's sensory judgments under strong evidence. The selected strengths were subsequently evaluated in 48 new base histories per circuit. Across the independent test histories, current decision commitment remained at 96.9--100\%. Strong-stimulus sensory responses were stable, with paired 95\% confidence intervals for changes in mean output lying within \(\pm0.02\) output units across all 36 comparisons (\hyperlink{main-figure-link-6}{Figure~6f} and Figure 6-5).

We first examined how changes in cortical operating states affect the expression of retained history. Both historical states received the same displacement, directed toward the interior of the PPC activation range according to their pair mean. This manipulation preserved their initial separation while changing their positions relative to the neuronal firing thresholds. Eight equal-norm random displacements provided a comparison for assessing the importance of displacement direction.

The inward displacements corresponded to 1--3\% of a reference shift that moved the mean PPC state halfway toward the interval \([2,18]\), within the linear activation range \([0,20]\). These shifts produced small changes in the subsequent historical contrast, with increases of up to 0.002 output units. In five of the nine circuits, the inward shift produced a larger change than the mean effect of the eight random directions (\hyperlink{main-figure-link-6}{Figure~6d}).

Cortical operating states thus modulate how an existing historical difference is expressed and propagated. Within the perturbation range that preserved current sensory judgments, this modulation exerted a modest influence on later history-guided choice.

\paragraph{Recurrent interactions sustain the influence of reconstructed history across decisions}

The preceding perturbation examined how the operating states of neurons influence the transformation of retained historical differences into neuronal activity. We next asked how these differences are maintained and transformed through the recurrent interactions of the circuit.

The two history-dependent trajectories generally receive different recurrent inputs because their population firing rates differ. To isolate the contribution of this history-dependent component, we partially reduced the difference between their recurrent drives while preserving the mean recurrent drive at each time point. This intervention lasted for 100 ms, after which the original recurrent dynamics resumed. The calibrated attenuation ranged from 1\% to 30\% across circuits.

Attenuating the history-dependent recurrent drive weakened the subsequent influence of sensory history in all nine circuits. The reduction in historical contrast ranged from 0.007 to 0.111 output units, with 95\% confidence intervals supporting a reduction in every circuit. For example, the historical contrast decreased by 0.040 [0.034, 0.046] in G20, 0.046 [0.041, 0.052] in G22, and 0.111 [0.097, 0.126] in S3 (\hyperlink{main-figure-link-6}{Figure~6e}).

The intervening sensory judgments remained stable under these perturbations (\hyperlink{main-figure-link-6}{Figure~6f} and Figure 6-5). Recurrent interactions thus contribute to sustaining the influence of reconstructed history across successive decisions.

\paragraph{Recurrent propagation explains the evolution of history-dependent output after sensory stimulation}

Our perturbation experiments showed that recurrent interactions contribute to the subsequent influence of a reconstructed history. As a complementary test, we asked whether these interactions could explain how the difference in decision output between two histories evolves over time after a sensory stimulus.

We constructed conditional propagation models using effective activation gains obtained by averaging finite-difference activation secants across 12 independent reference base histories at each sensory evidence strength (Materials and Methods; ET2). For each of 24 independent test base histories, we initialized the forecast with the measured neural state difference immediately after the fifth-trial sensory stimulus. We then predicted the evolution of the corresponding decision-output difference over the following 450 ms (90 time steps), comparing recurrent propagation with passive decay at matched activation gains (\hyperlink{main-figure-link-7}{Figure~7c,d}).

Combining evidence-specific activation gains with recurrent propagation explained 80.0--96.6\% of the pooled output-trajectory variance across the nine circuits. At matched activation gains, incorporating recurrence substantially improved prediction in every circuit. Root-mean-square error (RMSE) ranged from 0.11 to 0.27 with recurrent propagation and from 0.69 to 0.82 with passive decay alone (\hyperlink{main-figure-link-7}{Figure~7d}, left). These results reveal the contribution of recurrent interactions to the evolution of history-dependent output following sensory stimulation.

Current sensory evidence also establishes a characteristic output trajectory shared across different histories. To examine this common component, we compared the propagation models with a reference template constructed from the average output-difference trajectory of independent histories at each evidence strength. Across the pooled comparisons, the template achieved lower prediction error in six of the nine circuits, with paired confidence intervals favoring the template in five (\hyperlink{main-figure-link-7}{Figure~7d}, center). The performance of this simple template reveals the substantial contribution of sensory evidence to the common structure of the output trajectories.

We further examined prediction accuracy separately at each sensory evidence strength. The recurrent propagation model achieved its most consistent within-evidence fits under ambiguity, with positive \(R^2\) in eight of the nine circuits. Prediction accuracy varied more widely under strong sensory evidence, when history-dependent output differences were small (\hyperlink{main-figure-link-7}{Figure~7d}, right). The independent reference template supplies a separate comparison against the shared output-difference trajectory estimated from reference histories (Figure 7-3).

Together, these analyses reveal how current sensory evidence shapes the common structure of historical output trajectories and how recurrent dynamics transform the retained neural differences from which those trajectories emerge.

\newgeometry{left=17mm,right=17mm,top=20mm,bottom=22mm}
\begingroup
\mainfigurepage{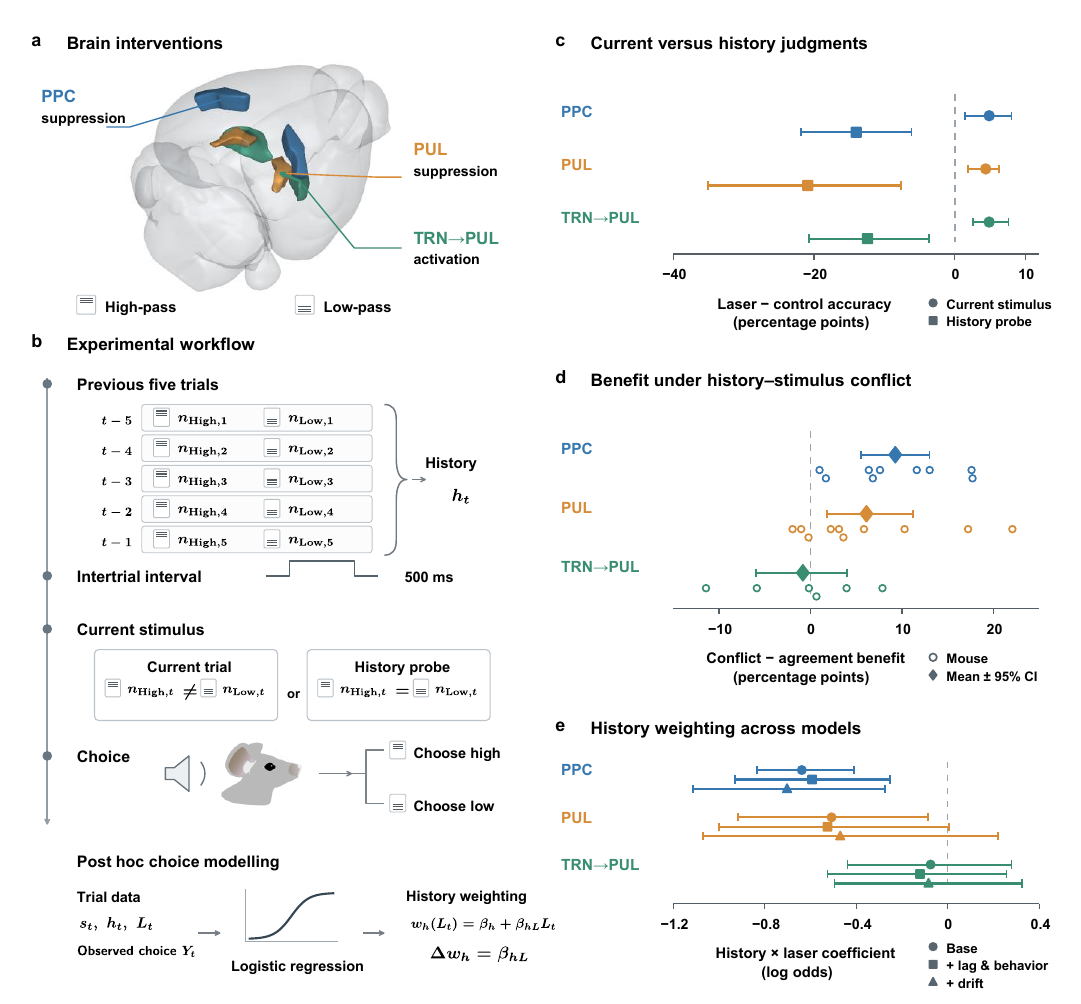}{The behavioral influence of sensory history depends on its relationship to current sensory evidence}{
\textbf{a,} Brain regions and pathways targeted in three separate mouse cohorts: suppression of posterior parietal cortex (PPC), suppression of pulvinar (PUL), and activation of the projection from thalamic reticular nucleus (TRN) to PUL. \textbf{b,} Behavioral task and analysis procedure. Regular trials require the animal to judge whether the current stimulus contains more high-pass or low-pass tone pips. History probes contain equal numbers of the two types of pips and are identified by the history-probe designation in the published dataset. Historical evidence is calculated from the preceding five trials. The 500-ms intervention occurs during the intertrial interval. Logistic choice models estimate how the intervention changes the influence of current and historical sensory evidence on behavior. \textbf{c,} Differences in accuracy between intervention and control trials for current sensory judgments and history probes, in the PPC (nine mice), PUL (ten mice), and TRN-to-PUL (six mice) cohorts. \textbf{d,} Difference in the accuracy benefit of intervention between trials with conflicting and agreeing sensory evidence. Open circles represent individual mice; diamonds represent cohort means. \textbf{e,} Changes in the fitted influence of history on choice under intervention, estimated using the base model and models that additionally account for prior intervention exposure, preceding behavior, and within-session drift. Negative coefficients indicate reduced historical influence. Bars represent 95\% mouse-bootstrap intervals. Extended Data Figures 1-1, 1-2, and 1-3 provide additional analyses of intervention timing, the contributions of fitted behavioral components, and dependence on sensory evidence strength.

The brain illustration uses Allen Institute for Brain Science Mouse Common Coordinate Framework surfaces (CCF 2017, structures 997, 22, 218 and 262). The mouse head is adapted from Alexandra Rother (2024), SciDraw, \href{https://doi.org/10.5281/zenodo.11431268}{doi:10.5281/zenodo.11431268} (CC BY 4.0).
}

\mainfigurepage{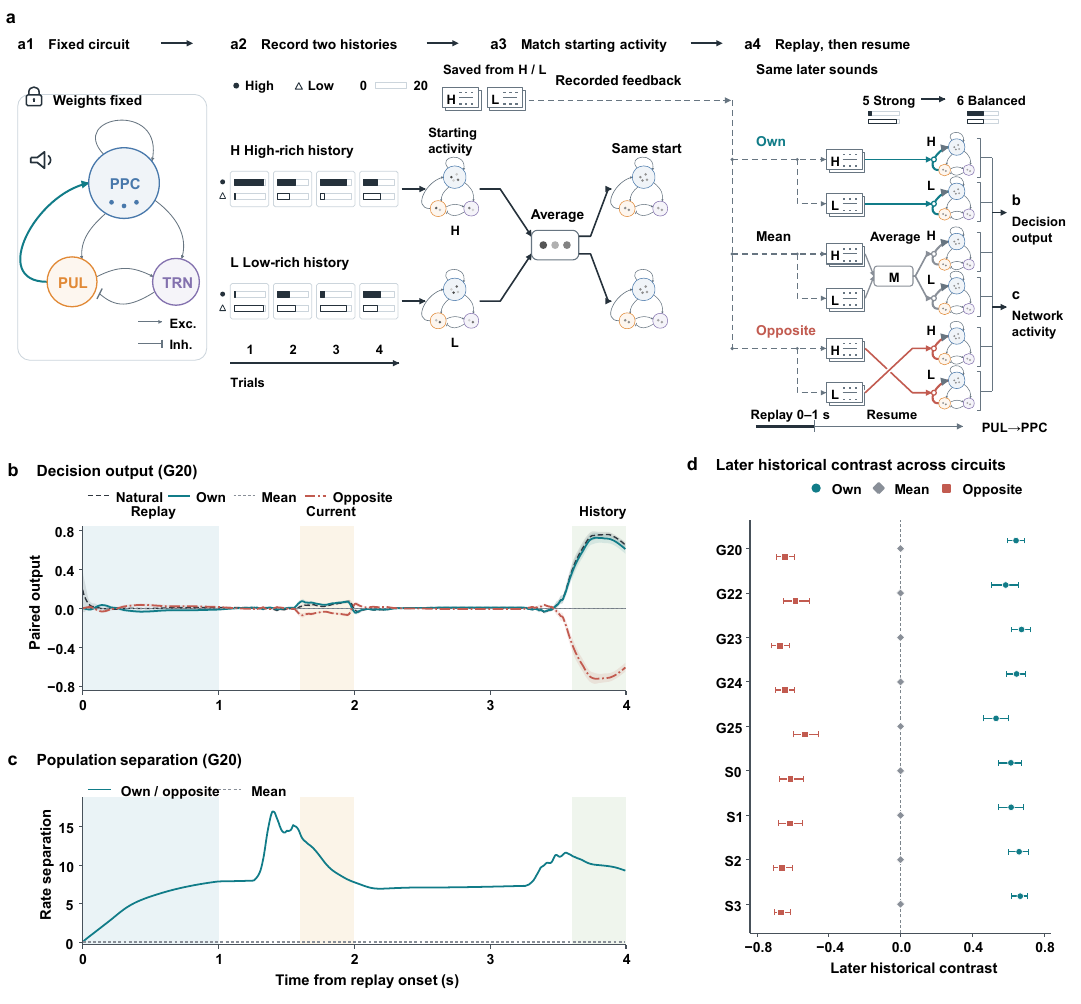}{Recorded thalamic input reconstructs sensory history in circuits initialized at identical neural states}{
\textbf{a,} Sequential reconstruction experiment following four historical trials. Two opposing sensory histories generate different patterns of current along the PUL-to-PPC projection in the same trained circuit. These patterns are recorded during natural evolution. At replay onset, both recipient circuits are initialized at the arithmetic mean of their complete 500-dimensional neural states. They then receive their own recorded input, the average of the two recorded inputs, or the input recorded from the opposing history. After one second of replay, the original PUL-to-PPC connection resumes. Both recipients receive the same subsequent strong fifth stimulus and balanced sixth stimulus. \textbf{b,} Time course of the historical output contrast, \(H_t=(y_t^+-y_t^-)/2\), in G20. Positive and negative values indicate opposite directions of historical influence. Shaded periods identify the replay interval and the two response windows. The natural reference follows the original, intact trajectories. \textbf{c,} Euclidean separation between the firing-rate vectors of the two histories across all 500 units. Own-history and opposite-history replay reconstruct the same magnitude of separation because the two trajectories are exchanged under this symmetric intervention, whereas pair-mean replay leaves them indistinguishable. \textbf{d,} Later historical contrast in the sequential reconstruction experiment across the nine trained circuits. The comparison uses 12 base histories combined with two strong fifth-trial stimuli. Shading in b and horizontal bars in d represent 95\% whole-base bootstrap intervals. Extended Data Figure 2-1 establishes the theoretical framework for reconstruction, and Figure 2-2 examines the respective contributions of input amplitude and spatial pattern.
}

\mainfigurepage{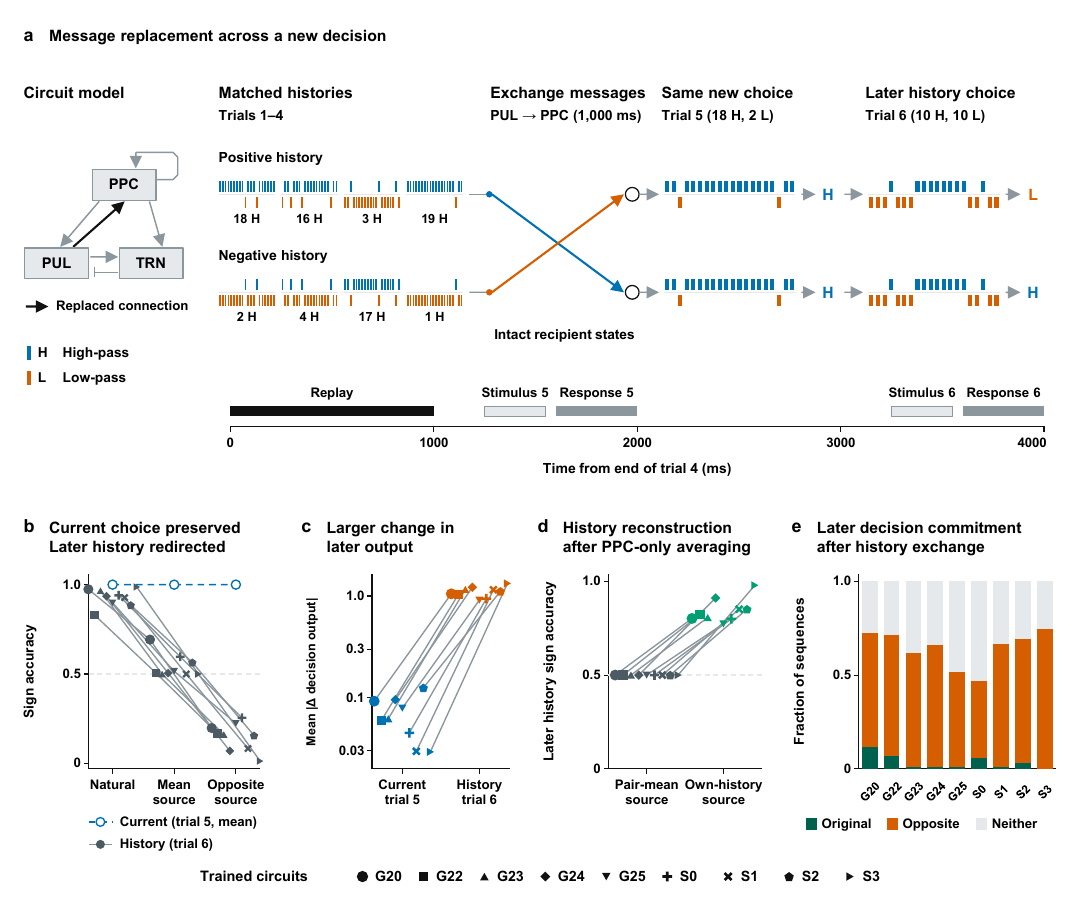}{Substituted sensory history remains available after an accurate intervening judgment}{
\textbf{a,} Sequential replay experiment. Two intact recipient circuits experience opposing four-trial sensory histories and then exchange their recorded PUL-to-PPC inputs for 1,000 ms. Both subsequently receive the same strong fifth stimulus and the same balanced sixth stimulus. In the illustrated G20 example, both circuits correctly judge the fifth stimulus, but their sixth-trial choices follow the substituted histories. \textbf{b,} Response-window mean-sign accuracy for the intervening sensory judgment and the subsequent history-guided judgment under natural evolution, pair-mean replay, and opposite-history replay. Sixth-trial accuracy is evaluated relative to the original history target. \textbf{c,} Absolute change in the response-window mean output produced by opposite-history replay, relative to natural evolution, during the fifth and sixth judgments. \textbf{d,} Later mean-sign accuracy after averaging the initial PPC states, comparing 500-ms pair-mean and own-history replay. \textbf{e,} Fraction of sequences producing decision commitment to the original history, the substituted history, or neither history after 1,000-ms opposite-history replay. Commitment requires the output to reach the corresponding threshold (\(\pm0.75\)) without crossing the opposite threshold during the response window. Markers link the same fixed circuits in b--d. Each condition uses 48 base histories crossed with four strong fifth stimuli and both history signs. Extended Data Figure 3-1 shows the current sensory responses across stimulus mixtures.
}

\mainfigurepage{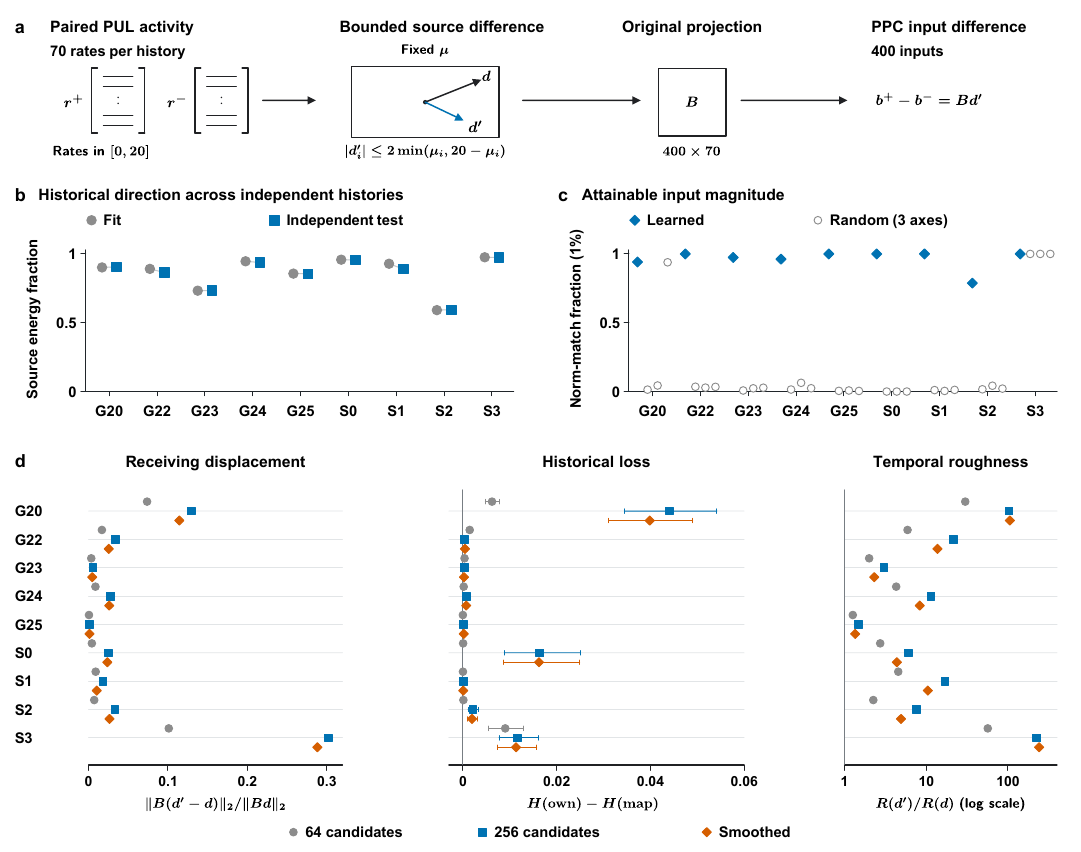}{The organization of pulvinar activity constrains the historical input patterns that can reach cortex}{
\textbf{a,} Geometry of source-constrained input. The two history-dependent PUL firing-rate vectors have a common mean, \(\mu_T\), and differ by \(d_T\). An alternative source difference, \(d_T'\), must keep all 70 PUL firing rates within \([0,20]\). The resulting difference in input to the 400 PPC neurons is obtained through the original trained projection, \(B\). \textbf{b,} Fraction of the PUL rate-difference energy captured by the leading history direction. The direction is estimated using 12 fitting histories and evaluated on 12 independently sampled test base histories per circuit. \textbf{c,} Fraction of evaluated time--context combinations in which a candidate pattern of PUL activity generates a cortical current difference within 1\% of its natural magnitude. Blue diamonds represent the learned source direction, and open circles represent three fixed random directions per circuit. Along each clipped source path, matching uses the first crossing of the natural receiving-current norm or the largest attainable norm when matching is infeasible. \textbf{d,} Source-constrained reallocations in 48 new base histories per circuit, crossed with four strong fifth stimuli and both history signs. The left panel measures the relative change in the cortical current-difference pattern. The center panel shows the reduction in later historical contrast relative to own-history replay. The right panel shows temporal roughness, measured from squared changes in the source difference between successive time steps and expressed relative to the natural signal on a logarithmic scale. The maps preserve the paired source mean, firing-rate bounds, receiving-current difference energy, and aggregate input differences to excitatory and inhibitory PPC targets. The 64- and 256-candidate maps share the same ordered candidate sequence. The smoothed map reduces rapid changes in the transformed-minus-natural signal over five time steps and restores receiving-current energy. Horizontal bars on historical loss represent paired 95\% whole-base bootstrap intervals. Extended Data Figures 4-1 to 4-6 examine signal magnitude, directional sensitivity, constrained source geometry, and temporal coherence in greater detail.
}

\mainfigurepage{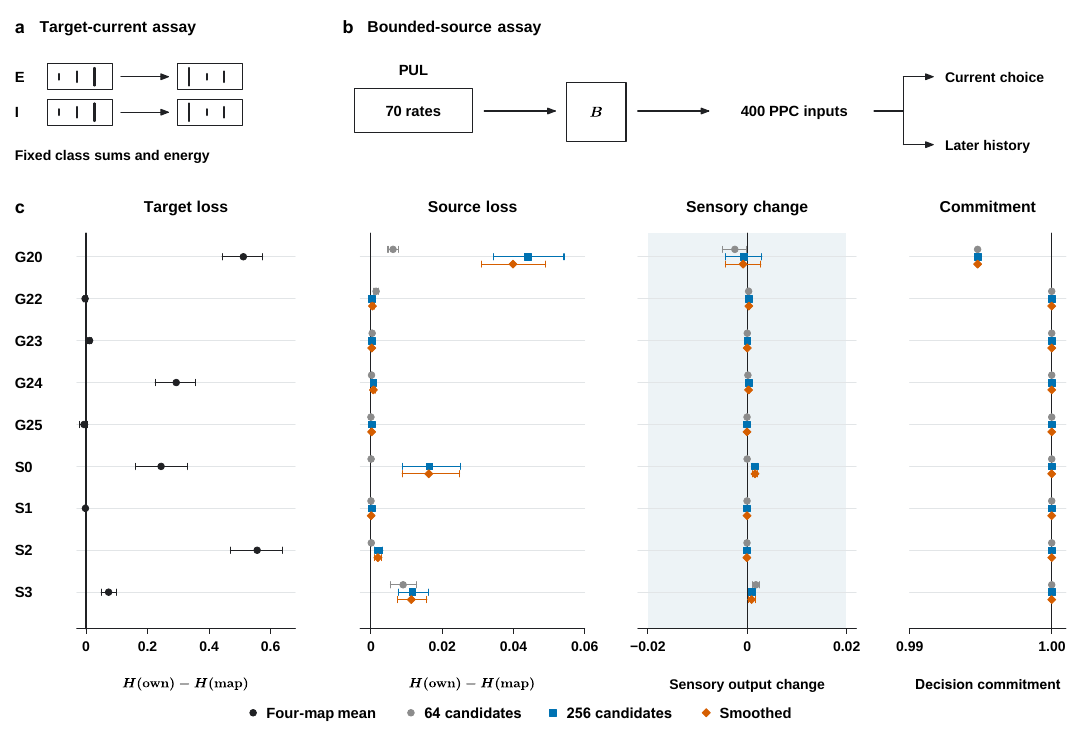}{The spatial allocation of thalamic input influences history-guided behavior under matched aggregate drive}{
\textbf{a,} Target-current redistribution. The history-dependent current difference is redistributed among neurons within each original excitatory or inhibitory PPC target class. The transformation preserves the paired mean, the differential sum and energy within each class, and the nonnegativity of both arriving currents. \textbf{b,} Source-constrained redistribution. Alternative patterns of activity in 70 PUL neurons generate input to 400 PPC neurons through the original trained projection. Both assays use one second of replay after complete-state averaging. \textbf{c,} The four panels compare the effects of redistribution on subsequent history and the intervening sensory judgment. The first panel shows the loss of historical contrast after direct target-current redistribution, averaged over four maps and 24 base histories. The second shows historical loss under the three bounded-source transformations. The third and fourth panels show the corresponding changes in signed sensory output and current decision commitment. The source analyses use the same 48 independent base histories crossed with four strong fifth stimuli. Historical loss is defined as the own-history replay contrast minus the transformed contrast. Horizontal bars represent paired 95\% whole-base bootstrap intervals; shading marks the predefined sensory equivalence range of \(\pm\)0.02 output units. Circles, squares, and diamonds represent the 64-candidate, 256-candidate, and smoothed source maps, respectively. Extended Data Figures 5-1 to 5-3 show the achieved redistribution, individual map effects, and sensory-output changes at each stimulus strength.
}

\mainfigurepage{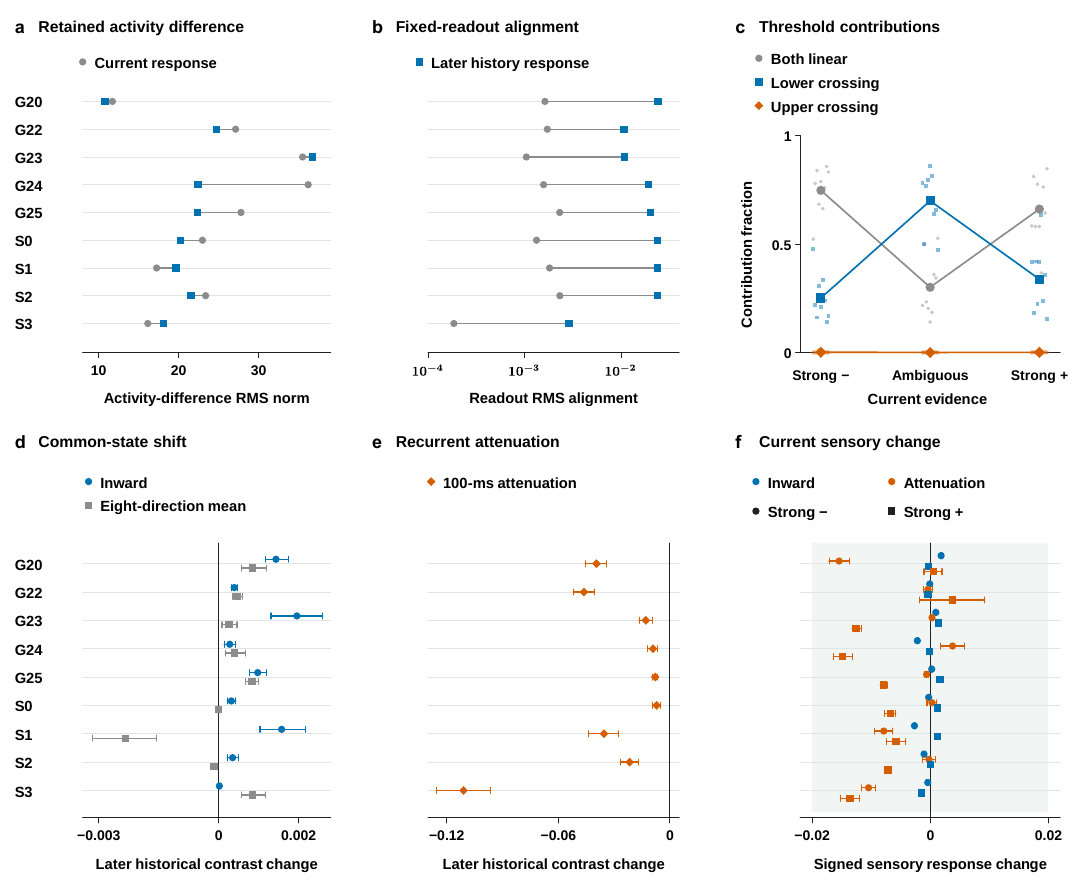}{Retained historical activity, neuronal thresholds, and recurrent interactions regulate the influence of history on choice}{
\textbf{a,b,} Magnitude of the population firing-rate difference and its RMS alignment with the trained decision readout (Equation 10), respectively, during the strong current judgment and the subsequent ambiguous history-guided judgment. Both quantities are measured after 500-ms own-history replay from intact states, using 16 base histories crossed with four strong stimuli. Lines connect the two response windows for each circuit. The alignment axis is logarithmic. \textbf{c,} Fractions of the summed absolute PPC readout contributions associated with different neuronal activation regimes. Paired states are classified according to whether they share the active segment of the firing-rate function or lie on opposite sides of its lower or upper threshold. Small points represent individual circuits; large connected points represent circuit means. \textbf{d,} Change in the subsequent historical contrast produced by a small common displacement of the two PPC states toward the active firing-rate range, compared with the mean effect of eight equal-norm random displacements. \textbf{e,} Change in subsequent historical contrast after partially attenuating the recurrent-drive difference for 100 ms while preserving the mean recurrent drive. \textbf{f,} Corresponding changes in current sensory output under strong negative and positive stimuli. Shading marks the predefined equivalence range of \(\pm\)0.02 output units. Panels c--f use 48 independent test base histories per circuit. Perturbation strengths in d--f were selected using 12 separate calibration histories. Bars in d--f represent paired 95\% whole-base bootstrap intervals. Extended Data Figure 6-1 illustrates evidence-dependent expression using minimal nonlinear models, Figure 6-2 decomposes historical readout contributions, and Figures 6-3 to 6-5 examine the effects and calibration of receiver perturbations.
}

\mainfigurepage{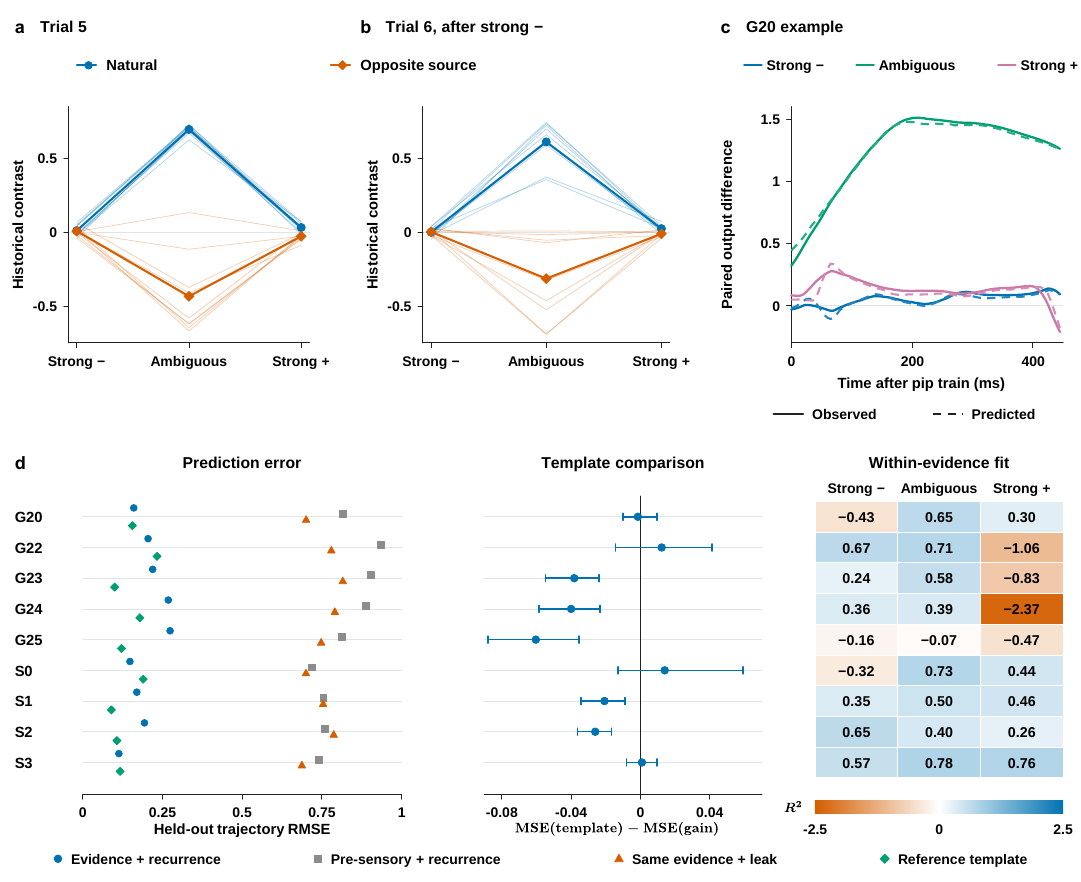}{Sensory evidence regulates historical expression and shapes the subsequent evolution of history-dependent output}{
\textbf{a,b,} Historical contrasts under strong negative, ambiguous, and strong positive sensory evidence on trial 5 and on trial 6 following a fixed strong-negative fifth stimulus, respectively. Changing evidence within a trial position allows its effect to be distinguished from the effect of sequence position. Natural evolution and 500-ms opposite-history replay are compared using 12 base histories per circuit. Thin curves represent circuit means, and thick curves represent their average. \textbf{c,} Mean observed (solid) and predicted (dashed) output differences between opposing histories in G20 over the 450 ms (90 time steps) following the sensory pip train. Colors identify the strength of current sensory evidence. \textbf{d,} Evaluation of conditional propagation models and the reference-response template on independent test histories. The left panel compares trajectory RMSE for recurrent propagation using evidence-specific or pre-sensory activation gains, leak-only propagation using the same evidence-specific gains, and the independent evidence-specific mean template. The center panel shows template-minus-propagation mean squared error, with positive values favoring recurrent gain propagation and negative values favoring the template. Horizontal bars give paired 95\% whole-base intervals. The right panel shows within-evidence \(R^2\) for evidence-specific recurrent propagation, using the single mean observed output difference across all test histories and time steps at each evidence strength as the baseline. Panels c and d use 12 reference histories and 24 independent test base histories. The propagation models are initialized from the measured state difference of each test pair, whereas the template uses the average output-difference trajectory of the reference histories at the corresponding evidence strength. Extended Data Figures 7-1 to 7-3 show retained state differences, the complete evidence-by-position comparison, and additional prediction benchmarks.
}

\restoregeometry
\endgroup
\subsection{Discussion}\label{discussion}

\subsubsection{From the maintenance of sensory history to its use in subsequent decisions}\label{a-causal-bridge-from-maintenance-to-use}

Previous studies have established that interactions between the cortex and thalamus contribute to the maintenance of cortical representations and to the use of recent sensory information in decision-making. Our results extend this account by showing that, in trained thalamocortical circuits, returning pulvinar input can also determine the content of the representation that is reconstructed. When the recipient circuits began from identical neural states, replaying inputs recorded from different sensory histories generated distinct subsequent trajectories. Replacing the original input with that recorded from the opposite history reversed the later historical response, and this reversal persisted through an accurate judgment about a new sensory stimulus. These findings establish a causal relationship between the information delivered through a particular thalamocortical pathway and the identity of a later choice.

Retaining information about the past and using that information to guide ongoing behavior are two complementary functions of a neural circuit. A retained representation may be useful for a subsequent decision even when its immediate influence on behavior must remain small. Distinct preparatory and movement-related subspaces in motor cortex illustrate how the same neural population can support different stages of behavior (Elsayed et al., 2016). In our circuits, substantial differences in neural activity associated with opposing histories persisted during strong sensory judgments, while their projections onto the decision readout remained small. Under ambiguous evidence, these activity differences became more strongly aligned with the readout and exerted a much larger influence on choice. The decomposition in Equation 11 provides a quantitative description of this relationship by separating the magnitude of the retained activity difference, its projection onto the decision readout, and the temporal consistency of that projection.

The behavioral data establish why such regulation is important. The influence of previous stimuli can support decisions when current evidence is weak, but it can reduce accuracy when past and present evidence favor opposing choices. Suppressing PPC produced its largest accuracy benefit in this latter condition, and the associated reduction in fitted history weighting persisted after accounting for prior intervention exposure, behavioral carryover, and within-session drift. PUL suppression showed a similar mean tendency, with greater variability across animals after sequential adjustment. These behavioral observations provide the functional context for the circuit mechanisms investigated here. They identify when reducing the influence of history benefits current behavior, while the replay experiments reveal how the content of thalamocortical input can determine which history subsequently influences choice.

\subsubsection{How the organization of thalamic input and receiving dynamics shape the use of history}\label{source-content-physical-reachability-and-receiving-computation}

What makes a pattern of thalamic activity an effective carrier of sensory history? Previous circuit models have demonstrated how retained information about past stimuli can generate recency and central-tendency biases in working memory (Boboeva et al., 2024). Our findings extend this perspective by examining how historical information is organized in the input transmitted between neural populations. Both changes in overall input amplitude and changes in the relative allocation of input across neurons were capable of influencing the reconstructed history, with different circuits relying on these features to different degrees.

The low-dimensional replay experiments further illustrate the relationship between signal content and signal strength. A single spatial direction captured much of the difference between the inputs associated with opposing histories and remained sufficient to reconstruct or reverse later history-guided choices. In the two circuits where dimensional reduction impaired the intervening sensory judgment, restoring the magnitude of the reduced input was sufficient to recover sensory performance while maintaining the reconstructed historical response. Thus, understanding the function of returning thalamic input requires considering both the information conveyed by its pattern and the strength with which that pattern drives the receiving circuit.

Low-dimensional descriptions of population dynamics have provided important insights into working memory and timing in trained networks and recorded neural populations (Sussillo and Barak, 2013; Cueva et al., 2020). Our input-manipulation experiments address a complementary question: how does the geometry of an incoming signal determine its ability to influence the dynamics of the receiving circuit? When differences in arriving current were matched in magnitude, the learned historical direction generally produced stronger history-dependent responses than the sampled alternatives. In several circuits, even redistributing input within excitatory and inhibitory target populations substantially weakened subsequent historical influence while preserving the intervening sensory judgment. These findings highlight the potential importance of the specific neurons receiving a signal, beyond the aggregate amount of input delivered to a population.

In the intact circuit, the patterns to which a receiving population is sensitive must also be considered in relation to the patterns that the source population can generate. The firing-rate bounds of the PUL neurons and the geometry of the trained projection constrain the attainable currents in PPC. These constraints favor the natural historical direction over many sampled alternative directions in most of the circuits examined here. Importantly, the range of attainable input patterns and the functional consequences of varying those patterns differ across trained solutions. S3, for example, permitted a substantial temporally coherent reallocation of its source signal while maintaining virtually unchanged historical behavior. Different neural circuits can therefore implement similar cognitive functions while exhibiting distinct degrees of sensitivity and tolerance to changes in their interregional communication patterns.

Once historical information has entered the circuit, its influence depends on how the receiving neurons and recurrent network transform the resulting state differences. Our analysis of activation thresholds revealed that neurons whose history-dependent states lay on opposite sides of the lower rectification threshold contributed strongly to the PPC readout difference under ambiguity. Small shifts in cortical operating states produced modest changes in subsequent historical influence, whereas partially attenuating the differences in recurrent input consistently weakened that influence while preserving current sensory judgments. These observations connect local neuronal nonlinearities and recurrent population dynamics to principles of context-dependent computation, output-weak population dimensions, and attractor-gated information flow (Mante et al., 2013; Kaufman et al., 2014; Finkelstein et al., 2021).

The trajectory-prediction results further clarify how sensory evidence and recurrent interactions contribute to this process. Current sensory evidence produces a substantial component of the output trajectory that is shared across histories, as reflected by the performance of the evidence-specific mean template. Recurrent propagation provides a mechanistic description of how the initial history-dependent differences subsequently evolve, particularly when the current stimulus is ambiguous. The influence of sensory history therefore emerges from the interaction between the incoming historical signal, the evidence-dependent operating state of the circuit, and the recurrent dynamics that transform retained information over time.

\subsubsection{Retaining information for flexible use across successive decisions}\label{from-neural-maintenance-to-sequential-cognitive-availability}

Previous studies of thalamocortical loops and latent working memory have demonstrated mechanisms through which neural systems can maintain and reactivate information over time (Guo et al., 2017; Rose et al., 2016; Wolff et al., 2017; Masse et al., 2019; Barbosa et al., 2020; Muhle-Karbe et al., 2021). Our replay experiments connect these processes to the identity of the information used in a sequence of decisions. A historical signal introduced through the PUL-to-PPC pathway can reconstruct a particular representation, and the subsequent recurrent dynamics can distribute its influence across multiple neural populations. The population through which information initially enters a circuit can therefore differ from the populations through which its influence is ultimately expressed in behavior.

Source-driven reconstruction can also operate at the level of intrinsic neuronal dynamics. In a separate conductance-based thalamocortical model (Russo et al., 2025; Krishnan et al., 2016), we examined how cortical feedback contributes to the reconstruction of a history-dependent difference in the inactivation gate of the thalamic T-type calcium current (Extended Theory ET7). This complementary analysis illustrates how synaptic input can reconstruct information retained in voltage-dependent neuronal variables, extending the mechanistic framework across different levels of neural organization.

Perturbing neural activity and examining its subsequent recovery provides an important approach to studying the maintenance of neural representations. Previous work in frontal cortex has connected recovery dynamics to redundancy across coupled populations and to discrete attractor dynamics (Li et al., 2016; Inagaki et al., 2019).

Our theoretical analysis identifies an additional geometric aspect of neural reconstruction. Reciprocal circuits with identical eigenvalue spectra and identical decay dynamics when either population is held fixed can exhibit different recovery trajectories following sequential erasure of their population states (Extended Theory ET6). The eigenvalues determine the timescales of the underlying modes, while the organization of those modes across neural populations determines how they contribute to recovery after a localized intervention. Manipulating the information transmitted through specific pathways therefore provides a complementary approach to understanding how neural representations are reconstructed.

More broadly, a central feature of flexible cognition is the ability to preserve information while regulating its influence on ongoing behavior. Information that has little relevance to one decision may become important for the next. The circuits studied here implement a particular form of this capability: a specific historical representation remains available through an intervening sensory judgment and can subsequently influence behavior when current evidence becomes ambiguous. This provides a concrete circuit-level mechanism for the selective availability of retained information across successive cognitive demands (Dehaene and Changeux, 2011).

Extending this framework to tasks involving flexible rule switching, selective retrieval of different remembered contents, and explicit reporting of retained information could reveal how thalamocortical interactions regulate the availability of information for different cognitive operations.

\subsubsection{Generalization and experimentally testable predictions}\label{mechanism-derived-tests-across-circuit-and-behavioral-scales}

The nine published circuit models provide multiple solutions to the same history-dependent decision task, revealing both shared computational mechanisms and diversity in their implementation. Testing newly generated histories allowed us to examine the generalization of these mechanisms within each trained circuit. The next step is to establish how the identified principles extend across independently trained model ensembles, different circuit architectures, and more diverse sensory and cognitive demands. In particular, a broader range of intermediate sensory strengths could clarify how the influence of retained history changes continuously with current evidence. Training circuits across related tasks may further reveal which mechanisms generalize across different forms of cognitive computation (Yang et al., 2019).

The results yield three experimentally testable predictions for pathway-specific investigations of history-guided behavior in vivo.

First, if returning pulvinar activity carries information capable of specifying a particular historical representation, then stimulation of PUL-to-PPC terminals should produce different subsequent effects depending on the information conveyed by the stimulation pattern. Following a transient intervention that reduces cortical history-dependent activity differences, patterned stimulation corresponding to congruent or opposing histories should selectively bias a later ambiguous choice. This bias should persist after stimulation has ended, including when the animal subsequently makes an accurate judgment about an intervening strong stimulus.

Second, if current sensory evidence regulates the behavioral expression of retained history, then increasing its strength should substantially weaken historical choice bias while preserving a larger fraction of the underlying history-dependent neural activity or state difference. Recording PPC and PUL populations while independently varying stimulus strength and trial position would allow these two components to be compared directly.

Third, if the efficacy of thalamic input depends on its fine spatial organization, changing the distribution of input across cortical neurons while preserving aggregate drive should alter subsequent history-guided behavior. The magnitude of this effect may vary across circuits, reflecting their sensitivity to the detailed organization of the incoming signal.

Together, these predictions provide a route from the causal mechanisms identified in trained circuits to experiments that examine how specific patterns of thalamocortical activity influence the retention and use of sensory history in the brain.

\subsubsection{Conclusion}\label{conclusion}

Our findings identify three complementary processes through which a thalamocortical circuit can support history-guided decision-making. Returning pulvinar input can specify which sensory history is reconstructed, recurrent interactions can sustain the influence of that history through subsequent neural processing, and current sensory evidence can regulate how strongly the retained information affects behavior. The effectiveness of these processes depends on the organization of the thalamic signal, the constraints imposed by its projection to cortex, and the nonlinear dynamics through which the receiving circuit transforms and expresses retained information. Together, these mechanisms allow information about the past to remain available for later decisions while the circuit responds accurately to the sensory demands of the present.

\subsection{Author contributions}

Yanzhe Zhao developed the theoretical mechanisms, designed and performed the computational experiments, analyzed the data, and wrote the original draft. Yongzhang Ji contributed to the theoretical analysis, led figure preparation, and contributed to writing and revising the manuscript.

\subsection{Funding}

This work received no specific funding.

\subsection{Competing interests}

The authors declare no competing interests.

\subsection{References}\label{references}
\begingroup
\setlength{\parskip}{4pt plus 1pt minus 1pt}

Akrami A, Kopec CD, Diamond ME, Brody CD (2018) Posterior parietal cortex represents sensory history and mediates its effects on behaviour. Nature 554:368--372.
\url{https://doi.org/10.1038/nature25510}

Barbosa J, Stein H, Martinez RL, Galan-Gadea A, Li S, Dalmau J, Adam
KCS, Valls-Solé J, Constantinidis C, Compte A (2020) Interplay between
persistent activity and activity-silent dynamics in the prefrontal
cortex underlies serial biases in working memory. Nat Neurosci
23:1016--1024. \url{https://doi.org/10.1038/s41593-020-0644-4}

Boboeva V, Pezzotta A, Clopath C, Akrami A (2024) Unifying network model links recency and central tendency biases in working memory. eLife 12:RP86725.
\url{https://doi.org/10.7554/eLife.86725}

Bolkan SS, Stujenske JM, Parnaudeau S, Spellman TJ, Rauffenbart C, Abbas AI, Harris AZ, Gordon JA, Kellendonk C (2017) Thalamic projections sustain prefrontal activity during working memory maintenance. Nat Neurosci 20:987--996.
\url{https://doi.org/10.1038/nn.4568}

Cueva CJ, Saez A, Marcos E, Genovesio A, Jazayeri M, Romo R, Salzman CD, Shadlen MN, Fusi S (2020) Low-dimensional dynamics for working memory and time encoding. Proc Natl Acad Sci USA 117:23021--23032.
\url{https://doi.org/10.1073/pnas.1915984117}

Dehaene S, Changeux JP (2011) Experimental and Theoretical Approaches to
Conscious Processing. Neuron 70:200--227.
\url{https://doi.org/10.1016/j.neuron.2011.03.018}

Elsayed GF, Lara AH, Kaufman MT, Churchland MM, Cunningham JP (2016) Reorganization between preparatory and movement population responses in motor cortex. Nat Commun 7:13239.
\url{https://doi.org/10.1038/ncomms13239}

Finkelstein A, Fontolan L, Economo MN, Li N, Romani S, Svoboda K (2021)
Attractor dynamics gate cortical information flow during
decision-making. Nat Neurosci 24:843--850.
\url{https://doi.org/10.1038/s41593-021-00840-6} Author correction:
\url{https://doi.org/10.1038/s41593-021-00869-7}.

Firth D (1993) Bias reduction of maximum likelihood estimates.
Biometrika 80:27--38. \url{https://doi.org/10.1093/biomet/80.1.27}

Fritsche M, Mostert P, de Lange FP (2017) Opposite effects of recent history on perception and decision. Curr Biol 27:590--595.
\url{https://doi.org/10.1016/j.cub.2017.01.006}

Guo ZV, Inagaki HK, Daie K, Druckmann S, Gerfen CR, Svoboda K (2017)
Maintenance of persistent activity in a frontal thalamocortical loop.
Nature 545:181--186. \url{https://doi.org/10.1038/nature22324}

Hosford PS, Mei H, Tagomori H, Hayde CP, Abdelaal MS, Nakajima M,
Schmitt LI (2026) Control of representation updating by higher-order
thalamus enables history-based decision-making. Neuron
114:3700--3720.e16. \url{https://doi.org/10.1016/j.neuron.2026.04.045}

Hwang EJ, Dahlen JE, Mukundan M, Komiyama T (2017) History-based action selection bias in posterior parietal cortex. Nat Commun 8:1242.
\url{https://doi.org/10.1038/s41467-017-01356-z}

Inagaki HK, Fontolan L, Romani S, Svoboda K (2019) Discrete attractor dynamics underlies persistent activity in the frontal cortex. Nature 566:212--217.
\url{https://doi.org/10.1038/s41586-019-0919-7}

Kaufman MT, Churchland MM, Ryu SI, Shenoy KV (2014) Cortical activity in
the null space: permitting preparation without movement. Nat Neurosci
17:440--448. \url{https://doi.org/10.1038/nn.3643}

Krishnan GP et al. (2016) Cellular and neurochemical basis of sleep stages in the thalamocortical network. eLife 5:e18607. \url{https://doi.org/10.7554/eLife.18607}

Li N, Daie K, Svoboda K, Druckmann S (2016) Robust neuronal dynamics in premotor cortex during motor planning. Nature 532:459--464.
\url{https://doi.org/10.1038/nature17643}

Libby A, Buschman TJ (2021) Rotational dynamics reduce interference between sensory and memory representations. Nat Neurosci 24:715--726.
\url{https://doi.org/10.1038/s41593-021-00821-9}

Mante V, Sussillo D, Shenoy KV, Newsome WT (2013) Context-dependent
computation by recurrent dynamics in prefrontal cortex. Nature
503:78--84. \url{https://doi.org/10.1038/nature12742}

Masse NY, Yang GR, Song HF, Wang XJ, Freedman DJ (2019) Circuit
mechanisms for the maintenance and manipulation of information in
working memory. Nat Neurosci 22:1159--1167.
\url{https://doi.org/10.1038/s41593-019-0414-3}

Muhle-Karbe PS, Myers NE, Stokes MG (2021) A Hierarchy of Functional
States in Working Memory. The Journal of Neuroscience 41:4461--4475.
\url{https://doi.org/10.1523/JNEUROSCI.3104-20.2021}

Murray JD, Bernacchia A, Roy NA, Constantinidis C, Romo R, Wang XJ (2017) Stable population coding for working memory coexists with heterogeneous neural dynamics in prefrontal cortex. Proc Natl Acad Sci USA 114:394--399.
\url{https://doi.org/10.1073/pnas.1619449114}

Rikhye RV, Gilra A, Halassa MM (2018) Thalamic regulation of switching
between cortical representations enables cognitive flexibility. Nat
Neurosci 21:1753--1763. \url{https://doi.org/10.1038/s41593-018-0269-z}

Rose NS, LaRocque JJ, Riggall AC, Gosseries O, Starrett MJ, Meyering EE,
Postle BR (2016) Reactivation of latent working memories with
transcranial magnetic stimulation. Science 354:1136--1139.
\url{https://doi.org/10.1126/science.aah7011}

Russo S et al. (2025) Thalamic feedback shapes brain responses evoked by cortical stimulation in mice and humans. Nat Commun 16:3627. \url{https://doi.org/10.1038/s41467-025-58717-2}

Saalmann YB, Pinsk MA, Wang L, Li X, Kastner S (2012) The pulvinar regulates information transmission between cortical areas based on attention demands. Science 337:753--756.
\url{https://doi.org/10.1126/science.1223082}

Schmitt LI, Wimmer RD, Nakajima M, Happ M, Mofakham S, Halassa MM (2017)
Thalamic amplification of cortical connectivity sustains attentional
control. Nature 545:219--223. \url{https://doi.org/10.1038/nature22073}

Shrikumar A, Greenside P, Kundaje A (2017) Learning Important Features
Through Propagating Activation Differences. Proc Mach Learn Res
70:3145--3153. \url{https://proceedings.mlr.press/v70/shrikumar17a.html}

Sussillo D, Barak O (2013) Opening the black box: low-dimensional dynamics in high-dimensional recurrent neural networks. Neural Comput 25:626--649.
\url{https://doi.org/10.1162/NECO_a_00409}

van den Berg M, Dryden EB, Kappeler T (2014) Isospectrality and heat
content. Bull Lond Math Soc 46:793--808.
\url{https://doi.org/10.1112/blms/bdu035}

Wimmer RD, Schmitt LI, Davidson TJ, Nakajima M, Deisseroth K, Halassa MM (2015) Thalamic control of sensory selection in divided attention. Nature 526:705--709.
\url{https://doi.org/10.1038/nature15398}

Wolff MJ, Jochim J, Akyürek EG, Stokes MG (2017) Dynamic hidden states
underlying working-memory-guided behavior. Nat Neurosci 20:864--871.
\url{https://doi.org/10.1038/nn.4546}

Yang GR, Joglekar MR, Song HF, Newsome WT, Wang XJ (2019) Task representations in neural networks trained to perform many cognitive tasks. Nat Neurosci 22:297--306.
\url{https://doi.org/10.1038/s41593-018-0310-2}\par
\endgroup

\clearpage

\newgeometry{left=13mm,right=13mm,top=17mm,bottom=17mm}
\begin{landscape}
\setlength{\tabcolsep}{4pt}
\AtBeginEnvironment{longtable}{\footnotesize\singlespacing}
\setlength{\LTpre}{2pt}
\setlength{\LTpost}{4pt}
\subsection{Table 1. Summary of evidence supporting the principal findings}\label{table-1.-evidence-for-the-main-findings}

{\def\LTcaptype{none} 
\begin{longtable}[]{@{}
  >{\raggedright\arraybackslash}p{(\linewidth - 8\tabcolsep) * \real{0.1800}}
  >{\raggedright\arraybackslash}p{(\linewidth - 8\tabcolsep) * \real{0.1450}}
  >{\raggedright\arraybackslash}p{(\linewidth - 8\tabcolsep) * \real{0.2250}}
  >{\raggedright\arraybackslash}p{(\linewidth - 8\tabcolsep) * \real{0.3800}}
  >{\raggedright\arraybackslash}p{(\linewidth - 8\tabcolsep) * \real{0.0700}}@{}}
\toprule\noalign{}
\begin{minipage}[b]{\linewidth}\raggedright
Scientific question or analysis
\end{minipage} & \begin{minipage}[b]{\linewidth}\raggedright
Sampling unit
\end{minipage} & \begin{minipage}[b]{\linewidth}\raggedright
Comparison
\end{minipage} & \begin{minipage}[b]{\linewidth}\raggedright
Main effect and uncertainty
\end{minipage} & \begin{minipage}[b]{\linewidth}\raggedright
Figure
\end{minipage} \\
\midrule\noalign{}
\endhead
\bottomrule\noalign{}
\endlastfoot
Behavioral cost of conflicting sensory history & 9 PPC, 10 PUL, and 6 TRN mice & Difference between
the intervention effects under conflict and agreement & PPC +9.3 pp
[5.5, 13.0] (\(p=0.004\), Holm \(p=0.012\)).
PUL +6.1 pp [1.8, 11.2] (\(p=0.027\), Holm \(p=0.055\)).
TRN -0.8 pp [-5.9, 4.0] & 1d \\
Intervention-related change in the influence of history & Mouse within each cohort & Choice
models with prior intervention, behavior, and session drift &
Adjusted \(\beta_{hL}\) in log odds per unit history. PPC -0.70
[-1.11, -0.27]. PUL -0.47 [-1.07, 0.22].
TRN -0.08 [-0.50, 0.32] & 1e \\
Reconstruction and reversal of sensory history by input replay & 96 base histories per circuit & Opposite-history
replay with matched future input & Later historical contrast from -0.64 to
-0.38, with all nine intervals below zero & Results, Fig.~2\textsuperscript{a} \\
Persistence of reconstructed history through a sensory judgment & 48 base histories per circuit & Opposite-history
replay followed by a strong stimulus and a balanced probe & Current
mean-sign accuracy 100\% and decision commitment 99.5--100\%.
Later historical contrast from -0.66 to -0.35 & 3 \\
Functional sensitivity to the spatial direction of input & 24 base histories per circuit & Learned
direction versus 16 random directions at matched difference energy &
Increase relative to random directions 0.32--0.65, with all nine intervals above zero.
Three of 144 alternatives reached the learned magnitude & 4-4 \\
Historical replay through the constrained PUL projection & 12 test base histories per circuit & Own, mean,
opposite, learned and three random source directions through \(B\) &
Own-history contrast 0.53--0.68. Learned-axis contrast 0.52--0.71.
Opposite replay reversed all nine circuits and preserved current
commitment at 100\%. Norm matching reached 78.8--100\% for the learned
axis and a median 2.1\% for random axes & 2, 4 \\
Source-constrained reallocation at matched receiving energy & 12 base histories per circuit & Two maps at fixed
source mean, receiving energy, and E/I totals & Realized changes at
21.1--100\% of time and context points. Relative receiving displacement
0.56--40.8\%. S3 historical loss 0.036 [0.014, 0.061] with 100\%
current commitment & 4-3 \\
Fine allocation of input within cortical target populations & 24 base histories per circuit &
Four maps at fixed class sums and difference energy & Historical loss
0.51 [0.44, 0.57] in G20, 0.29 [0.22, 0.36] in G24,
and 0.24 [0.16, 0.33] in S0 & 5 \\
Retention and behavioral expression of history-dependent activity & 16 base histories per circuit &
Compare current and later readout alignment across four strong stimuli
and both signs & Mean later-to-current RMS alignment ratio 6.2--17.9
across circuits & 6a,b, 6-2 \\
Effect of sensory evidence independently of trial position & 12 base histories per circuit & Three stimulus
strengths at each of two trial positions, starting from the same preceding
states & Fifth-trial output contrasts under strong evidence were
2.0--7.7\% of those under ambiguity. Corresponding state norms were
49.2--111.7\% & 7 \\
Generalization of source reallocations across histories & 48 base histories per circuit &
Three bounded-source maps, four strengths, and paired histories &
Historical loss 0.044 [0.034, 0.054] in G20, 0.016
[0.009, 0.025] in S0, and 0.012 [0.008, 0.016] in S3.
All 27 pooled 95\% confidence intervals for sensory-output change within \(\pm0.02\) & 4d, 5c, 5-3 \\
Temporally coherent alternative source patterns & 48 further base histories per circuit & Fixed spatial reflections with roughness at most
1.25 times the natural value & S3 reached 14.2\% receiving
displacement with historical loss -0.0001 [-0.0025, 0.0024] & 4-5 \\
Cortical operating-state and recurrent-drive perturbations & 12 calibration and 48 test base histories per circuit & Small inward and equal-norm random shifts, and 100-ms
recurrent attenuation & Later historical loss under attenuation
0.007--0.111. The 36 paired confidence intervals for changes in sensory output lay within \(\pm0.02\),
with 96.9--100\% current commitment & 6, 6-5 \\
Prediction of history-dependent output trajectories & 12 reference and 24 test base histories &
Evidence-dependent or pre-sensory gains with recurrence or leak, and a
reference template & At the same evidence-specific gain, RMSE was
0.11--0.27 with recurrence and 0.69--0.82 with leak. The reference template achieved lower pooled error than evidence-specific recurrent propagation in six circuits & 7d, 7-3 \\
\end{longtable}
}

{\footnotesize Intervals are 95\% confidence intervals obtained by resampling the relevant independent sampling units. Behavioral intervals resample mice within each intervention cohort, whereas computational intervals resample whole base histories within each trained circuit, preserving the associated stimulus conditions, history signs, and transformation variants. Ranges across trained circuits summarize the nine released trained circuits. Percentage-point differences are denoted by pp. Continuous historical contrasts and sensory-output changes are expressed in the original model's readout units. \textsuperscript{a}Figure~2 illustrates the sequential reconstruction assay.\par}

\clearpage

\end{landscape}
\restoregeometry

\clearpage
\section{Extended Data}\label{extended-data}
\subsection{Overview}\label{extended-overview}

The extended figures provide additional analyses and control experiments addressing the three principal questions of the article. Figures 1-1 to 3-1 examine the behavioral circumstances under which sensory history influences choice, the ability of returning thalamic input to reconstruct a particular historical distinction, and the selection of sensory stimuli for testing its persistence across decisions.

Figures 4-1 to 5-3 investigate how the organization of the thalamic signal contributes to its function. These experiments distinguish the effects of input magnitude and spatial pattern, characterize the restrictions imposed by the trained PUL-to-PPC projection, and test whether redistributing input among individual target neurons alters later history-guided behavior.

Figures 6-1 to 7-3 examine how reconstructed history remains available while the circuit processes new sensory evidence. They explore nonlinear activation, the relationship between retained neural activity and the decision readout, the effects of receiver perturbations, and the contribution of recurrent dynamics to post-stimulus output trajectories.

The accompanying Extended Theory (Figure 2-1) provides the mathematical definitions, exact decompositions, and theoretical constructions underlying these analyses.

Behavioral confidence intervals were estimated by resampling mice within each intervention cohort. Computational intervals resampled complete base histories within each trained circuit, preserving the associated stimulus conditions, history signs, and transformation variants. The analyses include all nine published trained circuits, together with the sampled perturbation directions and redistribution procedures.

\clearpage

\subsection{Figure 1-1. Temporal organization of intervention exposure and its implications for behavioral comparisons}\label{extended-figure-1-1.-intervention-schedules-carryover-and-treatment-overlap}

\begin{figure}
\centering
\includegraphics[width=0.98\linewidth,height=\textheight,keepaspectratio,alt={Intervention schedules and preceding exposure}]{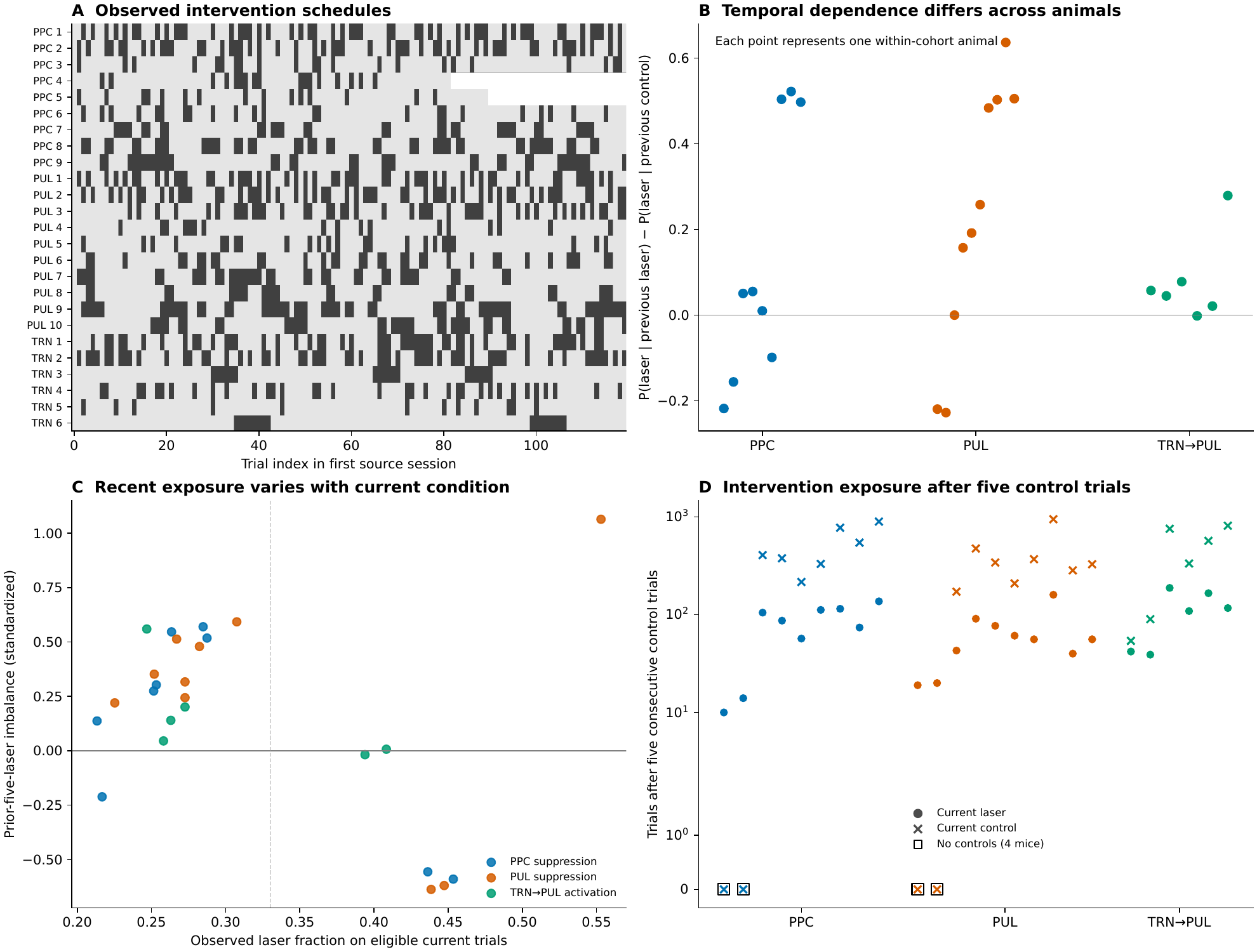}
\end{figure}

\textbf{Figure legend.} \textbf{A,} The first 120 records from the first
available session for each mouse. Dark cells mark intervention, gray
cells mark control, and white cells mark the end of shorter sessions.
\textbf{B,} The difference in current intervention probability after an
intervention versus a control trial. Consecutive trials are paired
within sessions. \textbf{C,} The eligible intervention fraction and the
standardized difference in preceding-five-trial exposure between current
intervention and control. The reference line marks the 33\% intervention
probability described in the source study. \textbf{D,} Counts by current
condition after five consecutive control trials. The symmetric-log axis
displays zeros. The four boxes identify mice whose eligible trials in
this stratum all received current intervention.

\subsubsection{Behavioral definitions and choice
models}\label{extended-behavioral-definitions-and-choice-models}

An intervention delivered on one trial may influence the neural or behavioral state present at the beginning of subsequent trials. Consequently, temporal differences in intervention exposure can complicate comparisons between intervention and control conditions. We first examined the intervention schedules in the released behavioral records to determine whether intervention exposure was temporally clustered and whether trials assigned to different current conditions differed in their recent exposure histories (Figure 1-1A--C). These analyses motivated the carryover and within-session drift adjustments used in the main behavioral analysis.

The analysis includes 14,611 PPC observations (nine mice, 83 sessions),
14,189 PUL observations (ten mice, 80 sessions), and 14,824 TRN→PUL
observations (six mice, 50 sessions). Accuracy is defined by the
recorded trial feedback (which matches choice--target agreement on
\(>99.7\%\) of trials eligible for the conflict analysis, differing in
only 18 PPC and 30 PUL trials).

This comparison establishes the correspondence between the recorded feedback and the accuracy measure defined from the observed choice and target.

Historical evidence is defined as the mean signed stimulus evidence across the preceding five recorded trials. History probes contain equal numbers of high-pass and low-pass pips and carry the history-probe designation in the published dataset. The conflict analysis includes regular trials in which current and historical evidence each favor a particular choice, the animal makes a binary response, and the recorded target agrees with current sensory evidence. Both correct and incorrect trials are included. The choice models use regular trials with recorded binary choices and targets, together with the preceding intervention condition and choice. All three model specifications use the same 13,539 PPC, 13,281 PUL, and 13,520 TRN-to-PUL observations. 

Let \(s\), \(h\), and \(L\) denote current evidence, historical
evidence, and current intervention. The base model includes a session
intercept and \(s,h,L,sL,hL\). For the centered preceding-trial exposure
\(p_1\) and the centered mean exposure at lags 2--5 \(p_{2:5}\), the
carryover model adds the following terms for each \(p\).

\[
p[\gamma_0+\gamma_s s+\gamma_h h+L(\delta_0+\delta_s s+\delta_h h)].
\]

Including these interactions allows recent intervention exposure to modify response bias, current sensory sensitivity, historical weighting, and the effects of the current intervention. Centering the exposure variables allows the history-by-intervention coefficient to be interpreted at their average levels.

Centered preceding choice and correctness enter with their main effects
and interactions with \(L\). The drift model adds centered linear and
quadratic session-position terms, interacting with \(1,s,h\). Session
position is scaled to \([-1,1]\). The history-by-intervention
coefficient \(\beta_{hL}\) is evaluated at the centered values of prior
exposure and behavior. When an unpenalized mouse fit failed the
stability criterion, the corresponding cohort and specification used the
same \(10^{-4}\) ridge penalty on all non-intercept terms.

All lag terms follow the trial order of the released sessions; the
reconstructed five-trial history matches the recorded history target on
95.0\% of PPC (575/605), 100\% of PUL (472/472), and 92.0\% of TRN→PUL
(873/949) probes.

\subsubsection{Comparison after five control
trials}\label{extended-comparison-after-five-control-trials}

We additionally examined trials preceded by five consecutive control trials, providing a common recent intervention history for comparing current intervention and control conditions. This restriction substantially reduced the available sample and treatment overlap. PPC mice 1 and 2 and PUL mice 1 and 2 contributed 10, 14, 19, and 20 eligible trials, respectively, all of which received the current intervention. Estimating a within-mouse treatment contrast requires observations under both current conditions; these four animals therefore contribute observations from a single condition in the restricted comparison. Population-level inference uses the carryover and drift models fitted to the full eligible samples, which provide broader coverage of intervention histories.

\clearpage

\subsection{Figure 1-2. Changes in the influence of past stimuli account for the behavioral benefit under conflict}\label{extended-figure-1-2.-historical-weighting-and-flexible-sensory-effects}

\begin{figure}
\centering
\includegraphics[width=0.98\linewidth,height=\textheight,keepaspectratio,alt={Historical and sensory components of the conflict effect}]{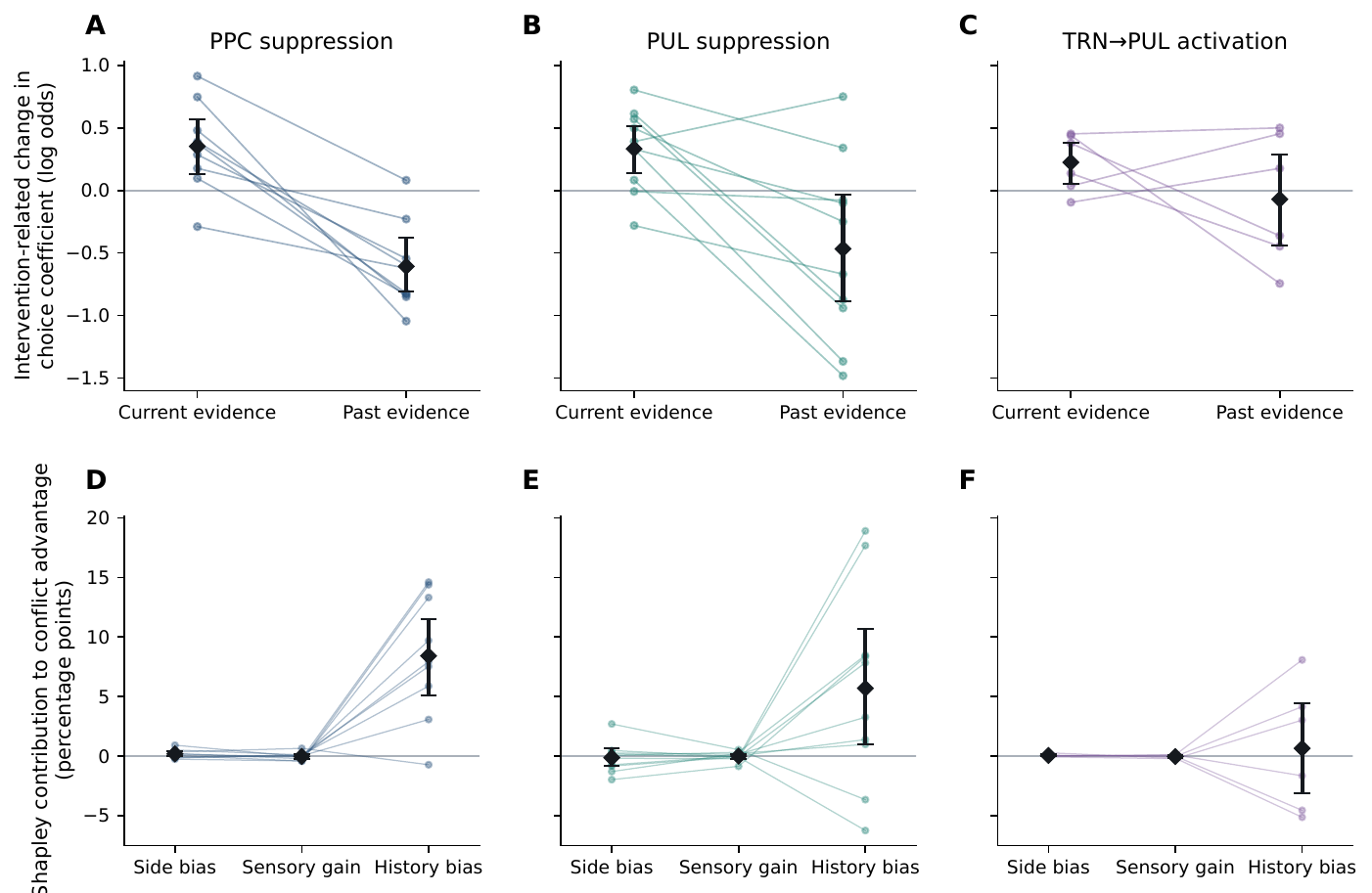}
\end{figure}

\textbf{Figure legend.} \textbf{A--C,} Intervention-related changes in
the fitted log-odds coefficients for current and past evidence under PPC
suppression, PUL suppression, and TRN-to-PUL activation, respectively.
\textbf{D--F,} Contributions of side bias, sensory gain, and history
weighting to the extra accuracy benefit of intervention on conflict
trials, in the same cohort order. Colored lines connect estimates for
individual mice. Black diamonds and bars show cohort means and 95\%
mouse-bootstrap intervals.

The decomposition distinguishes a change in the influence of history from changes in sensory sensitivity or general response preference that could also affect accuracy.

\subsubsection{Choice-model
decomposition}\label{extended-choice-model-decomposition}

An intervention may improve accuracy through several components of the decision process. It may change the animal's general response preference, the influence of current sensory evidence, or the influence of previous stimuli. Because these changes can interact through the nonlinear relationship between the fitted log odds and choice probability, we used a Shapley decomposition to estimate their individual contributions to the additional intervention benefit under history--stimulus conflict.

The decomposition evaluates fitted choice probabilities on a common set of observed covariates and averages the contribution of each component over all six possible orders in which the three components can be introduced. In the PPC cohort, changes in the history terms accounted for 8.40 [5.10, 11.52] percentage points of the additional benefit under conflict, compared with -0.04 points for the sensory terms.

To examine whether this conclusion depended on the functional form of the sensory response, we also used cubic B-splines with knots at -0.5, 0, and 0.5. The predominant contribution of history weighting was maintained across both the feedback-aligned and all-binary-choice samples. A similar pattern was observed in the PUL cohort, whereas TRN-to-PUL activation produced little history-related contribution.

\clearpage

\subsection{Figure 1-3. The effect of intervention on historical weighting varies with current sensory evidence}\label{extended-figure-1-3.-historical-weighting-across-current-evidence-strengths}

\begin{figure}
\centering
\includegraphics[width=0.98\linewidth,height=\textheight,keepaspectratio,alt={Evidence-dependent historical weighting in mouse behavior}]{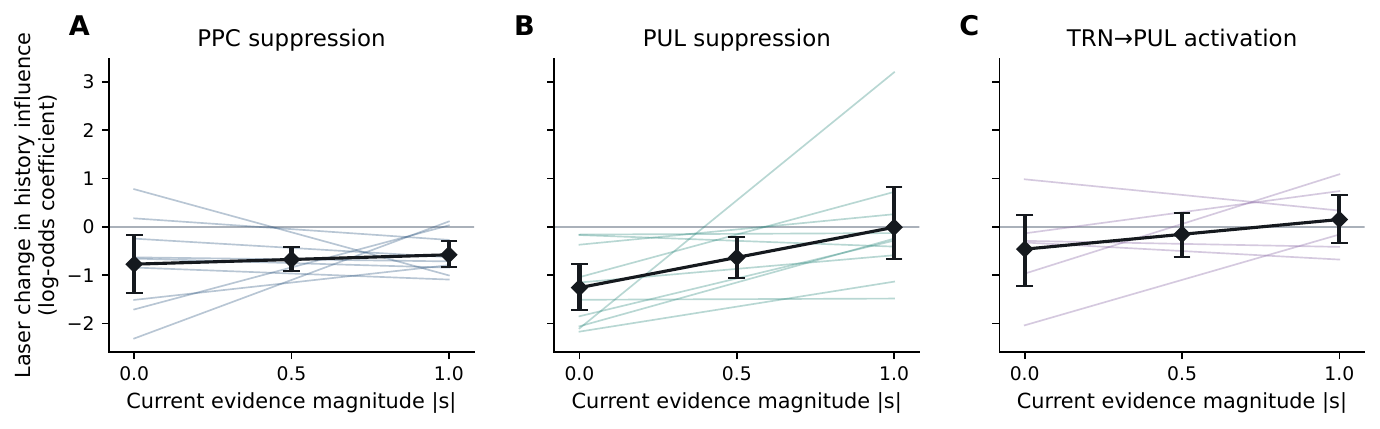}
\end{figure}

\textbf{Figure legend.} \textbf{A--C,} Intervention-related change in
the fitted history coefficient at current-evidence magnitudes
\(|s|=0,0.5,1\) under PPC suppression, PUL suppression, and TRN-to-PUL
activation, respectively. Thin colored lines represent individual mice.
Connected black diamonds and bars show cohort means and 95\%
mouse-bootstrap intervals. The horizontal zero line denotes an unchanged
history coefficient.

\subsubsection{Dependence on current
evidence}\label{extended-dependence-on-current-evidence}

The behavioral influence of history may depend on how informative the current stimulus is. To examine this possibility, we estimated intervention-related changes in history weighting across the magnitude of current sensory evidence, including eligible balanced trials for which historical evidence could be calculated.

PPC suppression reduced the fitted influence of history relatively uniformly across sensory strengths. In contrast, the reduction associated with PUL suppression was greatest when the current stimulus provided weak or ambiguous evidence and became smaller as sensory evidence increased. The PUL-by-evidence-magnitude interaction was +1.25 [0.46, 2.30] log-odds units. This analysis complements the regular-trial comparison in main \hyperlink{main-figure-link-1}{Figure~1e}, which accounts for preceding intervention exposure and within-session drift.

\clearpage

\subsection{Figure 2-1. Theoretical framework for historical reconstruction, retention, and behavioral expression}\label{extended-figure-2-1.-extended-theory-and-causal-definitions}

\textbf{Figure legend.} The Extended Theory develops a mathematical framework for distinguishing the information present in the initial neural state from the information conveyed by an arriving signal, and for explaining how the resulting historical differences influence behavior. ET1--ET3 establish the paired-history dynamics, derive the exact propagation of finite state and input differences, and characterize the input patterns that can be generated by a constrained source population. ET4--ET5 explain how substantial retained differences can produce weak or strong behavioral effects depending on their projection onto the decision readout and the nonlinear operating state of the circuit.

ET6 establishes how reciprocal circuits with identical eigenvalue spectra and identical conditional decay properties can exhibit different reconstruction dynamics following sequential population erasure. ET7 extends the analysis of source-driven reconstruction to the T-type calcium inactivation gate of thalamic relay cells in a conductance-based circuit. ET8--ET9 develop the mathematical basis of the constrained source transformations, receiver perturbations, and independent trajectory forecasts used in the computational experiments.

The feasible PUL currents form \(\{W_{CT}r:0\le r_i\le20\}\). This set describes all nonnegative currents that can be generated through the original trained PUL-to-PPC projection. Because the source contains 70 units, its linear span has dimension at most 70 in the 400-dimensional receiving-current space. Own-history, opposite-history, and pair-mean replay belong to this feasible set. Direct manipulations of arriving currents explore a broader set of nonnegative input patterns and thereby characterize the sensitivity of the receiving circuit independently of the source projection. Concrete linear bipartite examples
in ET6 confirm that paired-erasure recovery can diverge even when
conditional clamp curves and either the intact slow mode (14.7\%
difference in a four-state pair) or the complete eigenvalue spectrum
(0.23\% difference in an eight-state isospectral pair) are matched
identically.

These constructions demonstrate that the ability of a circuit to reconstruct a representation depends on how its internal dynamics are accessed through particular neural populations, in addition to the decay rates of the underlying modes.

\clearpage

\subsection{Figure 2-2. Both the strength and spatial distribution of thalamic input carry information about sensory history}\label{extended-figure-2-2.-amplitude-and-spatial-pattern-carry-historical-content}

\begin{figure}
\centering
\includegraphics[width=0.98\linewidth,height=\textheight,keepaspectratio,alt={Historical effects of exchanging current amplitude and pattern}]{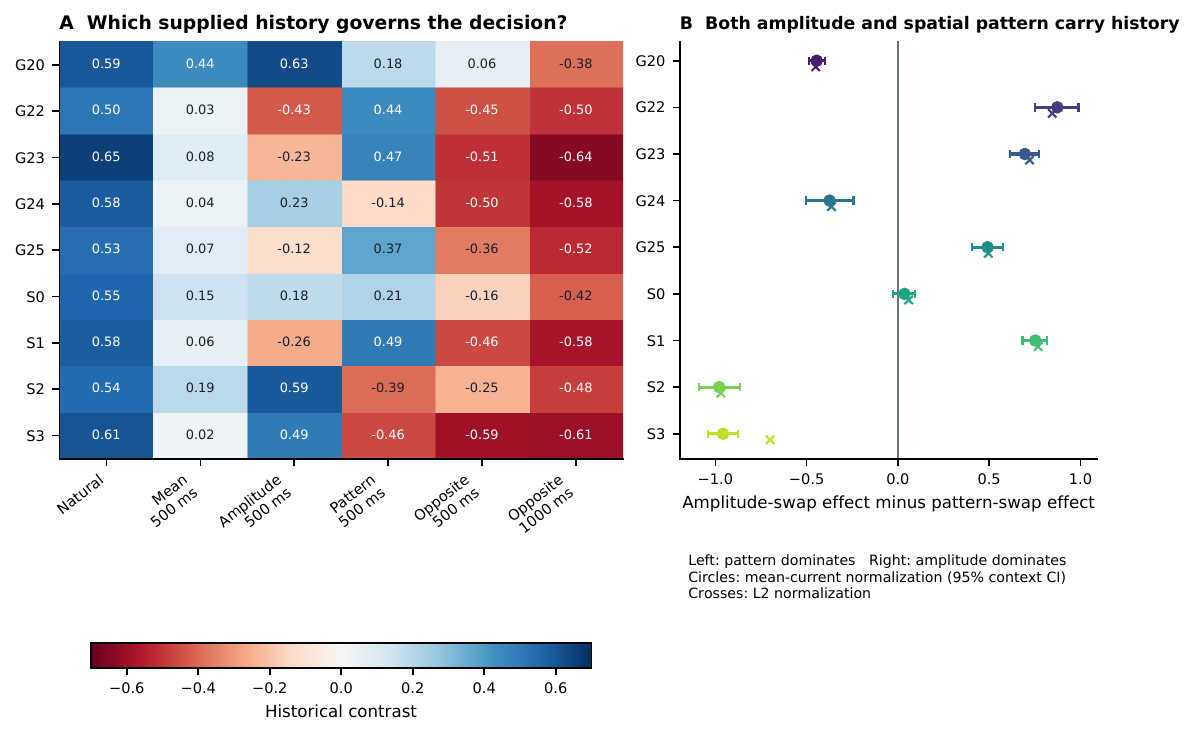}
\end{figure}

\textbf{Figure legend.} \textbf{A,} Signed historical contrasts for
the natural, pair-mean, amplitude-exchanged, pattern-exchanged, and
opposite-history arrival currents across the nine circuits.
\textbf{B,} Comparison of amplitude-exchange and spatial-pattern-exchange effects using mean-current and Euclidean-norm definitions of input amplitude. Error bars denote 95\%
bootstrap intervals across base histories within each circuit.
Amplitude exchange changes the overall input strength while preserving its normalized distribution across target neurons. Pattern exchange changes the normalized distribution while preserving the amplitude of the original input. Comparing these two manipulations reveals how different circuits use these complementary features of the returning signal.

Amplitude exchange produced larger historical effects in G22, G23, G25, and S1, while spatial-pattern exchange produced larger effects in G20, G24, S2, and S3. S0 exhibited more comparable effects of the two exchanges. The relative importance of amplitude and spatial pattern was similar under the mean-current and Euclidean-norm definitions. Main \hyperlink{main-figure-link-2}{Figure~2} illustrates the time course of historical reconstruction following complete-state averaging.

\clearpage

\subsection{Figure 3-1. Selecting strong sensory stimuli for the sequential decision assay}\label{extended-figure-3-1.-current-decisions-across-stimulus-mixtures}

\begin{figure}
\centering
\includegraphics[width=0.98\linewidth,height=\textheight,keepaspectratio,alt={Native sensory decisions across evidence strengths}]{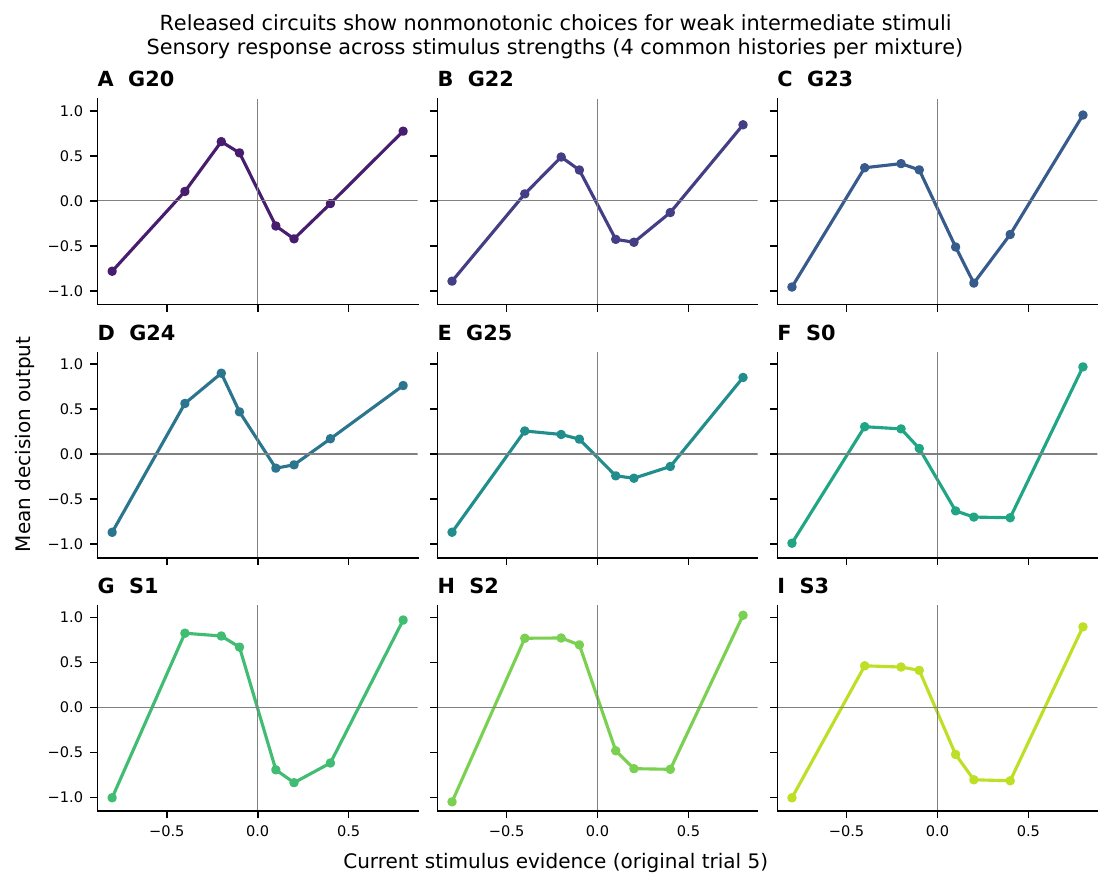}
\end{figure}

\textbf{Figure legend.} \textbf{A--I,} Mean trained decision readout over the
current response window as a function of signed stimulus evidence on the
original fifth trial. Each plot shows one circuit, with four common
histories per mixture. The reference lines mark balanced evidence and
zero mean output.

\subsubsection{Selection of sequential
stimuli}\label{extended-selection-of-sequential-stimuli}

To test whether a reconstructed history survives an accurate new sensory judgment, we first identified stimulus mixtures that the released circuits could judge reliably. Some circuits produced weak or nonmonotonic mean responses at intermediate mixtures in the history-dependent task. We therefore selected the four strongly discriminable fifth-trial mixtures containing 0, 2, 18, or 20 positive pips before evaluating new histories in the sequential experiments.

The comparison in main \hyperlink{main-figure-link-7}{Figure~7} serves a complementary purpose. By including balanced stimuli at the fifth-trial position, it examines how current evidence regulates the expression of history while holding sequence position and the corresponding preceding states fixed.

\clearpage

\subsection{Figure 4-1. Restoring differential-current magnitude recovers intervening sensory performance after dimensional reduction}\label{extended-figure-4-1.-restoring-current-magnitude-rescues-the-intervening-decision}

\begin{figure}
\centering
\includegraphics[width=0.98\linewidth,height=\textheight,keepaspectratio,alt={Source reduction, residual restoration, and energy matching}]{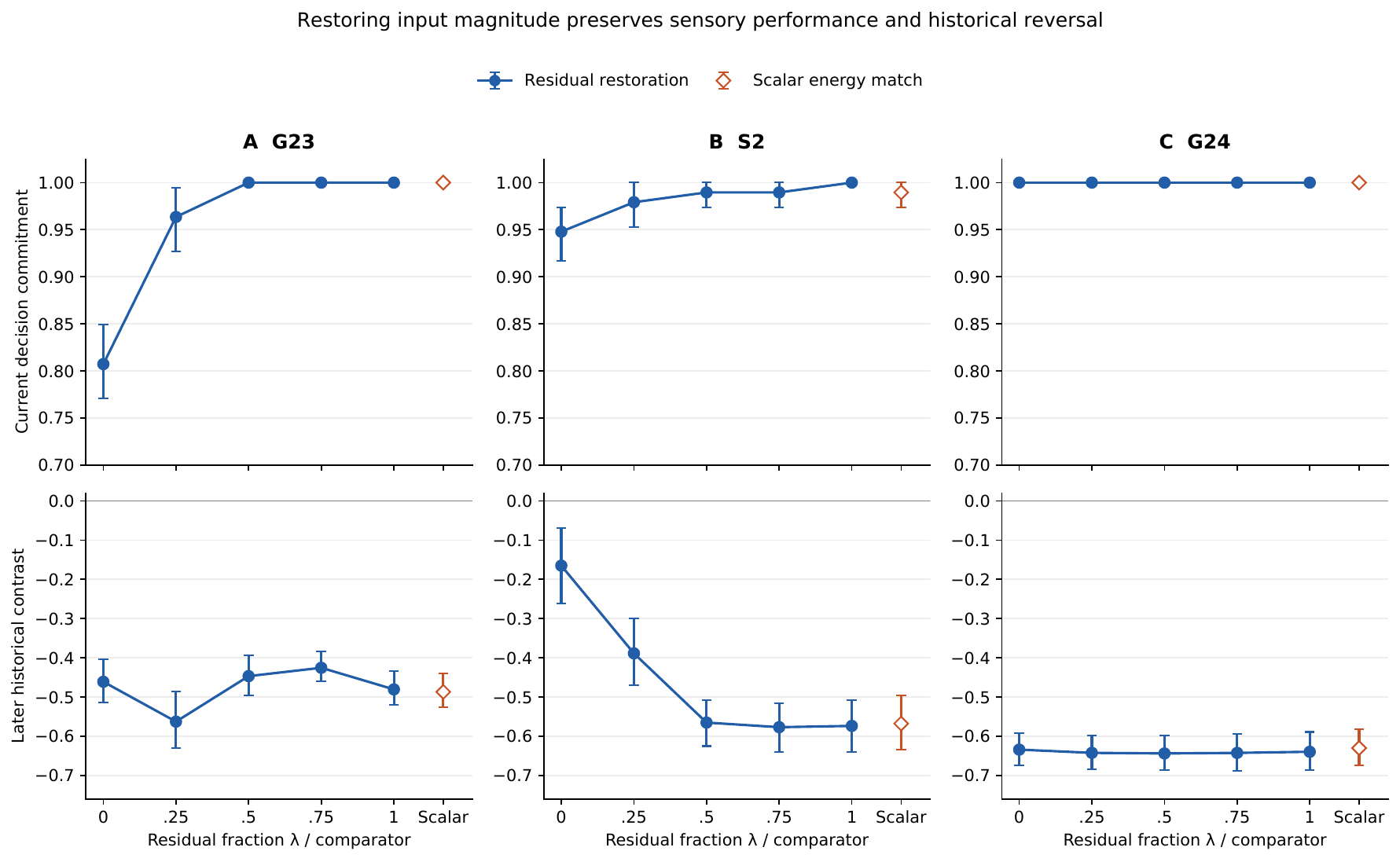}
\end{figure}

\textbf{Figure legend.} \textbf{A--C,} Current decision commitment (upper
row) and later signed historical contrast (lower row) in G23, S2, and
G24, respectively. Blue circles show the fraction of the residual
restored to the rank-one input; open orange diamonds show the scalar
energy-matched input. Points and bars show the means and paired 95\%
base-history bootstrap intervals across 24 held-out histories.

\subsubsection{Restoring input magnitude preserves sensory performance and historical reversal}\label{extended-restoring-input-magnitude}

Reducing the history-dependent current difference to its leading spatial direction changes both the dimensionality and the magnitude of the delivered signal. When this manipulation impairs sensory performance, the impairment could arise from the loss of particular spatial components or from the reduction in differential-current magnitude. We compared these possibilities by gradually restoring the discarded residual components or by rescaling the reduced signal to match the original current-difference norm.

The leading axis of the arriving-current differences was fitted on 12 histories and initially evaluated on eight independent histories.

Increasing the shared background current where needed by 0.2--15.9\% of the natural mean-current norm allowed the projected historical difference to remain rank one while preserving nonnegative arrival currents. Targetwise clipping preserves the natural mean, while global attenuation
can reduce retained difference energy to 0--12.5\% in circuits with
near-zero targets. In the held-out comparison, both restoration of the discarded components and scalar matching of the receiving-current norm recovered 100\% mean-sign accuracy. Scalar norm matching produced decision commitments on 192/192 and 190/192 trials in G23 and S2, respectively, compared with 192/192 trials under full residual restoration. The subsequent historical contrasts remained reversed at -0.49 and -0.57. G24 preserved decision commitment across the reduction and restoration conditions. These results indicate that restoring the differential-current magnitude is sufficient to recover the sensory performance lost under dimensional reduction in the tested circuits.

\clearpage

\subsection{Figure 4-2. Input geometry after clipping and the effectiveness of alternative directions}\label{extended-figure-4-2.-realized-geometry-and-effective-alternative-directions}

\begin{figure}
\centering
\includegraphics[width=0.98\linewidth,height=\textheight,keepaspectratio,alt={Geometry and historical effects of sampled arrival-current directions}]{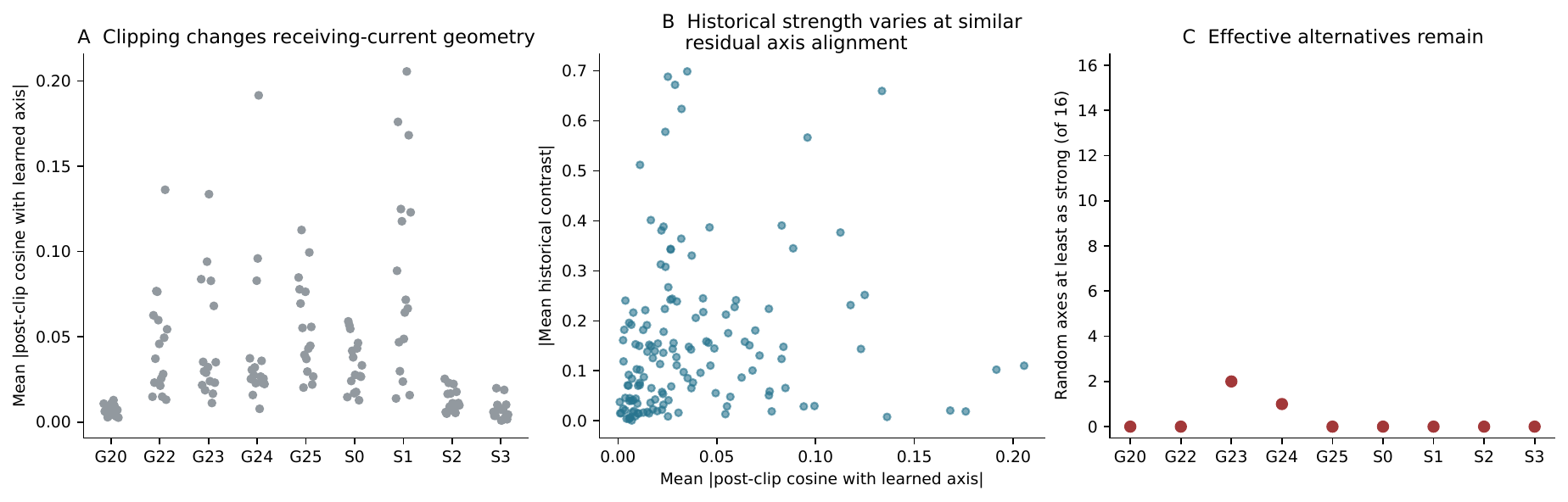}
\end{figure}

\textbf{Figure legend.} \textbf{A,} Mean absolute cosine between each
realized clipped difference and the independently estimated history
axis. \textbf{B,} Absolute context-averaged historical contrast at the
corresponding cosine. \textbf{C,} Number of alternatives among 16
directions per circuit that attain the learned direction's contrast
magnitude. The directions are sampled within the coordinates of PPC neurons receiving input from PUL and
made orthogonal to the history axis before clipping. Norm matching uses
the original scalar sign sequence. After clipping, the delivered
currents can change direction over time.

Coordinatewise clipping can rotate an initially orthogonal current difference because individual components reach their bounds at different magnitudes. We therefore measured the cosine between the actual delivered current difference and the estimated history axis after clipping.

Despite the changes in direction introduced by clipping, only 3 of the 144 sampled alternatives reached or exceeded the magnitude of the learned historical contrast. Some alternatives also impaired the intervening sensory decision. The principal comparison uses the absolute magnitude of the signed contrast averaged across 24 held-out base histories, with the candidate directions held fixed during bootstrap resampling. Within-context magnitude averages are additionally provided in the source tables.

\clearpage

\subsection{Figure 4-3. Alternative cortical input patterns generated through the bounded PUL projection}\label{extended-figure-4-3.-reallocating-input-through-the-bounded-pul-projection}

\begin{figure}
\centering
\includegraphics[width=0.98\linewidth,height=\textheight,keepaspectratio,alt={Source-space reflections at matched receiving energy and E/I totals}]{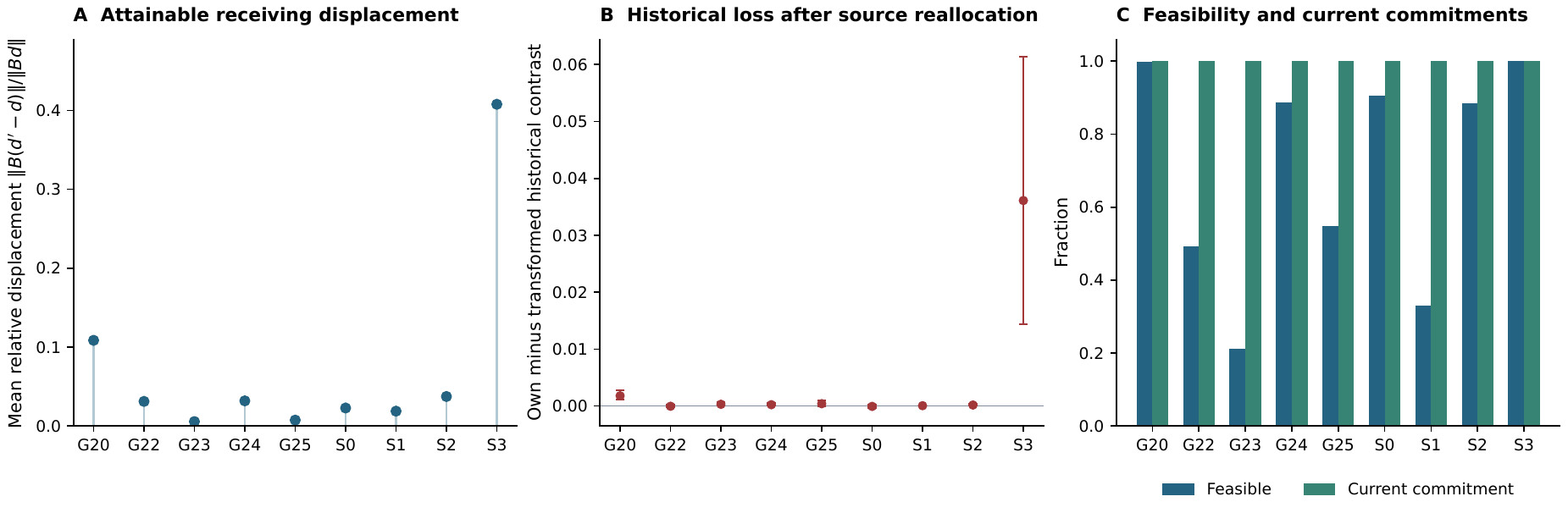}
\end{figure}

\textbf{Figure legend.} \textbf{A,} Relative displacement of the PPC
current difference attained by two source maps per circuit.
\textbf{B,} Historical contrast for own replay minus transformed replay,
with paired 95\% intervals across base histories. \textbf{C,} Fraction
of feasible source transformations and fraction of current decision commitment. The attained displacement is small in most circuits. S3
permits the largest reallocation, with a modest historical loss and
preserved current decisions.

\subsubsection{Source reflections at fixed receiving
energy}\label{extended-source-reflections-at-fixed-receiving-energy}

Direct redistribution of the current arriving at PPC explores input patterns that may extend beyond those realizable by the source population. To construct alternatives that can be generated by PUL, we instead varied the differences between the source firing-rate vectors while preserving their shared mean, the receiving-current difference energy, and the aggregate input differences to excitatory and inhibitory PPC targets.

Let \(d_T\) and \(\mu_T\) be the paired PUL rate difference and mean,
\(B\) the trained PUL-to-PPC projection, and \(S_{\rm EI}\) the map from
source differences to summed inputs for excitatory and inhibitory PPC targets. For \(S_{\rm EI}z=0\) and \(\|Bz\|_2>0\), the reflection
\(d'_T=d_T+\alpha z\) with

\[
\alpha=-2\,\frac{d_T^\top B^\top Bz}{z^\top B^\top Bz}
\]

preserves \(\|Bd_T\|_2\) and \(S_{\rm EI}d_T\). Keeping \(\mu_T\) fixed
preserves the common source activity. A candidate remains feasible when
\(|d'_{T,i}|\le 2\min(\mu_{T,i},20-\mu_{T,i})\) for every source
coordinate.

The two map searches explore feasible source transformations using this geometry. When no admissible alternative is identified, the circuit receives the original historical message. Extended Theory ET8 derives the corresponding reachable set and energy-preserving reflection.

\clearpage

\subsection{Figure 4-4. The naturally expressed historical input direction is more effective than most sampled alternatives}\label{extended-figure-4-4.-directional-sensitivity-of-ppc-arrival-currents}

\begin{figure}
\centering
\includegraphics[width=0.98\linewidth,height=\textheight,keepaspectratio,alt={Matched-energy comparison in arrival-current space}]{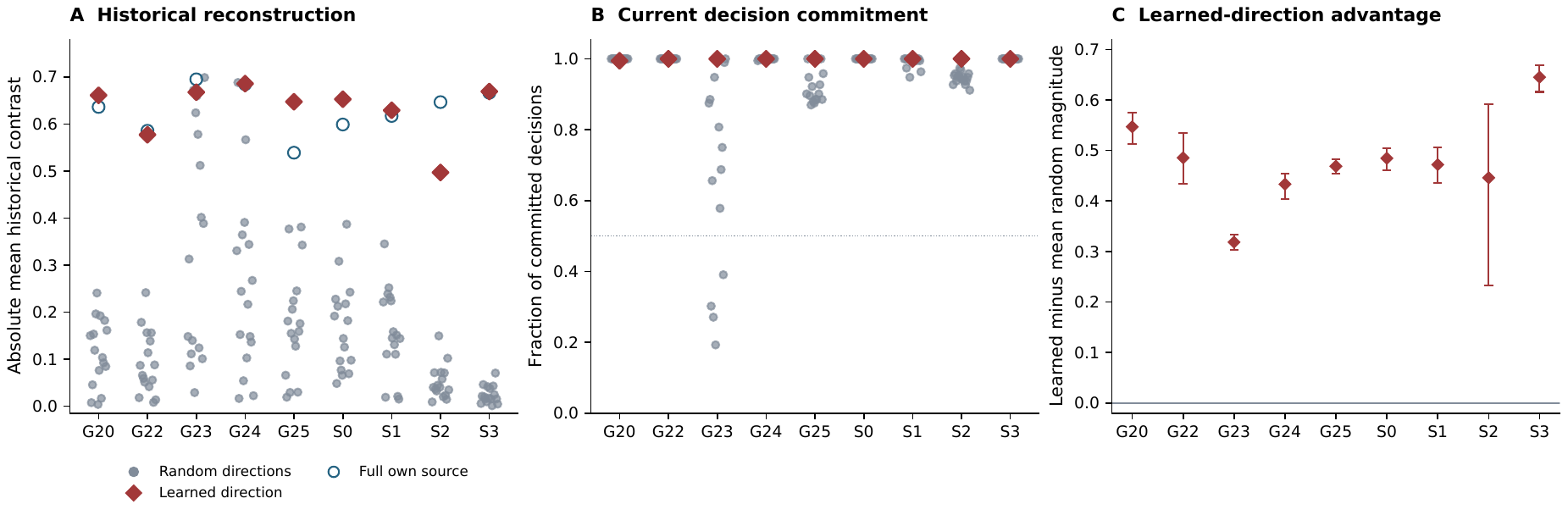}
\end{figure}

\textbf{Figure legend.} \textbf{A,} Absolute mean historical contrast
for the learned direction, 16 fixed random directions per circuit,
and full own-source replay. \textbf{B,} Current decision commitment under
the learned and random directions. \textbf{C,} Historical contrast produced by the learned direction minus the mean absolute historical contrast produced by the sampled random directions. Paired 95\% confidence intervals were obtained by resampling 24 base histories while retaining the same candidate directions. Each history is crossed with four strong stimuli and
both history signs. The learned axis exceeds the mean random alternative
in every circuit and preserves current commitment in circuits where
some alternatives impair it.

\subsubsection{Changing direction at matched receiving
energy}\label{extended-changing-direction-at-matched-receiving-energy}

This experiment examines whether the effectiveness of an input pattern depends on its spatial direction independently of the magnitude of the historical current difference. All candidate directions are evaluated under the same pair-mean background input.

For the natural pair \(b^\pm=\mu\pm d/2\), let \(v\) be the learned
axis. The random axes \(q\) are initially orthogonal to \(v\) within the coordinates of PPC neurons receiving input through the trained PUL projection. The matched replay uses
\(\widetilde b^\pm=\mu\pm\widetilde d/2\), with

\[
\widetilde d=\operatorname{clip}\bigl(k\,\operatorname{sign}(d^\top v)q,-2\mu,2\mu\bigr),
\qquad \|\widetilde d\|_2=\|d\|_2 .
\]

The learned-direction condition uses the independently estimated history axis, whereas the alternative conditions use the sampled random axes. Clipping ensures nonnegative arrival currents, and rescaling matches the differential-current norm within the attainable range. Both recipients begin from their complete pair-averaged state and receive one second of replacement input before encountering identical sensory stimuli. Figure 4-2 characterizes the actual directions produced after clipping, while main \hyperlink{main-figure-link-4}{Figure~4} examines whether comparable alternatives can be generated through the constrained PUL projection.

\clearpage

\subsection{Figure 4-5. Temporal coherence further constrains the alternative historical signals that PUL can generate}\label{extended-figure-4-5.-reallocations-that-preserve-temporal-organization}

\begin{figure}
\centering
\includegraphics[width=0.98\linewidth,height=\textheight,keepaspectratio,alt={Displacement, temporal roughness, and functional effects of trajectory-wide maps}]{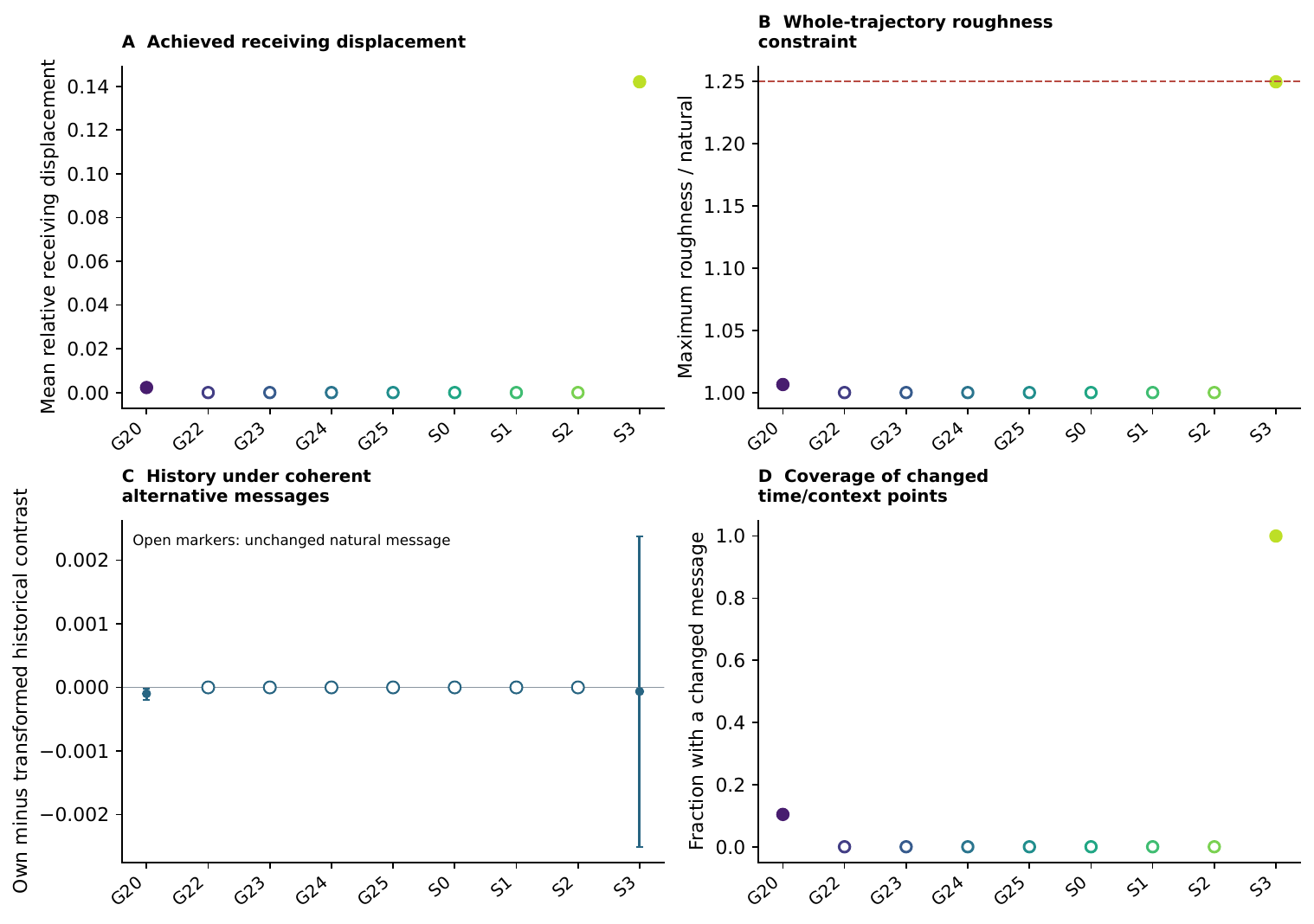}
\end{figure}

\textbf{Figure legend.} Source transformations selected independently at successive time points can introduce rapid fluctuations into the delivered signal. To investigate whether alternative source patterns can be maintained throughout replay, we required each candidate reflection to act along the same spatial direction over time steps 1400--1599. A reflection was accepted if it satisfied the firing-rate bounds at every time step, increased the integrated receiving-current displacement, and preserved temporal roughness within 1.25 times its natural value. We considered 256 candidate reflections in each context. Temporal roughness is \(\mathcal R(d)=\sum_t\|d(t+1)-d(t)\|^2\). Accepted reflections
compose one spatial transformation for the whole context. Each
circuit uses 48 base histories crossed with four strong stimuli.

\textbf{A--D,} Receiving displacement, relative roughness, historical
loss, and the fraction of time and context points with changed messages.
Filled markers show attained transformations and open markers show the
natural message. Under these joint spatial and temporal constraints, the sampled transformations retained the natural input in seven circuits. G20 achieved a receiving-current displacement of 0.23\%, while S3 permitted a displacement of 14.2\% with nearly unchanged historical influence.

\clearpage

\subsection{Figure 4-6. The leading source direction remains low dimensional and functionally effective under firing-rate constraints}\label{extended-figure-4-6.-reachability-and-geometry-along-clipped-source-axes}

\begin{figure}
\centering
\includegraphics[width=0.98\linewidth,height=\textheight,keepaspectratio,alt={Source-axis reachability and resulting historical output}]{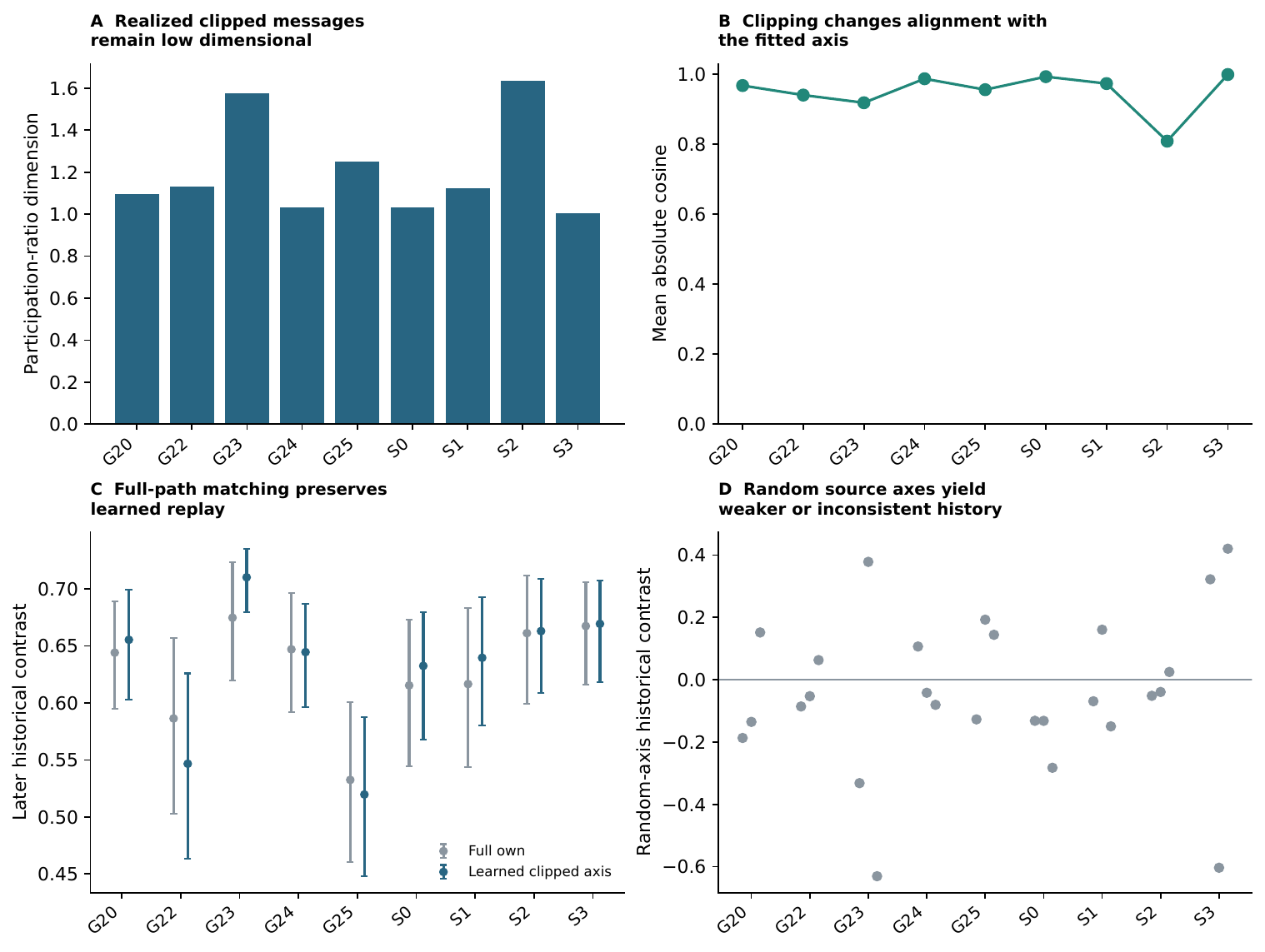}
\end{figure}

\textbf{Figure legend.} Scaling a candidate direction in the PUL firing-rate space can cause individual neurons to reach their firing-rate bounds. The resulting clipped signal may consequently deviate from the original direction and span more than one dimension. To examine these effects, we evaluated the complete clipped scaling path of the independently estimated source direction and three fixed random directions in 12 held-out base histories per circuit. The scaling rule uses the first crossing of the target norm or the largest attainable norm when the target is out of reach. \textbf{A, B,} Clipped messages along the learned source direction remain close to a one-dimensional representation, with participation-ratio dimensions of 1.0--1.6 and mean absolute alignment with the fitted axis exceeding 0.80. \textbf{C,} Full-path norm matching along the learned axis closely
reproduces full own-source historical contrasts (\(0.52\text{–}0.71\))
with 100\% current decision commitment. \textbf{D,} Fixed random source
axes produce weaker or inconsistent historical effects across circuits. The accompanying source data provide the receiving-current norms attained by each candidate direction.

We quantified effective dimensionality by the participation ratio \(D_{\mathrm{PR}}=(\sum_i\lambda_i)^2/\sum_i\lambda_i^2\), where \(\lambda_i\) are eigenvalues of the uncentered second moment of the realized PUL firing-rate differences, pooled across time and contexts for each axis and circuit. A rank-one difference has a participation ratio of one.

The learned direction thus retains a compact geometry after the source-rate constraints are imposed and remains sufficient to reproduce much of the natural historical influence. The random source directions reveal greater variability in attainable magnitude and behavioral effectiveness.

\clearpage

\subsection{Figure 5-1. Similar changes in input allocation can produce different historical effects across circuits}\label{extended-figure-5-1.-redistributing-input-within-native-recipient-classes}

\begin{figure}
\centering
\includegraphics[width=0.98\linewidth,height=\textheight,keepaspectratio,alt={Within-class current geometry and historical effects}]{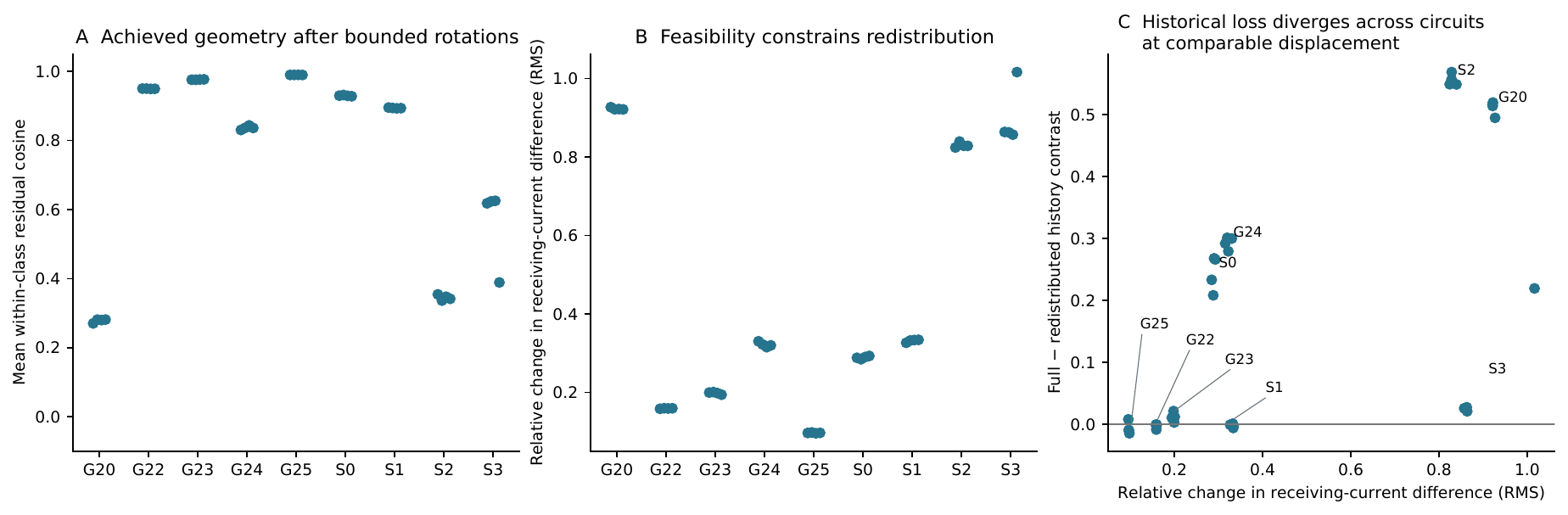}
\end{figure}

\textbf{Figure legend.} \textbf{A,} Cosine between the original and
transformed differences after centering within each excitatory or inhibitory PPC target class. \textbf{B,} Relative RMS change in the full arrival-current
difference. \textbf{C,} Historical loss versus the achieved change for
each of four maps per circuit.

The excitatory and inhibitory PPC neurons are interspersed across the network coordinates. We performed all redistributions within their original cell classes, preserving the total thalamic input delivered to each class.

Each map partitions recipients within a class into triplets and rotates
each triplet around its all-ones axis, preserving the class's
differential sum and energy. A rotation is applied at a time and context
only when the transformed difference fits the box \([-2\mu,2\mu]\), so
both arrivals remain nonnegative. Each sweep uses its own triplets and angles, with the same sequence of proposed rotations across all times and contexts. The input determines which rotations can be applied.

These bounded within-class rotations preserve the pair mean, total E and
I currents in each branch, and the summed squared norm across the pair
(while allowing individual branch norms to vary via \(\pm\mu^\top d\)).
These transformations therefore isolate changes in the spatial distribution of input within each class while retaining the aggregate input received by that class in each history.

The nonnegativity constraints limit the magnitude of the attainable redistribution in several circuits (A,B). Among circuits permitting substantial changes in input allocation, the functional consequences differ markedly: comparable displacements produce pronounced historical losses in G20 and S2 but relatively small losses in S3 (C).

\clearpage

\subsection{Figure 5-2. Effects of individual target-current reallocations on current and later choices}\label{extended-figure-5-2.-historical-losses-and-current-decisions-for-all-target-maps}

\begin{figure}
\centering
\includegraphics[width=0.98\linewidth,height=\textheight,keepaspectratio,alt={Target-current effects on historical and current responses}]{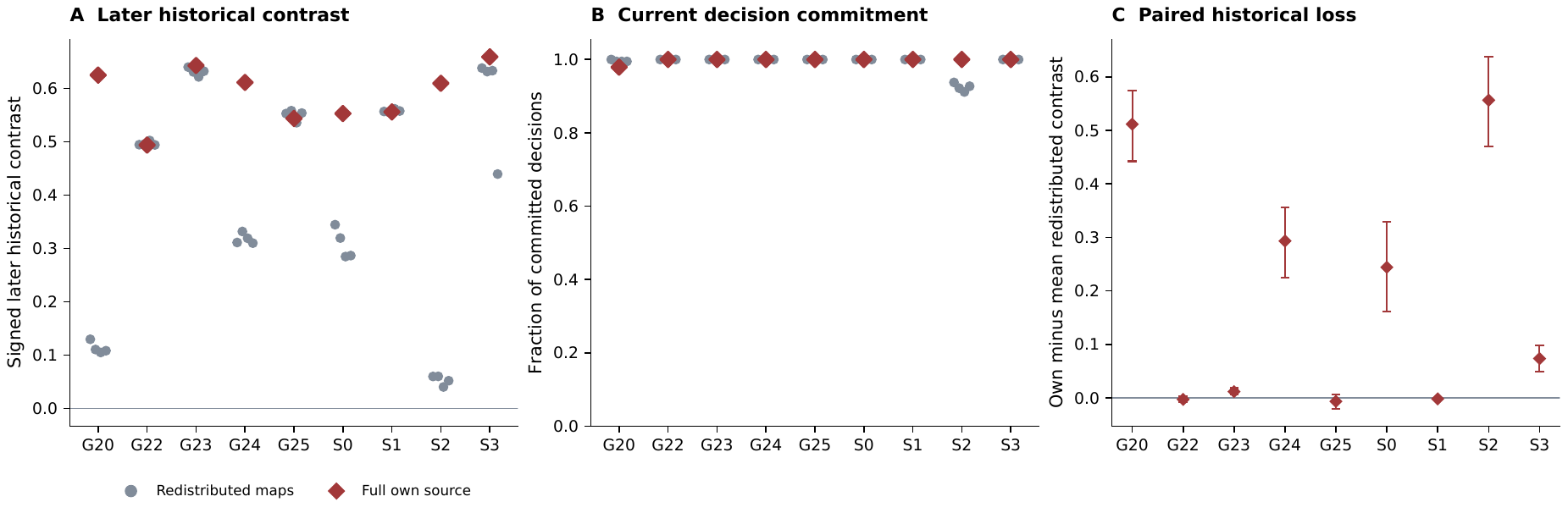}
\end{figure}

\textbf{Figure legend.} \textbf{A,} Signed later historical contrast
under full own-source replay and each of the four target-current maps.
\textbf{B,} Current decision commitment for the corresponding replays.
\textbf{C,} Own-source contrast minus the mean contrast across maps,
with paired 95\% base-history bootstrap intervals. Each circuit uses the same
24 base histories crossed with four strong inputs and two history signs.
Fine reallocation within excitatory and inhibitory target classes leaves
intervening commitments high in most circuits and produces pronounced
historical loss in selected circuits. Main Figure 5c, left, shows the
four-map loss.

\subsubsection{Rotation within excitatory and inhibitory target classes}\label{extended-rotation-within-native-target-classes}

To examine the contribution of fine input allocation, the transformations preserve the pair-mean current and change only the history-dependent difference. The redistribution is carried out separately within each excitatory or inhibitory PPC target class.

Write the natural arrivals as \(b^\pm=\mu\pm d/2\) and the transformed
pair as \(\mu\pm d'/2\). In each class
\(g\in\{\mathrm{E},\mathrm{I}\}\), triplet rotations preserve
\(\mathbf 1^\top d'_g=\mathbf 1^\top d_g\) and
\(\|d'_g\|_2^2=\|d_g\|_2^2\). A rotation is accepted when
\(|d'_i|\le2\mu_i\), preserving nonnegative arrivals in both histories.
At fixed mean and fixed within-class differential sums, each history
retains its total input to each class. Figure 5-1 describes the
resulting displacements.

The four maps provide repeated tests of whether alternative allocations of the same aggregate input can alter later historical influence. Across several circuits, the maps reduce historical contrast while preserving the ability to commit to the correct intervening sensory choice.

\clearpage

\subsection{Figure 5-3. Source-constrained input reallocations preserve sensory responses across strong stimulus mixtures}\label{extended-figure-5-3.-sensory-responses-to-bounded-source-maps-at-each-stimulus-strength}

\begin{figure}
\centering
\includegraphics[width=0.98\linewidth,height=\textheight,keepaspectratio,alt={Sensory changes by source map and stimulus strength}]{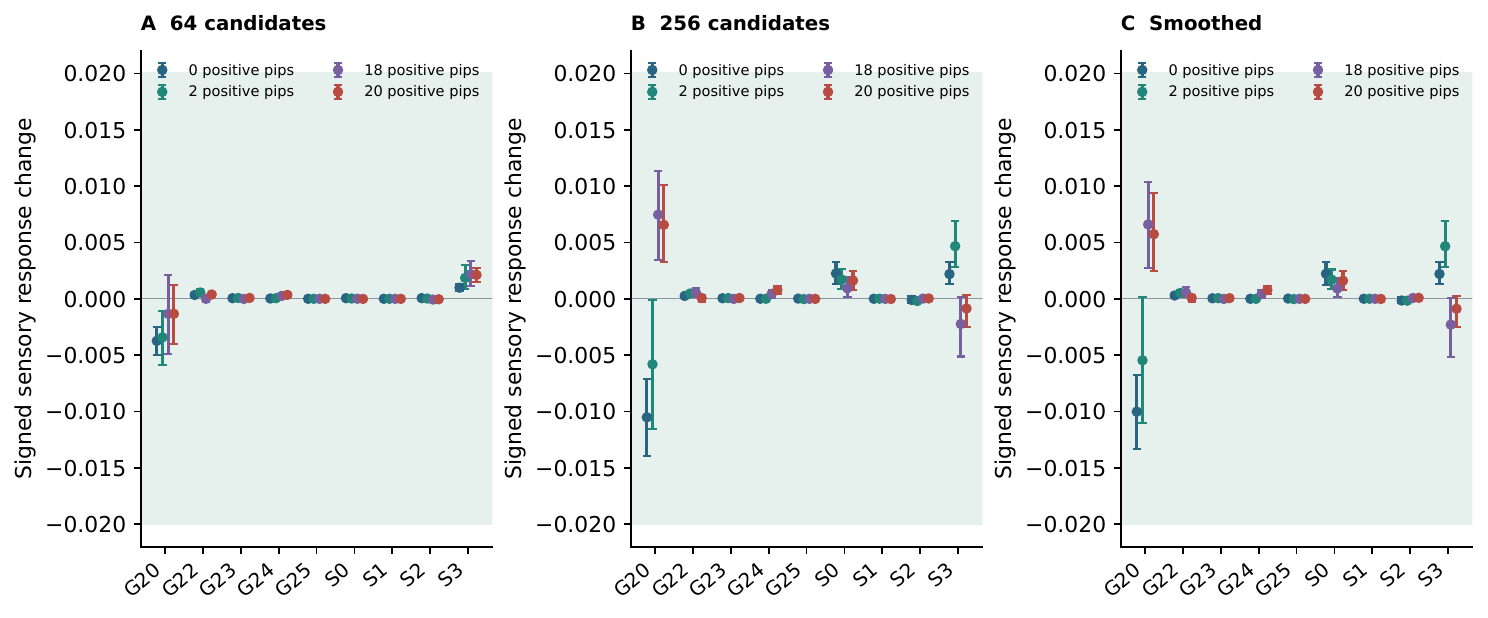}
\end{figure}

\textbf{Figure legend.} To establish whether the source transformations selectively affect historical influence, we examined current sensory output separately for each of the four strong stimulus mixtures. This prevents an effect at one stimulus strength from being obscured by averaging across different mixtures. \textbf{A--C,} Signed sensory-output change
relative to own replay for the 64-candidate, 256-candidate, and smoothed
source maps. Colors identify 0, 2, 18, or 20 positive pips. Points and
bars give the means and paired 95\% base-history bootstrap intervals. Each circuit uses 48 base histories. The third plot of main Figure 5c shows the pooled comparisons. Across the nine circuits, three source transformations, and four strong stimulus mixtures, all 108 paired 95\% confidence intervals for changes in mean sensory output lay within \(\pm\)0.02 output units. Thus, the intervening sensory response remained stable across the tested stimulus strengths, including in circuits where the transformations weakened later historical influence.

\clearpage

\subsection{Figure 6-1. A fixed nonlinear readout can suppress historical expression while retaining memory}\label{extended-figure-6-1.-saturation-changes-expression-at-fixed-memory}

\begin{figure}
\centering
\includegraphics[width=0.98\linewidth,height=\textheight,keepaspectratio,alt={Saturation and sequential judgments in a minimal receiver}]{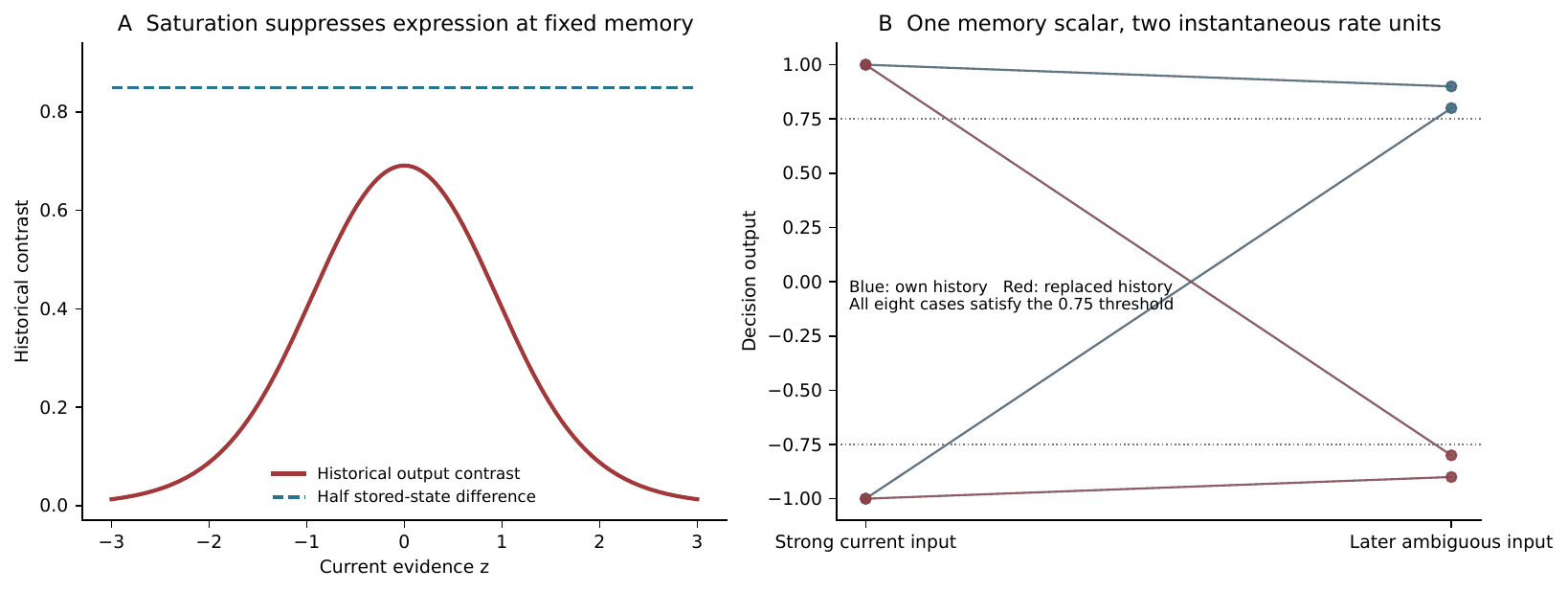}
\end{figure}

\textbf{Figure legend.} These minimal models illustrate how a circuit can retain the same information about the past while its behavioral influence changes with current sensory evidence. \textbf{A,} A fixed smooth readout gives the
paired contrast \(H(z)=\sinh(2ka)/[\cosh(2gz)+\cosh(2ka)]\), illustrated
at \(a=0.85\) and \(k=g=1\). The dashed line marks the constant half
difference in stored state. \textbf{B,} A receiver with retained history
\(h=\pm0.85\), two clipped rate units, and current stimulus \(s=\pm2\).
After the current judgment, \(h\leftarrow h+0.025s\). Later input is
zero. Across the eight history, source, and stimulus combinations, the
current output is the correct \(\pm1\). The later output follows the
written history with magnitude 0.80 or 0.90. Overlapping cases share the
same trajectory. Dotted lines mark the commitment thresholds at
\(\pm0.75\). Strong sensory input drives both historical conditions toward the same saturated output, whereas weak sensory input allows their retained memory difference to produce distinct responses. The models illustrate how a fixed nonlinear transformation can regulate the behavioral expression of a retained memory as current sensory evidence changes.

\clearpage

\subsection{Figure 6-2. Retained historical activity and its distributed contribution to the decision readout}\label{extended-figure-6-2.-retained-activity-readout-alignment-and-regional-contributions}

\begin{figure}
\centering
\includegraphics[width=0.98\linewidth,height=\textheight,keepaspectratio,alt={Finite historical differences and their projection onto the readout}]{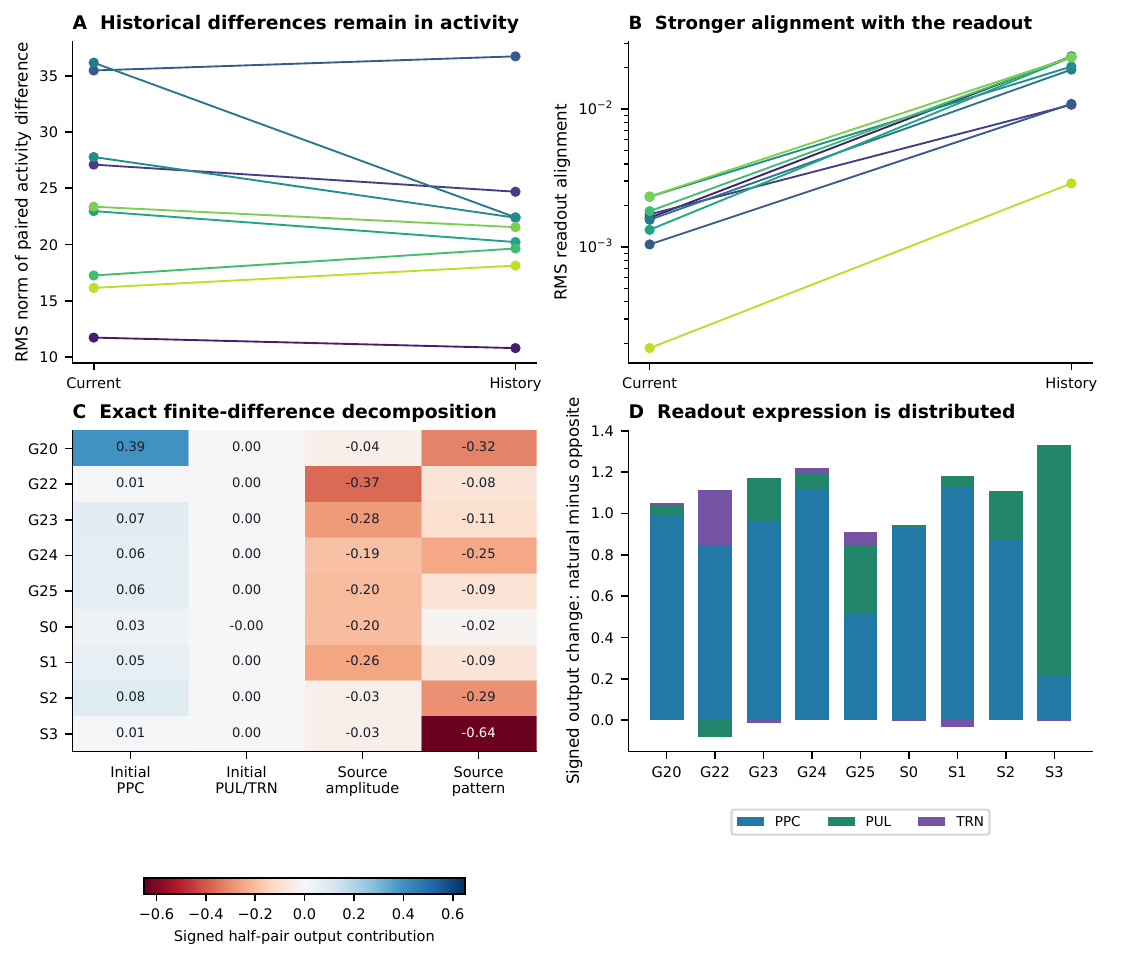}
\end{figure}

\textbf{Figure legend.} \textbf{A,} Historical activity separation
during the current and later judgments. \textbf{B,} RMS alignment between the history-dependent population firing-rate difference and the trained decision readout, as defined in Equation 10. \textbf{C,} Signed contributions to the paired response from
the initial PPC state, the initial PUL/TRN state, source amplitude, and
source pattern. \textbf{D,} Signed regional contributions to the output
change. Across all nine circuits, substantial differences in population firing rates remain present during both the current sensory judgment and the subsequent history-guided judgment (A). Their RMS alignment with the decision readout increases by approximately an order of magnitude during the later judgment (B). Thus, the behavioral influence of a retained activity difference depends on its relationship to the trained readout.

The exact finite-difference decomposition separates the contributions of the initial PPC state, the initial non-PPC state, the incoming amplitude, and the incoming spatial pattern to the realized historical contrast (C). Incoming amplitude and spatial pattern account for most of the reconstructed output. Decomposing the final output by neural population further shows that historical information introduced through one projection can ultimately influence behavior through activity in multiple populations (D). Main \hyperlink{main-figure-link-6}{Figure~6} investigates the contribution of neuronal operating points and recurrent propagation through targeted interventions.

\clearpage

\subsection{Figure 6-3. Large receiver perturbations can disrupt the intervening sensory judgment}\label{extended-figure-6-3.-current-responses-to-larger-receiver-interventions}

\begin{figure}
\centering
\includegraphics[width=0.98\linewidth,height=\textheight,keepaspectratio,alt={Current commitments, continuous output, and activity regimes}]{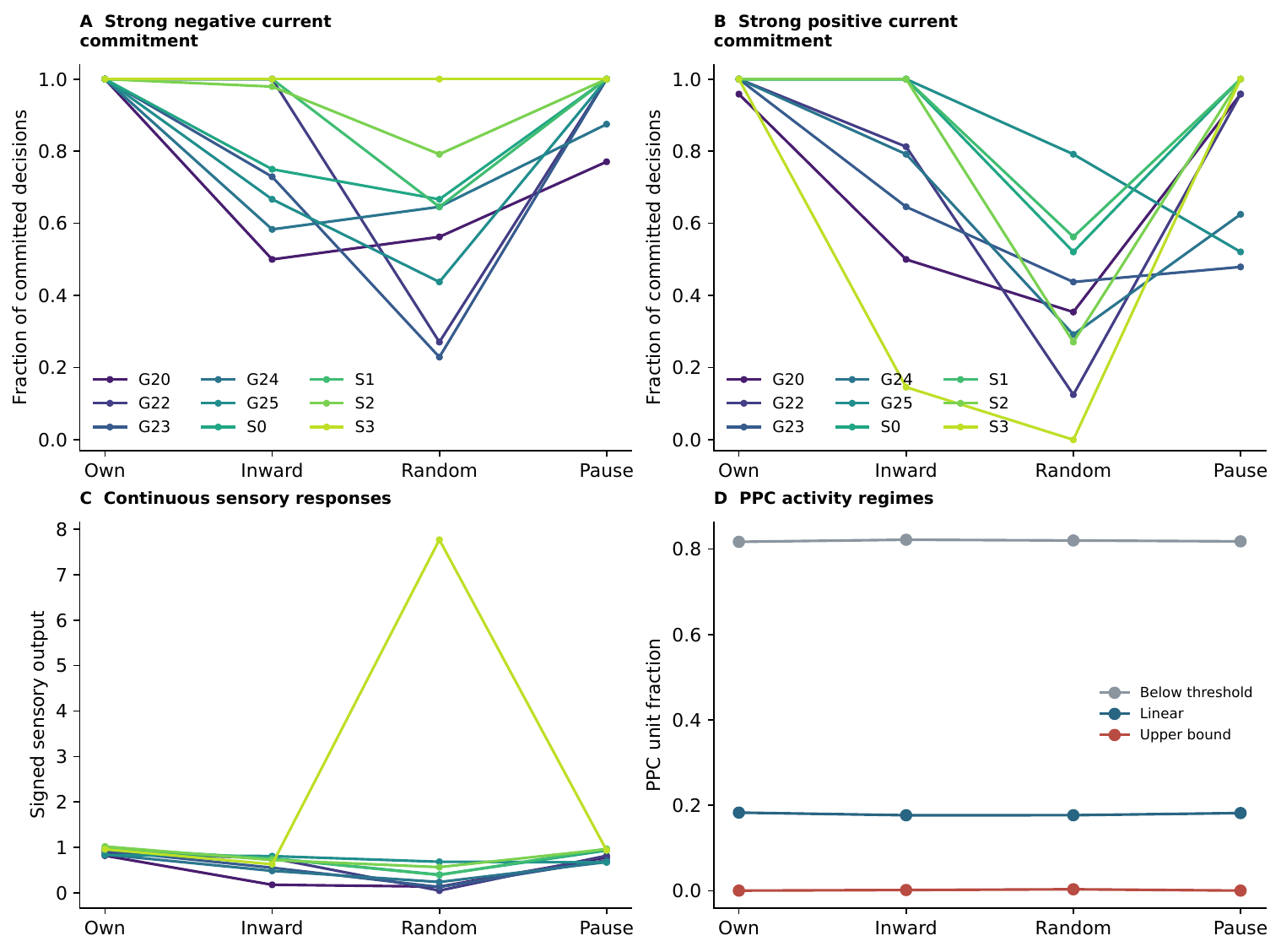}
\end{figure}

\textbf{Figure legend.} We initially examined the effects of relatively large perturbations to the receiving circuit in order to characterize the range of changes that can be produced by shifting cortical operating states or interrupting history-dependent recurrent propagation. The receiver interventions follow the shared
sensory pip train after the same history has been reconstructed. The
inward shift is half the common-state displacement toward the interval
\([2,18]\). The random shift has the same Euclidean norm. Pausing
recurrent-drive difference for 100 ms preserves the common recurrent
drive across all 500 units. \textbf{A, B,} Current decision commitment
under strong negative and positive stimuli (2 and 18 positive pips).
\textbf{C,} Mean signed sensory output across both stimulus signs. The large random displacement produces a pronounced increase in the mean sensory output of S3. \textbf{D,} Fraction of PPC units below threshold, in the linear regime,
or at the upper saturation bound. Several of these large perturbations substantially alter current sensory output and decision commitment, in addition to their effects on historical expression (Figure 6-4). These results motivate the independently calibrated smaller perturbations in main \hyperlink{main-figure-link-6}{Figure~6} and Figure 6-5, which are designed to examine historical mechanisms while preserving sensory performance.

\clearpage

\subsection{Figure 6-4. Effects of large receiver perturbations on historical activity and behavior}\label{extended-figure-6-4.-historical-responses-to-larger-receiver-interventions}

\begin{figure}
\centering
\includegraphics[width=0.98\linewidth,height=\textheight,keepaspectratio,alt={Historical contrasts and population geometry after larger receiver interventions}]{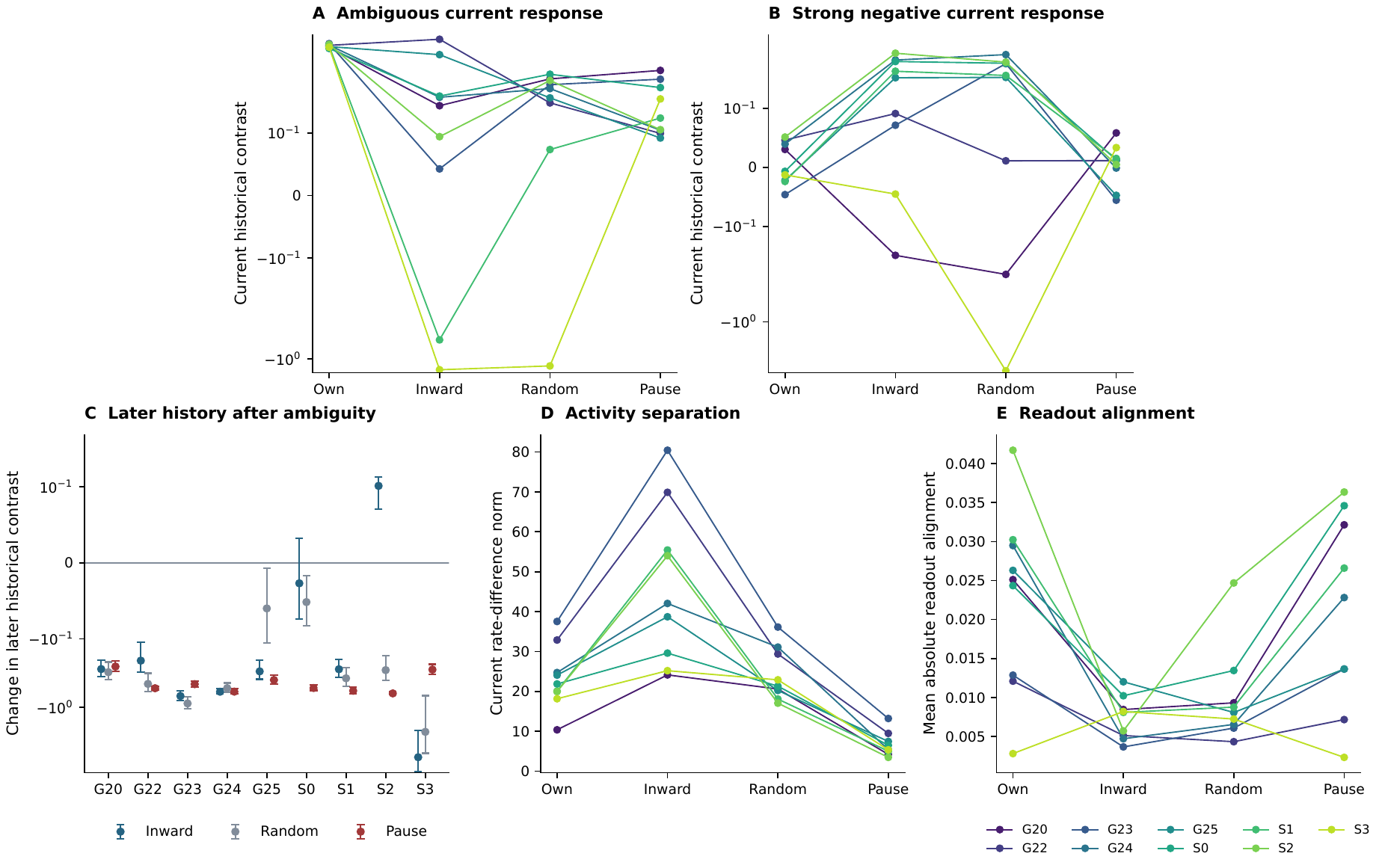}
\end{figure}

\textbf{Figure legend.} \textbf{A, B,} Current historical contrasts
under ambiguous and strong negative fifth-trial evidence. Lines connect
conditions for each circuit. \textbf{C,} Change in subsequent
historical contrast after an ambiguous fifth stimulus; markers show the
three perturbations relative to own replay, with paired 95\%
base-history bootstrap intervals. \textbf{D, E,} Current activity separation and
alignment with the fixed readout under ambiguity. The symmetric-log axes
have linear thresholds of 0.15 in A and B and 0.10 in C. The comparisons
use 24 base histories per circuit.

\subsubsection{Common-state shifts and the recurrent
pause}\label{extended-common-state-shifts-and-the-recurrent-pause}

All conditions receive the same reconstructed historical input and the same sensory pip train before the receiver is perturbed. The common-state shift displaces the two history-dependent states equally, preserving their initial difference while changing their positions relative to the neuronal activation thresholds. The inward shift moves the PPC pair mean halfway toward the active range [2,18], while the random shift has the same Euclidean magnitude. During the recurrent-difference pause, both histories receive the pair-mean recurrent drive for 100 ms, temporarily removing the difference between their recurrent inputs. Panels A--C show the change in historical contrast.

The activity-separation and readout-alignment measurements distinguish changes in the retained firing-rate difference from changes in its expression at the decision output. Figure 6-5 examines these relationships in 48 independent base histories using smaller perturbations calibrated to preserve current sensory judgments.

\clearpage

\subsection{Figure 6-5. Independently calibrated receiver perturbations distinguish historical effects from sensory impairment}\label{extended-figure-6-5.-calibrated-receiver-interventions-and-their-effects-across-evidence-strengths}

\begin{figure}
\centering
\includegraphics[width=0.98\linewidth,height=\textheight,keepaspectratio,alt={Calibration, sensory effects, and population geometry after smaller receiver interventions}]{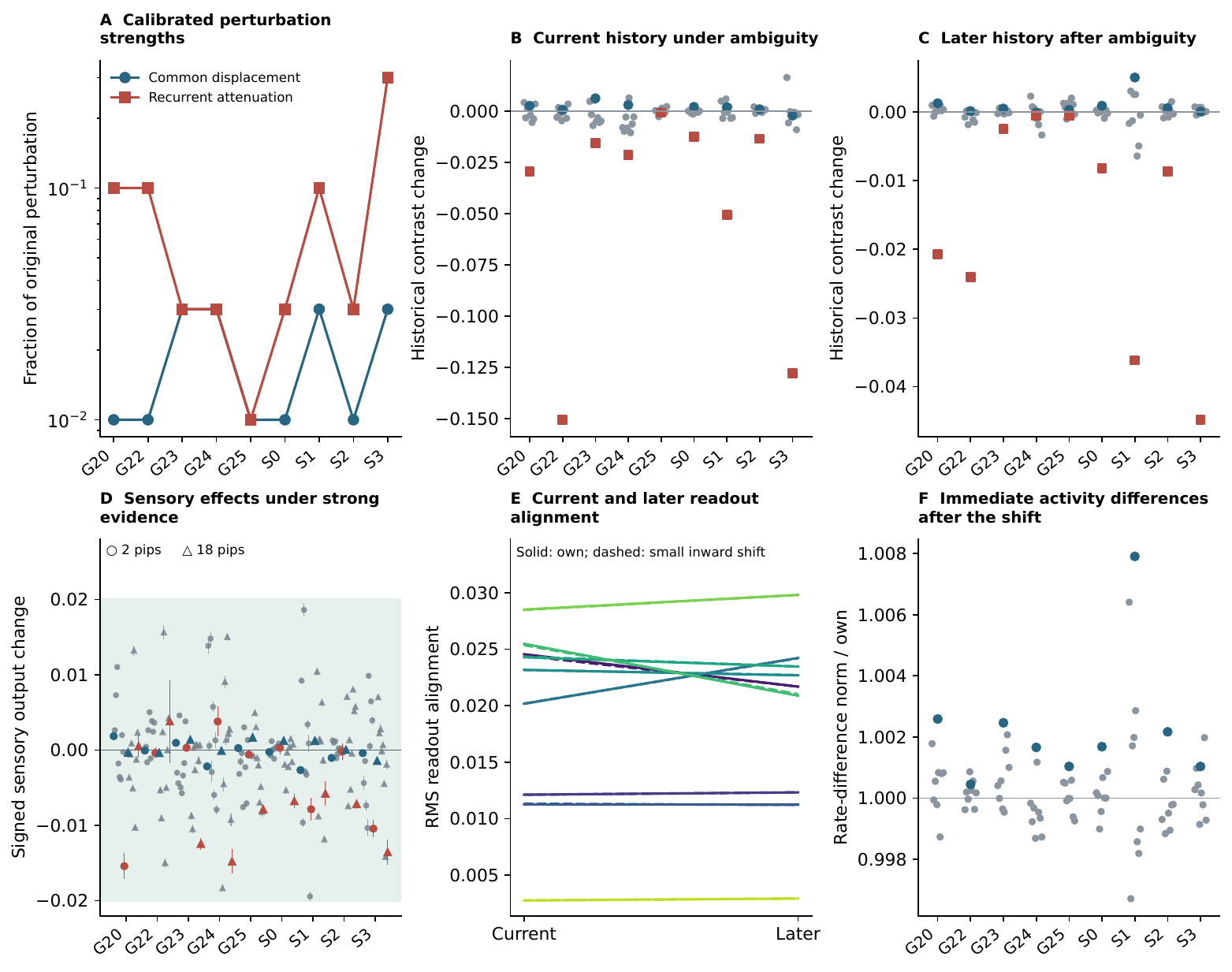}
\end{figure}

\textbf{Figure legend.} To examine historical processing while preserving strong-stimulus sensory judgments, we selected perturbation strengths using 12 independent calibration histories. We evaluated the selected common-state displacements and recurrent attenuation in 48 new base histories per circuit, combined with stimuli containing 2, 10, or 18 positive pips. The same eight random directions are
used across histories and evidence strengths within each circuit.
\textbf{A,} Selected perturbation strengths. \textbf{B, C,} Current and subsequent
historical contrasts relative to own replay under an ambiguous fifth
stimulus. \textbf{D,} Signed sensory changes and paired 95\% intervals
for the two strong stimuli. In B--D, blue marks the inward shift, red
marks recurrent attenuation, and gray marks random shifts. \textbf{E,}
Current and later readout alignment under own replay and the small
inward shift. \textbf{F,} Immediate rate-difference size relative to own
replay. The common shift preserves the paired state difference at onset
while changing the resulting rate difference.

Strong-stimulus sensory responses remained stable at the selected perturbation strengths. All inward-shift and recurrent-attenuation confidence intervals lay strictly within \(\pm0.02\) output units. Across the ten perturbation conditions, 179 of the 180 intervals lay strictly within this range, with one random-shift interval reaching the lower boundary [-0.0200, -0.0188]. Current decision commitment remained at or above the corresponding own-history replay condition.

Within this perturbation range, recurrent attenuation consistently weakened the influence of history during both the ambiguous fifth-trial judgment and the subsequent historical judgment. Common-state displacements produced smaller historical effects while subtly changing the conversion of internal state differences into firing-rate differences. Panels E and F illustrate the resulting relationships between population activity and its alignment with the decision readout.

\clearpage

\subsection{Figure 7-1. Strong sensory evidence suppresses historical output more than the underlying state difference}\label{extended-figure-7-1.-strong-evidence-reduces-expression-while-state-differences-persist}

\begin{figure}
\centering
\includegraphics[width=0.9\linewidth,height=\textheight,keepaspectratio,alt={State separation and historical expression under strong and ambiguous evidence}]{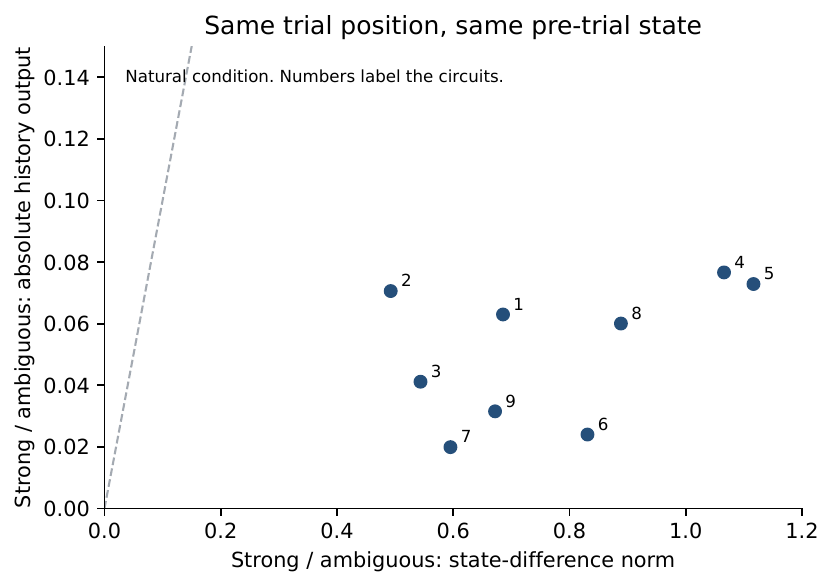}
\end{figure}

\textbf{Figure legend.} To distinguish the effect of sensory evidence from the effect of sequence position, we compared responses to strong and ambiguous stimuli delivered at the same trial position from the corresponding identical preceding states. The horizontal axis shows the ratio between the end-of-response state-difference norm under strong and ambiguous evidence. The vertical axis shows the corresponding ratio of the absolute historical output contrasts.

Across circuits, strong sensory evidence reduces historical output to 2.0--7.7\% of its ambiguous-condition value, while the norm of the 500-dimensional state difference remains at 49--112\% of the corresponding ambiguous-condition value. In G24 and G25, state separation is larger under strong input despite the substantial reduction in historical output. These measurements, obtained from 12 base histories per circuit, complement the response-window firing-rate measurements in main \hyperlink{main-figure-link-6}{Figure~6a} and demonstrate the separation between retained internal state differences and their behavioral expression. The numerical labels identify the nine circuits in the order G20, G22, G23, G24, G25, S0, S1, S2, and S3.

\clearpage

\subsection{Figure 7-2. Sensory evidence regulates historical influence at different positions in the decision sequence}\label{extended-figure-7-2.-history-across-evidence-strengths-and-sequence-positions}

\begin{figure}
\centering
\includegraphics[width=0.98\linewidth,height=\textheight,keepaspectratio,alt={Historical outputs and neural state differences across sequence positions}]{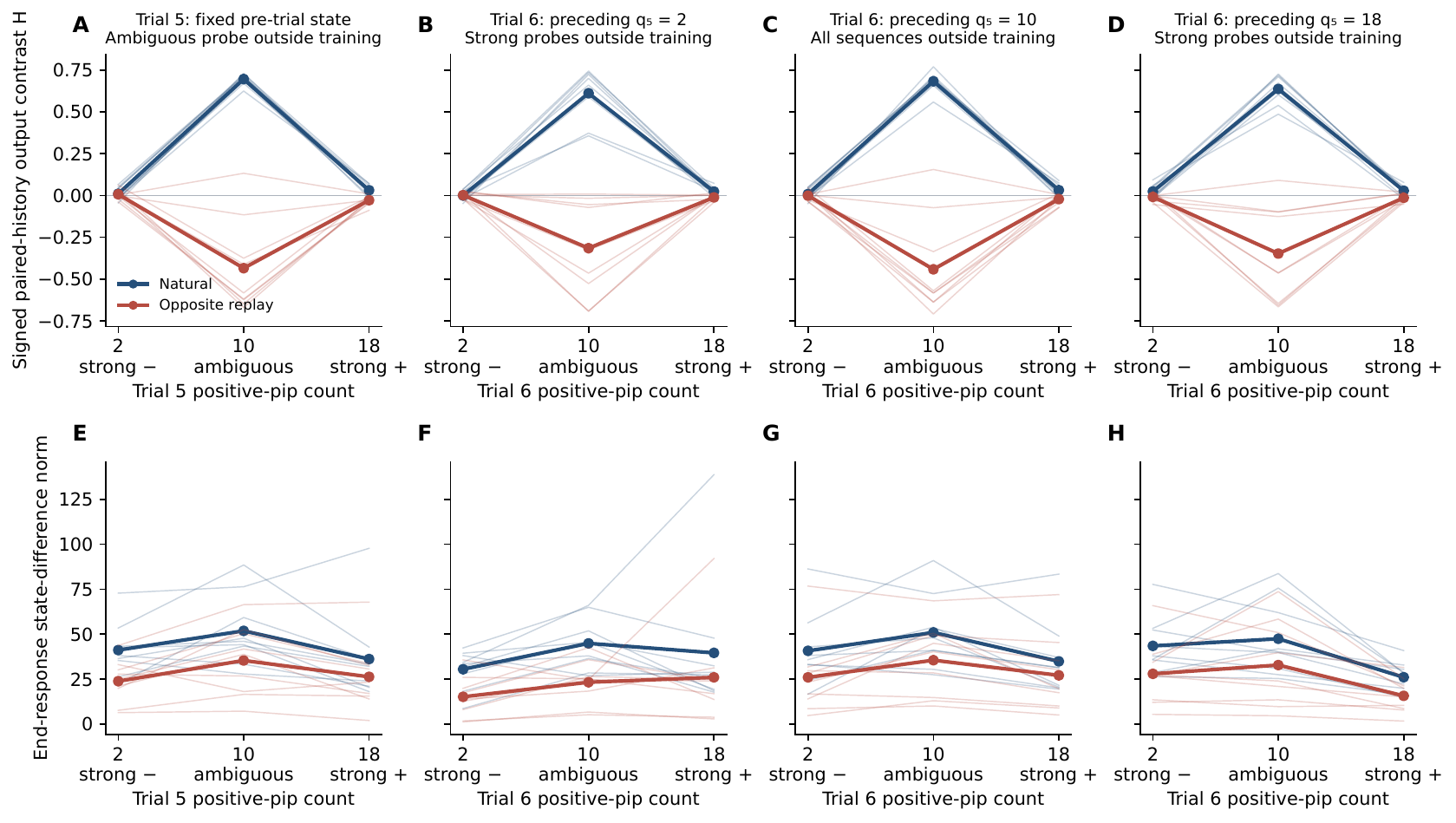}
\end{figure}

\textbf{Figure legend.} Current sensory evidence was varied independently on trials 5 and 6 under natural evolution and opposite-history replay. The comparison includes sensory sequences extending beyond the original training schedule, including ambiguous fifth trials and strong sixth trials. These conditions test whether evidence-dependent historical expression extends to new positions and combinations of sensory evidence. Panels A--D show the signed historical contrast for each combination of current and preceding sensory evidence, while panels E--H show the corresponding internal state-difference norms.

Across both trial positions and all preceding fifth-trial contexts, strong sensory evidence substantially reduces historical influence on the output while leaving substantial internal state differences. Under ambiguous evidence, the retained historical difference can again exert a strong influence on the decision. Main \hyperlink{main-figure-link-7}{Figure~7a},b presents a subset of these comparisons.

\clearpage

\subsection{Figure 7-3. Recurrent propagation and reference-template predictions capture different components of historical output dynamics}\label{extended-figure-7-3.-reference-templates-and-within-evidence-prediction}

\begin{figure}
\centering
\includegraphics[width=0.98\linewidth,height=\textheight,keepaspectratio,alt={Gain-by-recurrence comparisons and prediction relative to the reference response}]{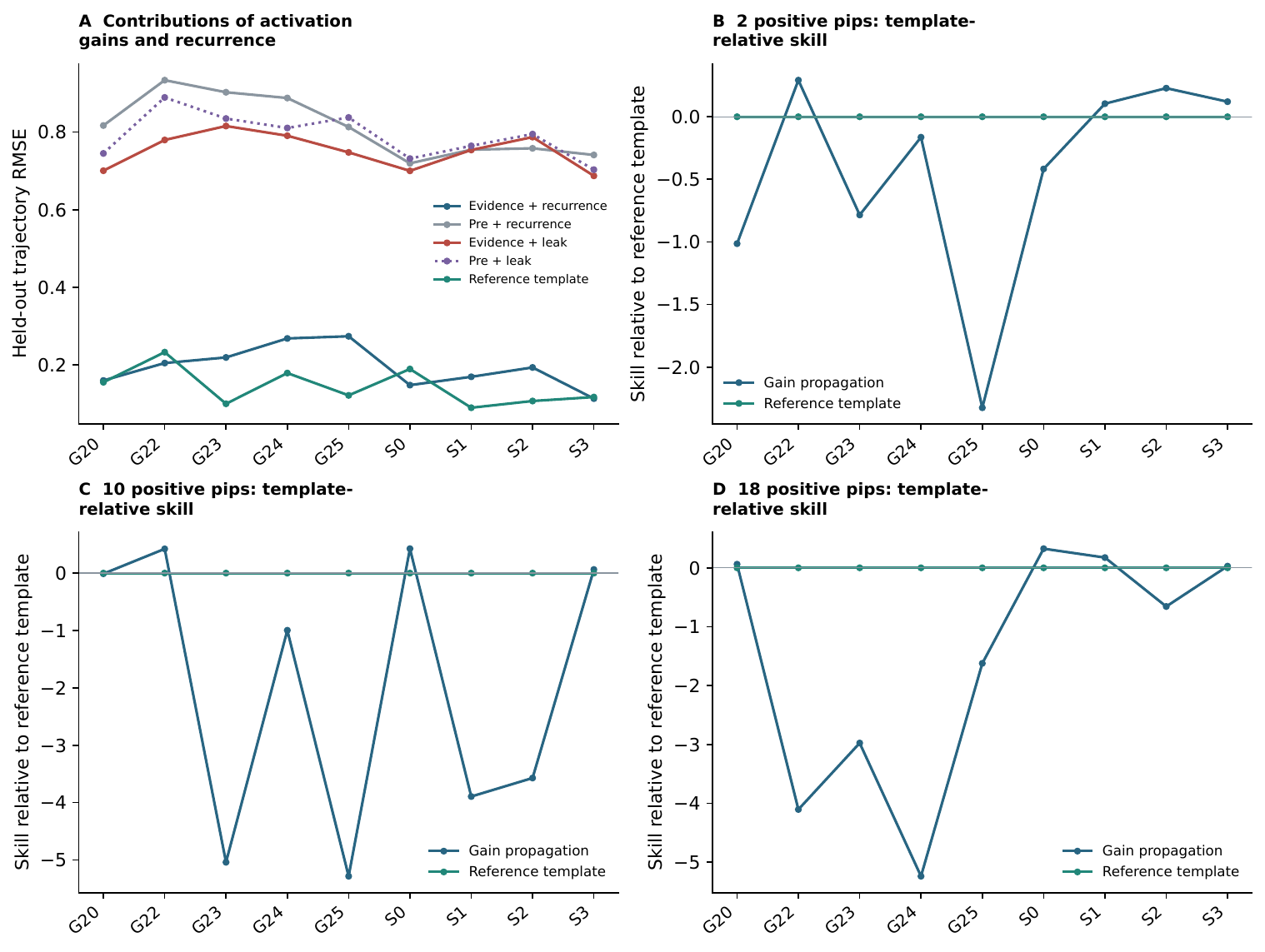}
\end{figure}

\textbf{Figure legend.} To determine how much of the post-stimulus historical trajectory can be explained by the circuit's dynamics, we constructed propagation models using activation gains estimated from 12 reference histories. Each model is initialized using the measured state difference of an independent test pair. We additionally constructed a reference-response template consisting of the mean output-difference trajectory for each evidence strength, estimated from the same reference histories. All forecasts are evaluated on the same 24 test
histories. \textbf{A,} Held-out error for evidence-specific or
pre-sensory gains crossed with recurrence or leak, alongside the
template. \textbf{B--D,} Skill relative to the template for 2, 10, and
18 positive pips. Skill is
\(1-\mathrm{SSE}_{\rm prediction}/\mathrm{SSE}_{\rm template}\). The
template has skill zero. Positive skill indicates lower squared error
than the template. Combining evidence-specific activation gains with recurrent propagation substantially reduces trajectory error compared with models using pre-sensory gains or passive leak alone (A). These comparisons support a role for both sensory-dependent activation and recurrent interactions in shaping the post-stimulus output difference.

The evidence-specific reference template achieves lower pooled error in six circuits. Its performance reflects the substantial structure shared by output trajectories at the same sensory evidence strength. Within individual evidence conditions, the template-relative skill of the propagation model varies across circuits and stimulus strengths and is frequently below zero (B--D). These comparisons characterize the predictive value of propagating the measured initial state difference using reference-averaged activation gains. Main \hyperlink{main-figure-link-7}{Figure~7d} shows within-evidence \(R^2\) relative to the single mean observed output difference across all test histories and time steps at each evidence strength.

\clearpage
\section{Figure 2-1. Extended Theory: Reconstruction, retention and expression of sensory history}\label{extended-theory}
\subsection{Overview}\label{theory-overview}

This theoretical framework addresses three related questions about the dynamics of history-guided decision-making. How can information about previous sensory stimuli be reconstructed by input arriving through a particular neural pathway? How do the resulting differences in neural state propagate through a recurrent circuit? And how can those differences remain present while their influence on behavior changes with current sensory evidence?

ET1--ET3 establish the dynamics of paired sensory histories, derive an exact decomposition of their state and input differences, and characterize the geometry of the signals that can be delivered through a constrained thalamocortical projection.

ET4--ET5 examine the relationship between retained neural differences and their behavioral expression. They show how the geometry of population activity and nonlinear activation functions can allow a historical representation to remain available while limiting its immediate contribution to the decision output.

ET6 investigates the information about reconstruction dynamics that can be obtained from measurements of stability and conditional decay. Through a constructive mathematical example, it shows that these measurements can leave important aspects of the circuit's response to localized erasure undetermined.

ET7 extends the source-driven reconstruction principle to an intrinsic neuronal recovery variable in a conductance-based circuit. ET8--ET9 establish the mathematical basis for the constrained source transformations, receiver perturbations, and conditional trajectory predictions used in the computational experiments.

\subsection{ET1. Circuit dynamics and the definition of a historical response difference}\label{theory-et1.-native-dynamics-and-the-historical-decision-contrast}

\subsubsection{State, rate and output}\label{theory-state-rate-and-output}

To examine how previous sensory stimuli influence a subsequent decision, we compare two trajectories of the same trained circuit that have experienced opposing histories. Because the subsequent external inputs are identical, differences between their later responses reflect the influence of the preceding histories and the internal or arriving signals through which that influence is transmitted. We first establish the dynamics and notation for these paired trajectories.

The model contains 400 posterior parietal cortex (PPC) units, 70
pulvinar (PUL) units, and 30 thalamic reticular nucleus (TRN) units.
Write their coordinate sets as \(C,T,R\) and use column vectors. The
maps \(E_C,E_T\) embed regional activity in the 500-dimensional state.
The activation functions are

\[
f_i(x)=
\begin{cases}
\operatorname{clip}(x,0,20),&i\in C\cup T,\\
\operatorname{clip}(10(x-9),0,20),&i\in R.
\end{cases}\tag{ET1}
\]

With the released recurrent weights \(W\), input weights \(W_{\rm in}\),
output weights \(w\), output bias \(b_0\), and heterogeneous update
factors \(A=\operatorname{diag}(\alpha_i)\), the 5-ms update is

\[
x_{t+1}=(I-A)x_t+A[Wf(x_t)+W_{\rm in}u_t],
\qquad y_t=w^\top f(x_t)+b_0.\tag{ET2}
\]

The first term in the state equation describes the persistence of the previous internal state, while the recurrent term transforms the activity of the network into new input to each unit. The external-input term describes the contribution of current sensory stimulation. The activation functions determine how internal states are converted into firing rates, and the trained linear readout determines how these rates contribute to the decision output.

The trained circuit specifies
\(W=\max(W_{\rm raw},0)\odot S_{\rm syn}\), where \(S_{\rm syn}\) sets
synaptic signs and zeros. The update factor for each unit is
\(\alpha_i=1/\tau_i\) in model update units. The interventions keep
these parameters fixed and measure choices through the trained readout
(Hosford et al., 2026).

We distinguish three levels at which opposing histories can differ: the internal states of the neurons, their firing rates, and the decision output. The transformation from internal state to firing rate depends on the neuronal activation function, while the contribution of population firing-rate differences to the decision output depends on their spatial organization relative to the trained readout.

Each context comprises two mirrored past histories, denoted \(+\) and
\(-\), followed by identical future input. Write \(\delta x=x^+-x^-\),
\(\delta r=f(x^+)-f(x^-)\), and \(\delta y=y^+-y^-\). For a response window \(\mathcal R\) containing \(L\) time steps, define

\[
H_{\mathcal R}=\frac{1}{2L}\sum_{t\in\mathcal R}\delta y_t.
\tag{ET3}
\]

The historical contrast \(H_R\) measures the signed half-difference between the response-window mean outputs of the two histories. It therefore quantifies the extent to which their different preceding experiences produce different subsequent decisions under common external input. Comparing this quantity across replay conditions allows the influence of a substituted historical signal to be measured continuously. A difference such as \(H^{\rm opposite}-H^{\rm own}\) compares the historical contrasts from two replay conditions.

The current and subsequent historical response windows span time steps \([1720, 1800)\) and \([2120, 2200)\), respectively. Outputs are
read out prior to the state update. The decision commitment criterion uses
response extrema: a right commitment requires \(\max y\geq0.75\) and
\(\min y\geq-0.75\); a left commitment requires \(\min y\leq-0.75\) and
\(\max y\leq0.75\). The exact decompositions developed below apply to continuous historical contrasts. Decision commitment is evaluated directly from the response extrema of the simulated trajectories.

\subsubsection{The intervened source}\label{theory-the-intervened-source}

A recurrent circuit can carry information about the past through both its ongoing internal state and the signals exchanged between its constituent populations. We seek to distinguish these contributions by controlling the current delivered through one specific pathway, while allowing the rest of the circuit to evolve normally.

The PUL-to-PPC current is

\[
b_t=W_{CT}E_T^\top f(x_t),\qquad W_{CT}=E_C^\top W E_T.
\tag{ET4}
\]

PUL is excitatory in this source block, so its delivered current is
nonnegative. Donor waveforms \(b_t^+,b_t^-\) are recorded from the
natural history pair. During a replay window, the endogenous edge is
removed and a specified donor current is substituted. With indicator
\(\chi_t\), define \(\widetilde W_t=W-\chi_tE_CW_{CT}E_T^\top\). The
recipient follows

\[
x_{t+1}^{\sigma}=(I-A)x_t^{\sigma}
+A\widetilde W_t f(x_t^{\sigma})
+AW_{\rm in}u_t+AE_C\chi_t\widehat b_t^{\sigma}.
\tag{ET5}
\]

The replay indicator \(\chi_t\) determines the source of the current delivered to PPC. Own-history replay uses \(\widehat b^\sigma=b^\sigma\), opposite-history replay uses \(b^{-\sigma}\), and pair-mean replay uses \(\mu=(b^++b^-)/2\). The endogenous PUL-to-PPC projection resumes when \(\chi_t=0\). All other recurrent connections remain active throughout the manipulation.

Let \(\bar x=(x^++x^-)/2\) at replay onset. Averaging only the PPC state removes the initial history-dependent difference within that population while leaving possible differences in the remaining populations. Averaging the complete network state removes the initial difference across all 500 units. The donor signals in both cases are obtained from the original, intact history trajectories.

In the complete-state-averaged condition, the two recipients begin from exactly the same state and receive identical subsequent external input. If they also receive the same pair-mean thalamic input, their deterministic trajectories remain identical. Supplying their original donor signals can instead create a difference between the recipients. Exchanging these signals exchanges the two resulting trajectories and reverses the sign of their historical contrast. This symmetry provides the basis of the content-replacement test.

\subsection{ET2. Exact transport of finite historical differences}\label{theory-et2.-exact-transport-of-finite-historical-differences}

\subsubsection{Finite activation
secants}\label{theory-finite-activation-secants}

A neuron can occupy different regions of its activation function under two opposing histories. Consequently, the difference between its firing rates depends on both the difference between its internal states and the locations of those states relative to activation thresholds. A conventional local derivative describes the response to an infinitesimal change near one state. Here, the histories may be separated by a finite distance and may cross a rectification or saturation boundary. We therefore use the secant of the activation function between the two realized states to describe their relationship exactly.

For the actual trajectory pair under each intervention, define a
diagonal matrix \(D_t\) by

\[
(D_t)_{ii}=\begin{cases}
\dfrac{f_i(x_{i,t}^+)-f_i(x_{i,t}^-)}{x_{i,t}^+-x_{i,t}^-},&x_{i,t}^+\ne x_{i,t}^-,\\
f_i'(x_{i,t}^+),&x_{i,t}^+=x_{i,t}^-.
\end{cases}\tag{ET6}
\]

For equal states at a clipping boundary we take the diagonal entry to be
zero. Away from the boundaries the derivative is 1 on the active PPC or
PUL segment, 10 on the active TRN segment, and zero on the flat
segments.

\textbf{Lemma ET1.} \(\delta r_t=D_t\delta x_t\).

At unequal states, the secant times the state difference equals the rate
difference by construction. At equal states, both differences vanish.
The relation applies to finite differences across thresholds and across
saturated regions. DeepLIFT uses a related divided-difference rule to
propagate activity differences (Shrikumar et al., 2017).

The diagonal entries of \(D_t\) can thus be interpreted as effective state-to-rate gains for the particular trajectory pair. They depend on the activation regimes occupied by the two states and can change over time as the network responds to sensory or recurrent input.

Having expressed the firing-rate difference as a transformation of the state difference, we can now substitute this relationship into the difference between the two network equations. This yields an exact description of how the existing historical state difference and the arriving historical input contribute to the next state.

Subtracting Equation ET5 gives the exact finite recursion

\[
\delta x_{t+1}=M_t\delta x_t+AE_C\chi_t\delta\widehat b_t,
\qquad M_t=I-A+A\widetilde W_tD_t.
\tag{ET7}
\]

The matrix \(M_t\) combines passive persistence with the recurrent transformation of the history-dependent difference at time \(t\). During replay, the endogenous contribution of the manipulated PUL-to-PPC connection is replaced by the difference between the two externally supplied donor currents.

Because both branches receive identical future input
(\(\delta u_t = 0\)), the external input term cancels (otherwise an
additional term \(AW_{\mathrm{in}}\delta u_t\) enters). To follow these contributions over multiple time steps, we define a propagation operator that composes the effective state transformations along the realized trajectory.

With \(\Phi(t,s)=M_{t-1}\cdots M_s\) for \(t>s\), and \(\Phi(s,s)=I\),
repeated substitution yields

\[
\delta x_t=\Phi(t,t_0)\delta x_{t_0}
+\sum_{k=t_0}^{t-1}\Phi(t,k+1)AE_C\chi_k\delta\widehat b_k.
\tag{ET8}
\]

Equation ET8 separates the later state difference into two contributions. The first is the part originating from the historical difference already present in the initial state. The second is the accumulated contribution of the differences introduced by the arriving input during replay. Each contribution is transformed by the intervening circuit dynamics before reaching the later time point.

In the complete-state-averaged condition, the initial state difference is exactly zero. Consequently, the entire subsequent state difference can be expressed in terms of the differences introduced by the replayed input and their propagation through the circuit. This provides a mathematical characterization of source-driven historical reconstruction.

Equation ET8 provides an exact decomposition along each realized trajectory pair. Its activation secants incorporate the neuronal activation regimes and threshold crossings encountered along those trajectories. Different interventions generate their corresponding trajectories and activation secants.

\subsubsection{Amplitude and spatial
pattern}\label{theory-amplitude-and-spatial-pattern}

The historical input difference can arise from changes in the overall strength of the arriving current, changes in how that current is distributed among target neurons, or a combination of the two. To distinguish these features, we represent each nonnegative current vector as the product of a scalar amplitude and a normalized spatial pattern.

For a positive source, use the physical mean-current normalization

\[
g^{\sigma}=\frac1{400}\mathbf1^\top b^{\sigma},\qquad
q^{\sigma}=b^{\sigma}/g^{\sigma}.
\tag{ET9}
\]

Then \(b=gq\) and \(q\) has unit mean. For the zero current, take
\(g=q=0\). Amplitude here is the mean delivered current. Recurrent gain
refers to the circuit transformation acting on a state difference.

The actual delivered pair obeys the symmetric product identity

\[
\delta\widehat b=\overline{\widehat q}\,\delta\widehat g
+\overline{\widehat g}\,\delta\widehat q.
\tag{ET10}
\]

The symmetric product identity separates the delivered current difference into an amplitude contribution, evaluated at the average spatial pattern, and a spatial-pattern contribution, evaluated at the average amplitude. Their sum is exactly the difference between the two delivered current vectors. Pattern exchange
\(g^{\sigma}q^{-\sigma}\) keeps the own donor's total current at each
time, while amplitude exchange \(g^{-\sigma}q^{\sigma}\) keeps its
normalized spatial pattern. These coordinate exchanges test whether
amplitude or spatial pattern alone suffices to drive the later
historical response; because pure exchanges can shift the paired mean,
the complementary current-matching experiments in ET3 hold the pair mean
fixed.

Figure 2-2B uses the Euclidean norm as an alternative amplitude
coordinate, \(g_2^{\sigma}=\|b^{\sigma}\|_2\), with pattern
\(q_2^{\sigma}=b^{\sigma}/g_2^{\sigma}\). Pattern exchange delivers
\(g_2^{\sigma}q_2^{-\sigma}\) and preserves the norm of the own current.
Amplitude exchange delivers \(g_2^{-\sigma}q_2^{\sigma}\). The recorded
donors have positive amplitude under both normalizations.

\subsubsection{Four signed components}\label{theory-four-signed-components}

Combining the propagation equation with the amplitude--pattern decomposition allows us to trace four sources of the realized historical difference: the initial PPC state, the initial state of PUL and TRN, the difference in arriving input amplitude, and the difference in its spatial distribution.

Let \(\Pi_C=E_CE_C^\top\). Initialize \(z^C=\Pi_C\delta x_{t_0}\),
\(z^O=(I-\Pi_C)\delta x_{t_0}\), and \(z^g=z^q=0\), where \(O\) includes
PUL and TRN. Propagate

\[
\begin{aligned}
z^C_{t+1}&=M_tz^C_t,&z^O_{t+1}&=M_tz^O_t,\\
z^g_{t+1}&=M_tz^g_t+AE_C\chi_t\overline{\widehat q}_t\delta\widehat g_t,\\
z^q_{t+1}&=M_tz^q_t+AE_C\chi_t\overline{\widehat g}_t\delta\widehat q_t.
\end{aligned}\tag{ET11}
\]

Equations ET7 and ET10 prove by induction that
\(\sum_jz_t^j=\delta x_t\), and hence

\[
H_{\mathcal R}=\sum_j\frac1{2L}\sum_{t\in\mathcal R}w^\top D_tz_t^j.
\tag{ET12}
\]

All four components follow the same trajectory-dependent propagator, with the endogenous PUL-to-PPC projection removed during replay and restored afterward. Their signed contributions sum exactly to the observed historical contrast, retaining the possibility of reinforcement and cancellation between components. The allocation among individual components depends on the chosen amplitude normalization and equal-state secant convention, while their sum remains invariant.

Figure 6-2C applies this four-component decomposition across the nine
trained circuits (500-ms replay, 64 paired contexts per circuit).

\subsection{ET3. Geometry of history-dependent input under current and source constraints}\label{theory-et3.-current-matching-constraints-and-source-dimensionality}

To investigate the function of the spatial pattern of historical input, we require a way to vary the difference between two arriving currents while controlling their shared background. Representing the pair in terms of its mean and difference provides a convenient coordinate system for constructing these controlled alternatives.

For natural pair mean \(\mu\) and difference \(d\), a general
mean-preserving replay has

\[
\widehat b^{\pm}=\mu\pm\widetilde d/2,
\qquad |\widetilde d_i|\leq2\mu_i.
\tag{ET13}
\]

Geometrically, the nonnegativity condition restricts each component of the historical difference to an interval determined by the corresponding component of the common current. A neuron receiving little shared background input therefore permits only a small difference between the two nonnegative historical currents.

A unit vector \(v\), estimated independently from the natural current differences, defines a candidate low-dimensional carrier of historical information. The scalar coefficient \(a=v^\top d\) measures how strongly this direction is expressed in a particular current difference. Restricting the replacement difference to \(av\) preserves a rank-one spatial structure across times and contexts.

At a fixed pair mean, however, even a single component with little available current can restrict the magnitude of this rank-one difference. Global attenuation preserves the direction exactly but may substantially reduce its magnitude. Coordinatewise clipping allows a larger feasible signal by adjusting each component separately, at the cost of potentially changing the realized direction. Thus, strict spatial dimensionality and the attainable magnitude of the reduced signal are distinct geometric properties. In the clipped construction, \(\widetilde d = \mathrm{clip}(av, -2\mu, 2\mu)\).

The preceding constraints characterize currents that could be supplied directly to the receiving population while preserving nonnegativity. The trained thalamocortical projection imposes a further restriction: the delivered current must be generated from the bounded firing rates of the PUL source neurons.

The trained PUL population generates the current set
\(\mathcal B=\{W_{CT}r:0\leq r_i\leq20\}\), whose span has dimension at
most 70 in 400 target coordinates. Own, opposite, and pair-mean replay
belong to this convex set.

This set is the linear image of the source firing-rate box. It describes the subset of the receiving-current space that can be accessed through the original PUL-to-PPC connections. The receiving circuit can exhibit functional sensitivity across a broader range of input directions than the set accessible through the original source projection. Within that accessible set, the attainable magnitude of each direction depends on the source firing-rate constraints.

Difference-energy matching rescales the scalar projection prior to
clipping so that \(\|\widetilde d\|_2 = \|d\|_2\). Matching separates
the contribution of the amplitude of the arriving difference from its
spatial form. The common waveform \(\mu(t)\) can vary across contexts in
many dimensions, while the signed source message is reduced to one
scalar coordinate.

Matching the shared current, the total input, and the differential-current energy imposes different constraints on the historical pair. The following identity makes the distinction between paired and individual-current energy explicit.

\[
\|\mu\pm d/2\|^2=\|\mu\|^2+\|d\|^2/4\pm\mu^Td.
\tag{ET14}
\]

At a fixed pair mean, preserving the norm of the current difference also preserves the combined squared current norm of the two histories. The squared current norm of each individual history additionally depends on the inner product \(\mu^\top d\). Preserving \(\mathbf 1^\top d\) separately within each cortical target class fixes the total current delivered to that class in each history. These constraints allow the spatial distribution of historical input to vary while controlling complementary measures of input strength. The model's 280 excitatory and 120
inhibitory PPC neurons are interspersed across the coordinates, and both
classes receive nonnegative PUL input.

To redistribute inputs, rotate triplets within each recipient class
about their all-ones axis. Such rotations preserve class sums and
energy. Applying a rotation only inside the box in Equation ET13
maintains nonnegative currents. The same triplets and angles are
proposed across contexts, and the local pair mean and difference
determine which rotations are feasible. Figure 5-1 relates the
displacement achieved by each map to its historical effect.

Together, these constructions allow the function of historical input to be examined at several levels: its low-dimensional structure, its magnitude, its distribution across cortical neurons, and its realizability through the trained source projection.

\subsection{ET4. Relating retained neural differences to the decision output}\label{theory-et4.-retained-separation-and-readout-expression}

The neural representation of a history-dependent difference and its influence on a decision are related but separable quantities. A difference between two population activity vectors may remain large even when their projections onto the decision readout are small. In addition, projections that change sign over time can cancel when averaged across the response window. To quantify these relationships, we separate the historical output contrast into the magnitude of the retained firing-rate difference, its alignment with the readout, and the consistency of its projection over time.

Define the response-window rate separation, RMS readout projection, and
temporal consistency by

\[
R_x = \sqrt{\langle \|\delta r_t\|^2 \rangle}, \qquad a_R = \frac{\sqrt{\langle \delta y_t^2 \rangle}}{\|w\| R_x}, \qquad c = \frac{\langle \delta y_t \rangle}{\sqrt{\langle \delta y_t^2 \rangle}},
\tag{ET15}
\] with \(c_R=|c|\). Angle brackets denote a time average within one
context and condition. Set ratios with zero denominators to zero.
Cauchy--Schwarz at each time and across time gives \(0\leq a_R\leq1\),
\(|c|\leq1\), and

\[
\boxed{2H=\|w\|\,R_x\,a_R\,c.}\tag{ET16}
\]

The quantity \(R_x\) measures the overall separation between the two firing-rate trajectories across the response window. The factor \(a_R\) measures how effectively this population difference projects onto the fixed decision readout, relative to its magnitude. The signed factor \(c\) measures the consistency of this projection over time, including its net direction. Equation ET16 shows that the historical contrast is exactly determined by the product of these factors and the fixed readout norm.

The decomposition therefore distinguishes three ways in which the behavioral influence of history can change. The underlying population activity difference may become smaller, its direction may become less aligned with the readout, or its projections at different times may cancel more strongly. The final two mechanisms allow substantial history-dependent neural activity to persist while producing a small time-averaged decision difference.

Taking magnitudes yields
\(2|H| = \|w\| R_x a_R c_R\) (Equation 11 of the main text). The net
readout alignment \(a_R c\) combines the last two factors, so its
magnitude can be small due to weak instantaneous readout projection,
temporal cancellation across the window, or both.

\(R_x\) is the temporal RMS of the Euclidean rate-difference norm for
the full population. The internal state snapshot \(\|\delta x\|\)
measures an instantaneous state difference. Ratios are formed within
contexts before averaging. This factorization relates the behavioral expression of retained history to the organization of population activity along output-potent and output-weak directions (Kaufman et al., 2014) and to the different functional states of retained information in working memory (Muhle-Karbe et al., 2021).

\subsection{ET5. Nonlinear activation can regulate historical expression at a fixed retained memory}\label{theory-et5.-saturation-supplies-a-concrete-expression-mechanism}

The preceding factorization establishes how a large retained activity difference can coexist with a small behavioral contrast. We next illustrate a simple mechanism that can generate such evidence-dependent expression. The key idea is that a nonlinear transformation can become relatively insensitive to historical information when driven strongly by current sensory evidence, while recovering that sensitivity as the sensory drive decreases.

Two saturating units illustrate how replacing the stored history can
reverse the later response while a strong current stimulus still
receives the correct choice. Set \(h = \pm 0.85\) and \(s = \pm 2\), and
use

\[
r_+=\operatorname{clip}(h+s,0,1),\qquad
r_-=\operatorname{clip}(-h-s,0,1),\qquad y=r_+-r_-.
\tag{ET17}
\]

After the current response, update \(h\leftarrow h+0.025s\) and set
\(s = 0\) for the later ambiguous response. Replacing \(h\) with its
opposite before the current judgment leaves current outputs at the
correct \(y=\pm1\). Later outputs follow the written history with
magnitudes 0.80 or 0.90. All eight history, source, and stimulus
combinations reach the decision commitment threshold of 0.75 (Figure
6-1B).

In this example, the strong current stimulus places the readout in the same saturated regime for both historical conditions, resulting in an accurate and history-insensitive current decision. After the sensory input is removed, the retained historical variable again determines the output. The separation between the two decisions is therefore produced by the changing influence of a persistent memory on a fixed nonlinear transformation.

To examine this relationship analytically, we consider a continuous-time model in which historical information and current sensory evidence evolve on different timescales. A slowly relaxing memory variable retains information about previous stimuli, while a faster sensory variable responds to current external input. Both variables contribute to the same fixed saturating readout. The equations are

\[
\tau_m\dot m=-m,\qquad \tau_z\dot z=-z+u,
\qquad y=\tanh(gz+km),\qquad \tau_m\gg\tau_z.
\tag{ET18}
\]

At a given time, let \(m^\pm=\pm a\). The historical contrast is then

\[
H(z)=\frac{\tanh(gz+ka)-\tanh(gz-ka)}2
=\frac{\sinh(2ka)}{\cosh(2gz)+\cosh(2ka)}.
\tag{ET19}
\]

At zero current evidence, the historical contrast is determined by the retained memory and equals \(\tanh(ka)\). As the magnitude of current evidence increases, the two historical responses are driven toward the same saturated region of the readout, and their output difference decreases. For fixed \(ka>0\) and nonzero sensory gain, the contrast is symmetric in \(z\) and decreases with \(|z|\). At fixed stored-state separation \(2a\), the historical contrast approaches \(2\sinh(2ka)e^{-2|gz|}\) as the magnitude of sensory evidence increases.

Because the sensory variable relaxes more rapidly than the memory variable, the influence of history can recover as the sensory drive subsides while the underlying memory remains largely present. These minimal models demonstrate a sufficient mechanism for separating the retention of historical information from its behavioral expression across changing sensory conditions.

\subsection{ET6. Identical decay properties can coexist with different reconstruction dynamics}\label{theory-et6.-stability-and-conditional-decay-leave-reconstruction-geometry-free}

The persistence of neural representations reflects the timescales of the dynamical modes through which information is maintained. The recovery of a representation after a localized perturbation also depends on how these modes are distributed across interacting neural populations and how they contribute to the measured output.

This relationship raises a geometric question about reconstruction: how much freedom remains in the recovery dynamics of circuits whose stability spectra and conditional decay properties are identical? We first establish exact identities for sequential erasure and then construct stable reciprocal systems in which the decay properties are fixed while the reconstruction dynamics vary.

\subsubsection{Exact affine erasure
identities}\label{theory-exact-affine-erasure-identities}

Consider a circuit partitioned into two interacting populations. We describe an erasure as a controlled operation that moves the states in one population toward a common reference, thereby reducing the difference between two historical conditions within that population. By applying such operations at two successive times, we can examine how information is redistributed between the populations during the intervening dynamics.

Let \(P_C+P_T=I\) be complementary coordinate projectors. Erasure of
block \(i\) with strength \(\rho\) maps \(x\) to
\((I-\rho P_i)x+\rho P_i r\), using the same reset value \(r\) for both
histories. The reset and any shared affine forcing cancel in the
difference. The erasure acts on the contrast through \(R_i=I-\rho P_i\).
The identities below apply when the histories share a linear propagator
between erasures.

For common linear propagator \(U(t,s)\), initial contrast \(d_0\), final
linear readout \(\ell\), and erasures at \(t_1<t_2\), define

\[
F_{ij}=\ell^TU(T,t_2)R_jU(t_2,t_1)R_iU(t_1,t_0)d_0.
\]

The first erasure changes the distribution of the historical difference across the circuit. During the subsequent propagation interval, interactions between the populations can reconstruct part of the erased difference. The second erasure probes how this reconstructed information has been redistributed at the later time.

Writing \(\kappa=2-\rho\), the identity \(R_C+R_T=\kappa I\) gives

\[
\sum_{i,j}F_{ij}=\kappa^2F_0,\qquad
F_{CT}-F_{TC}=\kappa[F_C(t_1)-F_C(t_2)].
\tag{ET20}
\]

Here \(F_i(t)\) is the single-erasure readout and \(F_0\) is the intact
readout. The indices name the erased block. The result applies to a
signed linear observation or a fixed linear time average over a complete
two-block partition.

The first identity constrains the sum of the responses across all four two-erasure sequences. The second relates the asymmetry between the two cross-population erasure orders to the difference between single-erasure effects measured at different times. These identities establish how the timing and location of localized perturbations interact with the propagation of retained information.

\subsubsection{A constrained isospectral
construction}\label{theory-a-constrained-isospectral-construction}

The preceding identities describe how erasure interacts with a given propagation operator. We now ask which aspects of this operator can be inferred from its stability spectrum and from decay measurements obtained when one population is isolated. To answer this question, we construct a family of simple reciprocal systems in which the within-population decay properties are determined by fixed marginal connection strengths, while the detailed organization of the cross-population coupling can vary.

Partition \(N=n+m\) coordinates into two blocks and take a symmetric
generator

\[
Q=\begin{pmatrix}-\operatorname{diag}(a)&B\\B^T&-\operatorname{diag}(d)\end{pmatrix},
\quad B\mathbf1=a,\quad B^T\mathbf1=d,\quad
M=Q-\epsilon I,\quad \epsilon>0.
\tag{ET21}
\]

The entries of \(B\) are strictly positive. Row and column sums of \(Q\)
vanish, so the intact uniform mode decays at \(-\epsilon\). The symmetric structure ensures reciprocal coupling between the two populations, and the shift by \(-\epsilon I\) makes the complete system strictly stable. The row- and column-sum constraints allow the overall coupling strengths of individual units to be held fixed while the allocation of connections between specific pairs of units is varied.

Holding the opposite block at a common reference gives intact-decay-normalized clamp
curves

\[
G_C(L)=N^{-1}\sum_i e^{-a_iL},\qquad
G_T(L)=N^{-1}\sum_j e^{-d_jL}.
\tag{ET22}
\]

Both conditional decay curves and their underlying block generators are determined by the row and column sums of the reciprocal coupling matrix. Fixing these sums therefore preserves the complete conditional decay dynamics while allowing the detailed organization of cross-population coupling to vary.

Sequential erasure probes a different aspect of the dynamics. After the first population has been erased, information initially retained in the other population can return through the cross-population connections. Erasing the second population at a later time then isolates the information that has been reconstructed in the first population during the intervening interval.

With full C erasure first and T erasure second, the
normalized recovery is

\[
H_{CT}(\tau)=N^{-1}\mathbf1_n^T[e^{Q\tau}]_{CT}\mathbf1_m.
\tag{ET23}
\]

Writing \(S=\sum_i a_i=\sum_jd_j\), direct multiplication gives

\[
\begin{aligned}
H'(0)&=S/N,\\
H''(0)&=-(\|a\|^2+\|d\|^2)/N,\\
H'''(0)&=[\sum_i a_i^3+\sum_jd_j^3+2a^TBd]/N.
\end{aligned}\tag{ET24}
\]

The fixed marginal connection strengths determine the initial slope and curvature of the recovery function. The third derivative additionally depends on the cross-coupling moment \(a^\top Bd\), which is sensitive to the allocation of reciprocal connections between individual units. Varying this moment while preserving the marginal strengths changes the higher-order recovery dynamics. The following construction establishes that this geometric freedom persists under the additional constraint of an identical complete eigenvalue spectrum.

\textbf{Theorem ET2.} There is a smooth local family of symmetric, strictly stable systems with strictly positive reciprocal cross-block couplings, identical complete spectra, identical conditional block generators and identical all-duration curves ET22, but differing recovery functions ET23.

The theorem identifies a family of reconstruction dynamics compatible with the same complete stability spectrum and the same conditional decay dynamics in both populations. The proof constructs a local family in which these measured properties remain fixed while a coefficient of the recovery function varies.

\textbf{Proof.} Work first with dimensionless rates

\[
a_0=(10,20,30,40)^T,\quad d_0=(15,20,25,40)^T,\quad
B_0=\begin{pmatrix}2&2&3&3\\3&5&5&7\\4&6&8&12\\6&7&9&18\end{pmatrix}.
\]

The nine matrices \(V_{ij}=(e_i-e_4)(e_j-e_4)^T\), \(i,j\leq3\),
preserve row and column margins. Each of these basis perturbations changes the internal allocation of the cross-population coupling while preserving all row and column sums. They therefore provide local coordinates for varying the coupling structure without changing the conditional block generators.

Set
\(B(\theta)=B_0+\sum\theta_{ij}V_{ij}\), and form
\(\widetilde Q(\theta)\) by Equation ET21. Within this margin-preserving parameterization, we additionally require the complete eigenvalue spectrum to remain fixed. We impose this requirement through the spectral moments of the generator.

Impose the six spectral trace
constraints
\(\operatorname{tr}(\tilde{Q}(\theta)^k) = \operatorname{tr}(\tilde{Q}(0)^k)\)
for \(k = 2, \dots, 7\). Their Jacobian \(J\) has rank 6 at
\(\theta = 0\). The appended gradient is

\[
L = \nabla_\theta \bigl(2 a_0^\top B(\theta) d_0\bigr)\big|_{\theta=0} =
(1500, 1200, 900, 1000, 800, 600, 500, 400, 300).
\]

Appending this gradient raises the rank to 7, as verified by the nonzero
minors

\[
\begin{aligned}
\det J_{:, 1:6} &= 150676075875044966400000,\\
\det [J;L]_{:, 1:7} &= 353574528412802457600000000.
\end{aligned}
\]

The rank increase shows that the gradient of the cubic recovery moment is independent of the gradients of the imposed spectral constraints. Consequently, there exists a local direction that preserves all six spectral moments while changing the cubic coefficient of the recovery function.

The implicit-function theorem therefore supplies a three-dimensional
local constraint manifold on which the cubic moment varies. Fixed trace
and traces of powers 2 through 7 determine the first seven
characteristic-polynomial coefficients by Newton identities; the last
coefficient is zero because row sums vanish. Hence the full
characteristic polynomial is fixed. Scaling by
\(10^{-3}\,\mathrm{ms}^{-1}\) and shifting by \(-\epsilon I\) preserves
these equalities and gives strict stability. Positivity of the cross
edges persists in a neighborhood. Equation ET24 then guarantees
different analytic recovery functions. \(\square\)

The construction thus preserves the decay rates of all dynamical modes, the conditional generators of both populations, and their full conditional decay curves, while changing the way information is transferred and reconstructed following sequential erasure.

The eigenbasis provides a direct geometric interpretation of why the recovery functions can differ despite identical spectra. If
\(Qu_r=-\mu_ru_r\), \(u_0=\mathbf1/\sqrt N\), and \(c=P_C\mathbf1\),
then

\[
H_{CT}(\tau)=\frac{nm}{N^2}-\sum_{r=1}^{N-1}\omega_r e^{-\mu_r\tau},
\qquad \omega_r=\frac{(u_r^Tc)^2}{N}\geq0.
\tag{ET25}
\]

In this expression, the eigenvalues \(\mu_r\) determine the rates at which the dynamical modes decay. The coefficients \(\omega_r\), in contrast, determine how strongly those modes are accessed by the particular population partition used in the erasure experiment. Fixing the spectrum therefore fixes the available decay rates while leaving some freedom in their contributions to the measured recovery.

The fixed marginal coupling strengths additionally constrain the zeroth, first, and second moments of these mode weights. The constructed family changes higher moments by redistributing the reciprocal cross-population connections. Because the generators are symmetric and the complete system satisfies \(\|e^{M\tau}\|_2=e^{-\epsilon\tau}\) for \(\tau\ge0\), the differences in reconstruction arise within a nonamplifying stable system.

Related phenomena occur in isospectral systems with different heat responses (van den Berg et al., 2014). The present construction establishes that this geometric freedom can persist even when both conditional generators and their complete clamp curves are also held fixed in a positive, symmetric bipartite system.

\Needspace{10\baselineskip}

\subsubsection{Finite examples illustrating unconstrained recovery
geometry}\label{theory-finite-examples-illustrating-unconstrained-recovery-geometry}

{\def\LTcaptype{none} 
\begin{longtable}[]{@{}
  >{\raggedright\arraybackslash}p{(\linewidth - 4\tabcolsep) * \real{0.3000}}
  >{\raggedright\arraybackslash}p{(\linewidth - 4\tabcolsep) * \real{0.3000}}
  >{\raggedleft\arraybackslash}p{(\linewidth - 4\tabcolsep) * \real{0.4000}}@{}}
\toprule\noalign{}
\begin{minipage}[b]{\linewidth}\raggedright
Construction
\end{minipage} & \begin{minipage}[b]{\linewidth}\raggedright
Quantities matched
\end{minipage} & \begin{minipage}[b]{\linewidth}\raggedleft
Peak recovery difference
\end{minipage} \\
\midrule\noalign{}
\endhead
\bottomrule\noalign{}
\endlastfoot
Four-state reciprocal pair & Intact slow mode and both conditional-clamp
curves & 0.028 at 151 ms (14.7\% relative to the smaller response) \\
Eight-state isospectral symmetric pair & Complete spectrum, both
conditional generators and both clamp curves & \(5.4 \times 10^{-4}\) at
111.5 ms (0.23\% relative to the smaller response) \\
\end{longtable}
}

The four-state example illustrates the recovery differences that can arise when the slowest intact mode and both conditional decay curves are matched, while the remaining eigenvalues differ by up to \(4.8\times10^{-3}\,\mathrm{ms}^{-1}\). The eight-state example provides a more restrictive numerical realization in which the complete eigenvalue spectrum and both conditional block generators are identical. Its distinct recovery trajectory illustrates the existence result established by Theorem ET2.

Together, these constructions distinguish the timescales of neural dynamics from the geometry through which those dynamics are accessed by population-specific perturbations. This geometric distinction provides a theoretical motivation for manipulating the information transmitted through particular pathways when investigating neural reconstruction.

\subsection{ET7. Incoming synaptic drive can reconstruct history-dependent intrinsic neuronal states}\label{theory-et7.-a-neuronal-recovery-variable-receives-historical-drive}

The preceding analyses describe how arriving input can reconstruct history-dependent differences in a rate-based neural circuit. A related principle can apply to intrinsic neuronal variables in a more detailed conductance-based model. For example, voltage-dependent recovery variables can retain information about earlier activity, while subsequent synaptic input influences their evolution through changes in membrane voltage and channel kinetics. We therefore examine how differences in such intrinsic variables can be reconstructed by history-dependent driving signals.

For a voltage-dependent gate \(h_s\) in history \(s\), write
\(\dot h_s=\alpha_s-k_sh_s\), with \(k_s>0\). Comparing the gate variables of two histories allows us to distinguish the relaxation of an existing intrinsic state difference from the contribution of differences in the voltage-dependent driving terms.

Arithmetic means and
differences give the exact identity

\[
\delta\dot h=-\bar k\,\delta h+\delta\alpha-\bar h\,\delta k.
\tag{ET26}
\]

The first term in Equation ET26 describes relaxation of the existing difference between the two gate variables. The remaining terms describe how differences in the driving parameters can create or modify this gate difference. Following a common reset, the initial gate difference is zero, but subsequent history-dependent differences in the drivers can reconstruct it.

When these driving trajectories are prescribed, the gate equations define scalar conditional dynamics that can be solved using integrating factors. In a recurrent conductance-based circuit, however, changing an intrinsic state can also modify membrane voltage, synaptic activity, and subsequent feedback. A conditional prediction must therefore specify the driving trajectories on which it depends.

We used the awake stimulation configuration of Russo et al. (2025), in which 500 cortical excitatory neurons and 100 cortical inhibitory neurons interact with 100 thalamic relay cells and 100 reticular neurons. Cortical excitation reaches relay cells through AMPA and AMPA-D2 conductances, while reticular inhibition reaches them through GABA\textsubscript{A} and GABA\textsubscript{B} conductances. The measured gate, \(h_T\), is the inactivation variable of the relay cell's low-threshold T-type calcium current. The conditional forecast integrates 12 intrinsic state variables for each thalamic relay cell: membrane voltage \(V\), intracellular calcium concentration \(\mathrm{Ca}\), and the remaining ten variables describing the channel and intracellular dynamics of the original conductance model, \((m_T,h_T,o_1,p_1,o_2,m_{\mathrm{Na}},h_{\mathrm{Na}},n_K,m_A,h_A)\). We used the original model equations (Russo et al., 2025; Krishnan et al., 2016), a time step of 0.02 ms, and the recorded incoming synaptic conductances.

At 75 ms after cortical stimulation, each relay cell received the 12 intrinsic coordinates from the opposite history. In the prescribed-source condition, its own-history cortical AMPA and AMPA-D2 conductances were then replayed until 125 ms. Reticular input and the rest of the network continued to evolve. For each simulation replicate, we measured the endpoint projection as \(q(125)=d^\top d_N/\|d_N\|_2^2\), where \(d\) is the 100-cell \(h_T\) contrast at 125 ms and \(d_N\) is the corresponding intact natural contrast. This normalized projection measures the component of the reconstructed gate-difference vector along the natural historical contrast. A value of one corresponds to recovering the full natural contrast along that direction, while a value of zero corresponds to a vanishing projection onto it. With endogenous cortical feedback, the mean endpoint projection was 0.56. Prescribing the cortical input changed it to 0.41, closely matching the conditional prediction of 0.41. The comparison illustrates how feedback generated by the interacting circuit contributes to the reconstruction of intrinsic thalamic states.

Separate tests used half-amplitude cortical history differences and a later stimulation onset. Under these prescribed inputs, median normalized RMSE for the \(h_T\) contrast from 76 to 125 ms was \(2.6\times10^{-4}\) at half source amplitude and \(8.1\times10^{-5}\) at the later onset. Normalization uses the RMS magnitude of the observed paired gate difference over cells and time. These simulations illustrate source-driven reconstruction at the level of an intrinsic neuronal variable. The trained decision circuits examine how the same general relationship between arriving input and reconstructed states can support subsequent history-guided behavior. Together, the two models provide complementary descriptions of history-dependent reconstruction at cellular and circuit levels.

\subsection{ET8. Constrained source transformations}\label{theory-et8.-constrained-source-transformations}

\subsubsection{Reachable source geometry at fixed mean (Figure
4-3)}\label{theory-reachable-source-geometry-at-fixed-mean-figure-4-3}

A pattern of arriving current that is effective at influencing the cortical circuit may be difficult for the thalamic population to generate. To characterize this distinction, we derive the set of receiving-current differences that can be produced by the trained projection when the source firing rates are bounded and their pair mean is preserved.

Let \(W_{CT} \in \mathbb{R}^{400 \times 70}\) be the trained PUL-to-PPC
projection and \(r^{\pm} \in [0, 20]^{70}\) a pair of admissible PUL
firing-rate vectors, with pair mean \(\mu = (r^+ + r^-)/2\) and
difference \(d = r^+ - r^-\). At fixed common source activity \(\mu\),
the set of reachable historical current differences in PPC is the
zonotope

\[\mathcal{Z}(\mu) = \bigl\{\, W_{CT} d' : |d'_i| \le 2\min(\mu_i, 20 - \mu_i)\ \text{for all } i \,\bigr\},\]

which automatically preserves the receiving pair mean. Let
\(S_{\mathrm{EI}} \in \mathbb{R}^{2 \times 70}\) sum the rows of
\(W_{CT}\) separately over excitatory and inhibitory PPC targets,
and let \(G = W_{CT}^\top W_{CT}\). The feasible source differences form a box whose coordinatewise bounds depend on the common source activity. Applying the trained projection maps this box into a zonotope in the receiving-current space. This geometric object describes the historical current differences that are reachable from the specified source mean.

When a PUL neuron operates close to either firing-rate bound, its available difference range is small. The source population's operating state therefore influences the range of alternative historical signals that the pathway can produce.

The condition \(S_{\mathrm{EI}}z=0\) identifies directions in source space that leave the aggregate input differences to excitatory and inhibitory PPC targets unchanged. The quadratic form \(d^\top Gd=\|W_{CT}d\|_2^2\) measures the energy of the source firing-rate difference after its transmission through the trained projection.

Starting from the natural source difference, we search along an aggregate-preserving direction for another point that produces the same receiving-current energy. Expanding the energy along this line yields a quadratic equation in the displacement magnitude.

For any direction
\(z \in \ker(S_{\mathrm{EI}})\) with \(z^\top G z > 0\), expanding the
receiving difference energy along \(d + \alpha z\) gives

\[\|W_{CT}(d + \alpha z)\|^2 - \|W_{CT} d\|^2 = 2\alpha\, d^\top G z + \alpha^2\, z^\top G z.\]

The energy-preserving displacement \(\alpha_* = -2(d^\top G z)/(z^\top G z)\) defines the linear reflection

\[d' = Q_z d, \qquad Q_z = I - 2\frac{z z^\top G}{z^\top G z},\]

which simultaneously preserves aggregate E/I input
(\(S_{\mathrm{EI}} d' = S_{\mathrm{EI}} d\)) and receiving difference
energy (\(d'^\top G d' = d^\top G d\)), and is feasible whenever \(d'\)
remains inside the rate-difference box. Candidate directions
\(z \in \ker(S_{\mathrm{EI}})\) are generated either by restricting Gaussian directions to source coordinates with available room for changes in the paired firing-rate difference and imposing the aggregate-input constraint or by taking the cross
product of the two rows of \(S_{\mathrm{EI}}\) within source triplets.
When \(d^\top Gz\neq0\), the reflection generates a distinct point on the same receiving-energy surface; when \(d^\top Gz=0\), it preserves the original source difference.

Direct rotations of the 400-dimensional receiving-current vector probe the sensitivity of the cortical circuit to a relatively broad range of nonnegative input patterns. Source-space reflections explore the smaller set of patterns that can be generated through the original trained projection at the specified PUL mean. Comparing these experiments distinguishes the receiving circuit's functional sensitivity from the flexibility of the source population in generating alternative signals.

\subsubsection{Temporal coherence and clipped source axes (Figures 4-5
and
4-6)}\label{theory-temporal-coherence-and-clipped-source-axes-figures-4-5-and-4-6}

Selecting different feasible source transformations at successive time steps can introduce rapid fluctuations into the historical signal. To preserve its temporal organization alongside aggregate target input and receiving-current energy, we considered two complementary transformations.

First, we smoothed the displacement from the natural source difference, \(d'(t)-d(t)\), over time. Because the displacements satisfy the linear aggregate-input constraint, temporal smoothing preserves this constraint. The energy-preserving intersection then restores the instantaneous receiving-current norm wherever the firing-rate bounds permit.

Second, we constructed transformations that remained fixed throughout the replay interval. Each accepted reflection contributes to a time-invariant composition of operators \(Q_z\). A reflection was accepted when it satisfied the rate bounds at every time step, increased the integrated receiving-current displacement, and preserved temporal roughness \(\mathcal{R}(d')=\sum_t\|d'(t+1)-d'(t)\|^2\) within \(1.25\mathcal R(d)\).

A related question concerns the maximum receiving-current magnitude attainable along a particular source direction. As the magnitude of a source difference increases, some coordinates reach their firing-rate bounds before others, changing the relationship between the scaling coefficient and the resulting cortical input.

Along a fixed source axis \(q\) with rate-difference bounds
\(c_i = 2\min(\mu_i, 20 - \mu_i)\), the clipped path
\(d(k) = \mathrm{clip}(kq, -c, c)\) (\(k \ge 0\)) is piecewise linear
between clipping breakpoints, \(d(k) = f + ku\). Its receiving energy
\(E(k) = \|W_{CT} f\|^2 + 2k(W_{CT} f)^\top (W_{CT} u) + k^2 \|W_{CT} u\|^2\)
is convex on each interval, so its global maximum occurs at a breakpoint
or full saturation. Continuity from \(E(0)=0\) ensures that every attainable target magnitude has a first crossing along the clipped path. The maximum over the clipping breakpoints and the fully saturated endpoint determines the attainable strength of the candidate direction. We therefore use the first target crossing when matching is feasible and the maximal attainable magnitude otherwise. The sign of the learned projection coefficient determines the orientation of the delivered historical difference over time.

\subsection{ET9. How receiver dynamics transform and propagate retained historical differences}\label{theory-et9.-receiver-perturbations-and-conditional-trajectory-propagation}

\subsubsection{Operating-point shifts and recurrent attenuation (Figures
6-3 to
6-5)}\label{theory-operating-point-shifts-and-recurrent-attenuation-figures-6-3-to-6-5}

After replay has reconstructed a historical state difference, its subsequent influence depends on two transformations. The neuronal activation functions convert internal state differences into firing-rate differences, and the recurrent connections propagate those firing-rate differences into future neural states. The following equations separate these two contributions.

In the unperturbed receiver following replay, the finite historical
state difference evolves as

\[\delta x_{t+1} = (I - A)\delta x_t + A W D_t \delta x_t, \qquad \delta y_t = w^\top D_t \delta x_t,\]

where \(A = \mathrm{diag}(\alpha_i)\) contains the integration factors
and \(D_t\) is the trajectory-dependent activation secant from Equation
ET6. The activation secant \(D_t\) specifies how the current state difference is expressed in neuronal firing rates. The recurrent weights \(W\) determine how that activity difference contributes to the next internal state. This structure motivates separate perturbations of the neuronal operating points and the recurrent propagation of historical differences.

The first perturbation moves both historical states by the same vector, \(x^\pm\mapsto x^\pm+s\). The initial difference between the states remains fixed, while their positions relative to the activation thresholds change. The corresponding activation secants can therefore change, altering how the retained state difference is expressed in firing rates and propagated through the circuit. Equal-norm random displacements provide a comparison for assessing the directional specificity of this effect.

The second perturbation regulates the recurrent propagation of history-dependent differences. At each time step, the pair-mean recurrent drive is preserved and the recurrent-drive difference is attenuated by a fraction \(\gamma\in(0,1]\). During the 100-ms intervention, the historical state difference evolves according to

\[\delta x_{t+1} = (I - A)\delta x_t + (1 - \gamma) A W D_t \delta x_t,\]

The attenuation parameter determines the contribution of recurrent propagation to the difference dynamics. A complete pause, corresponding to \(\gamma=1\), reduces these dynamics to passive leak. Endogenous recurrence resumes after the intervention. Perturbation strengths are selected using independent calibration histories to preserve strong-stimulus sensory responses.

\subsubsection{Factorized propagation forecasts (Figure
7-3)}\label{theory-factorized-propagation-forecasts-figure-7-3}

The receiver perturbations establish how changing activation or recurrent propagation affects the subsequent influence of history. A complementary question is whether these mechanisms can predict the evolution of history-dependent output in a new trajectory pair. To examine this, we construct conditional forecasts that begin with the measured post-stimulus state difference of an independent test history and estimate its subsequent evolution using gains obtained from reference histories.

These models cross two choices
of activation secant \(D_t\)---averaged over reference histories under
either the matched sensory evidence or the pre-sensory regime---with two
dynamics rules (full recurrence versus leak-only decay), keeping \(A\)
and \(w\) fixed.

The comparison between evidence-specific and pre-sensory gains tests the importance of the changes in neuronal activation induced by current sensory evidence. The comparison between recurrent and leak-only propagation, at the same activation gains, tests how recurrent interactions transform the retained state difference beyond its passive persistence or decay.

The evidence-specific mean template captures the output trajectory shared across histories encountering the same sensory evidence. Each propagation model additionally uses the measured initial state difference of the individual test pair, allowing its predictions to vary between histories. We quantify prediction accuracy relative to the template using

\[ 1-\frac{\mathrm{SSE}_{\mathrm{pred}}} {\mathrm{SSE}_{\mathrm{template}}}. \]

Positive values indicate improved predictive accuracy relative to the template, and negative values indicate larger prediction errors.

\subsection{References and data}\label{theory-references-and-data}

Together, ET1--ET9 connect the content of an arriving historical signal, the dynamics through which it reconstructs and propagates a neural state difference, and the conditions under which that difference influences the decision output. The framework separates source geometry, recurrent transformation, and behavioral expression while retaining their interdependence within the complete neural circuit.

ET6 provides the finite reciprocal-system construction and its spectral constraints. Extended Data documents the corresponding source and receiver comparisons in the trained decision circuits. ET7 uses the published conductance-based simulation of Russo et al. (2025), adapted from Krishnan et al. (2016), to examine source-driven reconstruction at the level of intrinsic neuronal dynamics.

\begin{itemize}
\item Russo S et al. (2025). Thalamic feedback shapes brain responses evoked by cortical stimulation in mice and humans. \emph{Nature Communications} 16:3627. \url{https://doi.org/10.1038/s41467-025-58717-2}
\item Krishnan GP et al. (2016). Cellular and neurochemical basis of sleep stages in the thalamocortical network. \emph{eLife} 5:e18607. \url{https://doi.org/10.7554/eLife.18607}
\tightlist
\item
  Hosford PS et al.~(2026). Control of representation updating by
  higher-order thalamus enables history-based decision-making.
  \emph{Neuron}. \url{https://doi.org/10.1016/j.neuron.2026.04.045}
\item
  Kaufman MT, Churchland MM, Ryu SI, Shenoy KV (2014). Cortical activity
  in the null space: permitting preparation without movement.
  \emph{Nature Neuroscience} 17:440--448.
  \url{https://doi.org/10.1038/nn.3643}
\item
  Muhle-Karbe PS, Myers NE, Stokes MG (2021). A hierarchy of functional
  states in working memory. \emph{Journal of Neuroscience}
  41:4461--4475. \url{https://doi.org/10.1523/JNEUROSCI.3104-20.2021}
\item
  Shrikumar A, Greenside P, Kundaje A (2017). Learning important
  features through propagating activation differences. \emph{Proceedings
  of Machine Learning Research} 70:3145--3153.
  \url{https://proceedings.mlr.press/v70/shrikumar17a.html}
\item
  van den Berg M, Dryden EB, Kappeler T (2014). Isospectrality and heat
  content. \emph{Bulletin of the London Mathematical Society}
  46:793--808. \url{https://doi.org/10.1112/blms/bdu035}
\end{itemize}

\end{document}